\documentclass[floatfix,aps,prb,reprint,showpacs,superscriptaddress]{revtex4-2}
\usepackage[left=1.5cm,right=1.5cm,top=2cm,bottom=2cm]{geometry}
\usepackage{xcolor}
\usepackage{graphicx,amsmath,wasysym}
\usepackage{dcolumn}
\usepackage{bm}
\newcommand{\blue}[1]{\textcolor{blue}{#1}}

\usepackage[colorlinks=true,linkcolor=blue,allcolors=blue]{hyperref}%
\usepackage{natbib}
\usepackage{geometry} 
\usepackage{ifthen}
\usepackage{slashed}
\usepackage{graphicx}
\usepackage{amssymb}
\usepackage{physics}
\usepackage{dsfont}
\usepackage{upgreek}
\usepackage{hyperref}
\usepackage{booktabs}       
\usepackage{array}

\usepackage{hhline}

\usepackage{tikz,xcolor,hyperref}

\definecolor{lime}{HTML}{A6CE39}
\DeclareRobustCommand{\orcidicon}{\hspace{-4pt}
\begin{tikzpicture}
\draw[lime, fill=lime] (0,0)
circle [radius=0.16]
node[white] {\hspace{0.1mm}{\fontfamily{qag}\selectfont \tiny ID}};
\draw[white, fill=white] (-0.07,0.1)
circle [radius=0.01];
\end{tikzpicture}
\hspace{-3.2mm}
}

\foreach \x in {A, ..., Z}{\expandafter\xdef\csname
orcid\x\endcsname{\noexpand\href{https://orcid.org/\csname orcidauthor\x\endcsname}
{\noexpand\orcidicon}}
}

\begin{document}
\title{Thermal Hall response of an abelian chiral spin liquid at finite temperatures}

\author{{Avijit Maity}\orcidA{}}
\affiliation{Department of Theoretical Physics, Tata Institute of
Fundamental Research, Homi Bhabha Road, Colaba, Mumbai 400005, India.}

\author{{Haoyu Guo}\orcidB{}}
\affiliation{Laboratory of Atomic and Solid State Physics, Cornell University,\\ 
142 Sciences Drive, Ithaca, New York 14853-2501, USA}

\author{{Subir Sachdev}\orcidC{}}
\affiliation{Department of Physics, Harvard University, Cambridge, Massachusetts 02138, USA}

\author{{Vikram Tripathi}\orcidD{}}
\affiliation{Department of Theoretical Physics, Tata Institute of
Fundamental Research, Homi Bhabha Road, Colaba, Mumbai 400005, India.}


\begin{abstract}
Thermal Hall transport has emerged as a valuable tool for probing the fractionalized excitations in chiral quantum spin liquids. Observing quantized thermal Hall response, expected at temperatures below the spectral gap, has been challenging and controversial. The finite temperature behavior, especially in the quantum critical regime above the spectral gap, can provide useful signatures of the underlying topological order. In this context, we study the spin-$1/2$ Heisenberg antiferromagnet on a kagome lattice that is believed to be a U$(1)$ Dirac spin liquid over a wide intermediate energy range. 
Scalar spin chirality perturbations turn this into a gapped abelian chiral spin liquid (CSL) with semionic topological order. 
Using a recently developed large-$N$ technique [Guo {\it et al.\/}, \href{https://doi.org/10.1103/PhysRevB.101.195126}{Phys. Rev. B {\bf 101}, 195126 (2020)}], we obtain explicit expressions for the thermal Hall conductivity $\kappa_{xy}$ at finite temperatures taking into account both matter and gauge fluctuations. At low temperatures below the spectral gap, the quantized thermal Hall response agrees with that expected from conformal field theory and gravitational anomaly arguments. Our main finding is that in a large temperature window spanning the spectral gap and the Curie temperature scales where quantum critical fluctuations dominate, $\kappa_{xy}/T$ obeys a power-law with logarithmic corrections. Our analysis also provides a route to understanding the thermal Hall response at higher temperatures in the quantum critical regime.
\end{abstract}

\maketitle

\section{Introduction}
Following Anderson's introduction of the resonating valence bond (RVB) state concept \cite{Anderson_RVB}, substantial progress has been made in unraveling the mysteries of quantum spin liquids (QSLs), an unconventional category of quantum phases that go beyond the traditional Landau symmetry-breaking framework \cite{Lee-Nagaosa-Wen_HighTc-review, Wen_ManyBodyQFT_Book, Savary_QSLReview_2016, Zhou_QSL_RevModPhys,SS_QPM}. These states not only hold conceptual importance and relevance in experimental contexts but also intertwine with intricate issues, spanning topological order, strongly coupled gauge theories, and high-temperature superconductivity. These highly entangled quantum states are defined by charge-neutral spinons-excitations which are coupled with emergent gauge fields and carry fractional values of the integer spin. 

The interest in chiral spin liquid (CSL), a subclass of QSLs that break time-reversal and parity symmetries, originated from the contributions of Kalmeyer and Laughlin \cite{Kalmeyer-Laughlin_RVB-FQH_PRL1987, Kalmeyer-Laughlin_SL-Heisenberg_PRB1989}. They suggested that frustrated magnets could host bosonic analogs of fractional quantum Hall states, giving rise to anyonic quasiparticles in the bulk and chiral gapless modes on the edge. These edge modes are described by $(1+1)$-dimensional conformal field theories (CFT).

Thermal transport measurements offer an attractive means to probe the nature of spin liquid states as heat currents in insulating quantum magnets at low temperatures can get significant contributions from emergent and charge-neutral modes \cite{zhang-Chen2023thermalReview}. In this context, the thermal Hall effect is sensitive to the nature of elementary excitations (e.g. fermions or bosons) as well as the topological properties associated with their respective energy bands, in particular the Berry curvature of elementary excitations \cite{Katsura-Lee_kxy_Magnet_PRL2010, Murakami_thermalHall_magnon}. For example, thermal Hall measurements have been used to reveal intrinsic topological properties of quantum Hall liquids such as the non-Abelian nature of the $\nu = 5/2$ fractional quantum Hall liquid \cite{Banerjee_thermalHall_Nature2018}. 

The distinctive feature expected from the abelian CSL state is the quantization of thermal Hall conductivity in the zero temperature limit in units of $(\pi k_B^2 /6\hbar)\, T$. This quantization arises from the chiral central charge of the chiral edge theory, which is an integer for abelian CSLs \cite{Wen:1990se,Kane_Fisher_1997,Cappelli_2002}. In contrast, for non-abelian CSLs, the chiral central charge can have a more complex expression \cite{DiFrancesco:1997nk,Guo_GaugeThermalHall_PRB2020}. The Kitaev model is an exactly solvable model with exact spin liquid ground-state \cite{Kitaev_Anyon_2006}; under the influence of a magnetic field, this model exhibits a gapped phase characterized by topological order, accommodating non-Abelian Ising anyons \cite{Nayak_NonAbleianAnyon_RMP2008}. The presence of a chiral Majorana fermion edge state suggests an anticipated response of half-quantized thermal Hall conductivity at $T \rightarrow 0$. Such a conductivity has been claimed to be observed in $\alpha$-RuCl$_3$ \cite{Kasahara_Kitaevkxy_Nature2018}, but this remains controversial \cite{Czajka-Ong_KitaevThermalHall_Nature2021, Czajka-Ong_KitaevThermalHall_Nature2022, Taillefer_RuCl3}. Another recent discovery of an enhanced thermal Hall conductivity in cuprates \cite{Grissonnanche2019, Taillefer_Chen} has sparked considerable interest in unraveling its underlying mechanism: this was initially proposed to be associated with topological spin excitations \cite{Samajdar_kxy_SchwingerBoson_PRB2019, Samajdar_Emhamcement_kxy_Nature2019}, but now appears more likely to be associated with phonons \cite{Kivelson_phonons,Guo_phonons}.

Detecting a quantized thermal Hall effect, expected theoretically at low temperatures, has been both technically challenging and widely debated in experiments. At higher temperatures, especially in the quantum critical regime above the spectral gap, the thermal Hall response shows different behavior, creating a chance to observe topological order in ways that don’t appear as clearly at lower temperatures. At these low temperatures, thermal Hall conductivity in topological phases is linked to chiral edge states and gravitational anomaly, which connects edge and bulk properties. However, this interpretation doesn’t apply at finite temperatures above 2+1-dimensional quantum critical points. In this work, we examine the thermal Hall effect of fermionic matter interacting with emergent gauge fields, representing a low-energy effective theory for the CSL phase stabilized on a kagome lattice.

Quantum Heisenberg antiferromagnets on two-dimensional kagome lattices offer an ideal platform for investigating exotic spin liquid states and have been used to model materials like Volborthite and Herbertsmithite \cite{Helton_Kagome_exp_PRL2007,Matsuda24}. Understanding the ground state of the antiferromagnetic spin-$1/2$ Heisenberg model on a kagome lattice remains a highly challenging problem in quantum magnetism. Numerous theoretical proposals have been suggested, with the most promising contenders currently being the gapped $\mathbb{Z}_2$ spin liquid \cite{NRSS91, Wen_MFT_GappedQSL_PRB1991, Subir_KagomeTriangle_Bosonic_PRB1992} and the gapless U$(1)$ Dirac spin liquid (DSL) \cite{Hastings_KagomeDSL_PRB2000}. Initial investigations using density matrix renormalization group (DMRG) techniques indicated the presence of a gapped $\mathbb{Z}_2$ spin liquid \cite{Yan_Huse_white_KagomeZ2_DMRG_Science2011, Depenbrock_KagomeZ_DMRG_PRL2012, Jiang_Balents_KagomeZ2_DMRG_Nature2012}. More recent work supports U$(1)$ DSL over a significant energy range. Notably, variational Monte Carlo \cite{Ran_KagomeDSL_PRL2007, Iqbal_Kagome_PRB2011, Iqbal_Kagome_PRB2013, Iqbal_Kagome_PRB2014, Iqbal_Kagome_PRB2015}, more recent DMRG \cite{He-Zaletel-Pollman_KagomeDSl_DMRG_PRX, Zhu_KagomeDSL_Entanglement_Science}, and two-dimensional tensor network studies \cite{Liao_KagomeDSL_Tensor_PRL, Jiang-Ran_KagomeDSL_Tensor_Scipost} are consistent with a DSL, but an eventual transition at the lowest energies to a $\mathbb{Z}_2$ spin liquid with a small gap is not ruled out \cite{Imai_kagome1, Imai_kagome2, Wen_kagome}. Direct experimental observation and characterization of QSLs remain challenging. However, their fractionalized excitations can be probed using conventional techniques. For instance, in the specific case of the U$(1)$ DSL on the spin-$1/2$ kagome lattice, Raman spectroscopy has been proposed as a tool to detect the spinon continuum as well as fluctuations of the emergent gauge field \cite{Raman_DSL_Ko_Lee_2010}.

 In kagome antiferromagnets, a CSL was stabilized by the addition of scalar chirality terms to the Hamiltonian \cite{Bauer_KagomeCSL_Nature2014} that explicitly breaks parity and time-reversal symmetry but not spin rotation symmetry. Utilizing DMRG, the authors of Ref.~\cite{Bauer_KagomeCSL_Nature2014} demonstrated that, particularly under a strong enough time-reversal symmetry breaking term, the ground state is a CSL state, sharing the same topological class (semionic topological order) as the $\nu=1/2$ Laughlin state. It is also known \cite{Gong-Sheng_CSL_Z2_Nature2014} that such CSL states can appear as a result of spontaneous symmetry breaking in kagome Heisenberg antiferromagnets with further-neighbor interactions. CSLs also have been discovered in diverse theoretical studies for different models in more recent investigations \cite{Yin-Chen-He_CSL_Kagome_PRL2014, Yin-Chen-He_XXZ_Kagome_PRL2015, Sheng_VMC_CSL_kagome_PRB2015, Gong-Balents-Sheng_GlobalPhase_PRB2015, Bieri-Lhuillier_Gapless_CSL_PRB2015, Yin-Subhro-Pollmann_PRL2015, Kumar-Fradkin_CSLKagome_PRB2015, Niu_KagomeCSl_iPess}.

The thermal Hall response has been observed in the spin-liquid phase of kagome antiferromagnets, such as volborthite \cite{Watanabe_KagomeExp_volborthite_PNAS2016} and Cakapellasite (Ca-K) \cite{Doki_KagomeExp_CaK_PRL2018, Akazawa_KagomeExp_CaK_PRX2020}. Previous studies have extensively explored the theoretical aspects of the thermal Hall effect in insulating phases on kagome lattices \cite{Katsura-Lee_kxy_Magnet_PRL2010, Mook_KagomeMagnon_PRB2016, Owerre_KagomeMagnon_PRB2017, Gao-Chen_KagomeSpinon_Scipost2020}. For example, in weak Mott insulating U$(1)$ spin liquids with spinon Fermi surface, it has been shown that external magnetic fields induce internal U$(1)$ gauge flux, impacting neutral spinons and contributing to the thermal Hall effect \cite{Katsura-Lee_kxy_Magnet_PRL2010}. The thermal Hall effect in the strong Mott regime of QSL originates from the effective generation of an internal U$(1)$ gauge flux for spinons induced by an external magnetic field through the Dzyaloshinskii-Moriya interaction, as demonstrated in Ref.~\cite{Gao-Chen_KagomeSpinon_Scipost2020}. 

In this paper,  we analyze the thermal Hall response of the Heisenberg spin model on the kagome lattice, offering an effective description of the low-energy characteristics of the Hubbard model, when subjected to a time-reversal symmetry-breaking term. The spin liquid state on the kagome Heisenberg antiferromagnet model, found to be close to a chiral spin liquid stabilized by longer-ranged spin exchange interactions \cite{Yin-Chen-He_CSL_Kagome_PRL2014, Gong-Sheng_CSL_Z2_Nature2014}, suggests a potential continuous transition \cite{Yin-Chen-He_XXZ_Kagome_PRL2015}; this transition is easily understood when considering the ground state of the Heisenberg antiferromagnet on the kagome lattice as a gapless U$(1)$ DSL. In this study, rather than directly addressing a specific numerical solution for a spin model on the kagome lattice, we assume the system stabilizes a gapless U$(1)$ DSL and illustrate the emergence of quantized thermal Hall conductivity with the introduction of a time-reversal symmetry-breaking term. The key new ingredient distinguishing our approach from earlier mean-field studies of the kagome antiferromagnets based on partons or spin waves is that we take into account the effect of gauge fluctuations following Guo {\it et al.\/} \cite{Guo_GaugeThermalHall_PRB2020}, and show that it changes the quantized value of the low-temperature thermal Hall response and also strongly affects the finite temperature behavior.

We now describe our main findings. Our analysis is consistent with semionic topological order in the CSL phase. The (chiral) central charge inferred from the U$(1)$ Dirac Chern-Simons theory has the value $c_{-}=1$ consistent with recent DMRG-based numerical analysis \cite{Bauer_KagomeCSL_Nature2014}. On the other hand, neglecting gauge contributions would have given us (erroneously) a chiral central charge of $c_{-}=2$ based on purely mean field parton theory. The peculiar distribution of the Chern numbers of the occupied parton bands results in non-monotonic temperature dependence of the free parton thermal Hall response, reminiscent of the observed experimental behavior in some of the kagome antiferromagnet systems \cite{Watanabe_KagomeExp_volborthite_PNAS2016, Doki_KagomeExp_CaK_PRL2018, Akazawa_KagomeExp_CaK_PRX2020}. For finite temperature properties that account for the gauge fluctuation effects, we take a large-$N_f$ linear response approach.  The leading order gauge fluctuation effects in the thermal Hall response are governed by Aslamazov-Larkin type processes \cite{Guo_GaugeThermalHall_PRB2020} which are $\mathcal{O}(1)$ in the large-$N_f$ expansion. Remarkably, higher order corrections vanish in the zero temperature limit, which means the contribution of gauge fluctuations to the chiral central charge is independent of $N_f$ in contrast with the non-abelian counterpart \cite{Guo_GaugeThermalHall_PRB2020}. At finite temperatures, it is crucial to consider the effects of matter and gauge excitations along with their interactions. We present explicit expressions for the finite-temperature thermal Hall response, specifically computing the Aslamazov-Larkin diagram at finite temperature while using a temperature-dependent Maxwell-Chern-Simons action for the gauge field, where we observe a logarithmic behavior at very high temperatures. More generally, our approach should be relevant for arbitrary abelian chiral spin liquids.

The rest of the paper is organized as follows. We begin in Sec.~\ref{sec:model_parton} by setting up the mean-field formalism for the CSL state on the kagome lattice using the Abrikosov fermion (parton). Here we discuss the continuum limit of the mean-field state and the effect of gauge fluctuation. Sec.~\ref{sec:kxy_parton_kagome} evaluates the thermal Hall effect
in the CSL state. We describe the thermal Hall response of the U$(1)$ Dirac Chern-Simons theory in the $1/N_f$ expansion in Sec.~\ref{sec:kxy_DCS}.

\section{The model and spin fractionalization}
\label{sec:model_parton}

Consider a spin $S = 1/2$ antiferromagnet on the kagome lattice described by the Hamiltonian
\begin{equation}
\label{eq: H_full_kagome}
    H = H_1 + H_{\chi},
\end{equation}
where
\begin{eqnarray}
       H_1 & = & J_1 \sum_{\langle \bm{i j} \rangle} \bm{S}_{\bm{i}} \cdot \bm{S}_{\bm{j}}, \label{eq: H_Heisenberg_kagome}\\
    H_\chi & = & J_\chi \sum_{\bm{i j k} \in \Delta, \nabla} \bm{S}_{\bm{i}} \cdot (\bm{S}_{\bm{j}} \times \bm{S}_{\bm{k}})\label{eq: H_chi_kagome},
\end{eqnarray}
$\bm{S}_{\bm{i}}$ are the spin-$1/2$ degrees of freedome on the sites $\bm{i}.$ $H_1$ describes nearest-neighbor Heisenberg interaction which preserves the global SU$(2)$ rotation in spin space, time-reversal, and all the discrete lattice symmetries on the kagome lattice. $H_\chi$ describes scalar spin chirality interaction on the elementary triangles, which breaks time-reversal symmetry and parity but preserves global SU$(2)$ spin rotation symmetry. $J_\chi$ can originate, for example, in the presence of an external magnetic field: in this case, its magnitude is directly proportional to the sine of the magnetic flux through each elementary triangle of the kagome lattice \cite{Diptiman_Chiral_PRB1995}. An external magnetic field also generates a Zeeman coupling which will break the spin rotational symmetry and degeneracy. For small external fields that do not cause phase transitions in the model of Eq.~(\ref{eq: H_full_kagome}), we may safely disregard this term in the rest of the analysis.

\begin{figure}[tb]
    \includegraphics[width=\linewidth]{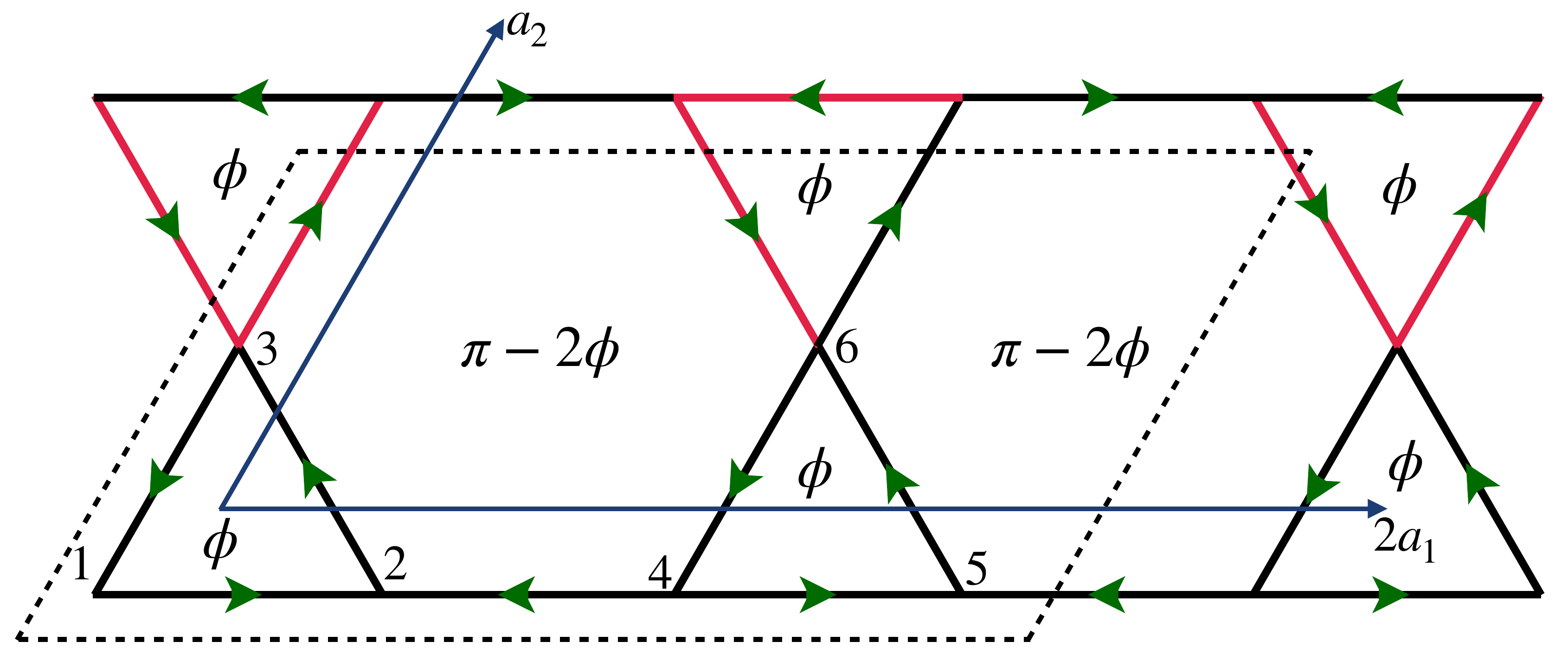}
    \caption{The enlarged unit cell is utilized for diagonalizing the mean-field spinon Hamiltonian in Eq.~(\ref{eq: Hspin_MF_compact_kagome}). Black lines denote positive hoppings, while red lines signify negative hoppings, defining the U(1) DSL. The addition of phases $\theta$ to this ansatz results in the formation of the CSL. The additional bond-dependent phase factor takes the form of $e^{i\theta}$ when the spinon hops along the arrows, and it becomes $e^{-i\theta}$ when the spinon hops in the opposite direction of the arrows. In every upward and downward triangle, the flux is $\phi = 3 \theta$, while within each hexagon, the flux amounts to $\pi - 2\phi= \pi - 6 \theta$.}
    \label{fig:Kagome lattice}
\end{figure}

\subsection{Parton mean-field theory}
\label{subsec:parton_MFT_kagome}
To characterize the spin liquid state, we utilize a parton construction wherein we decompose the spin operator into spin-$1/2$ charge-neutral fermionic spinons $f_{\bm{i} \alpha}^{}$, ($\alpha = \uparrow, \downarrow$), commonly referred to as Abrikosov fermions:
\begin{equation}
\label{eq: Spin_to_parton}
    {\bm S}^{\phantom\dagger}_{\bm{i}} = \frac{1}{2}\, f_{\bm{i} \alpha}^\dagger {\bm \sigma}^{\phantom\dagger}_{\alpha \beta} f^{\phantom\dagger}_{\bm{i} \beta},
\end{equation}
where $\bm{\sigma}$ denotes the Pauli matrices. The transformation from the spin-$1/2$ Hilbert space to the fermionic counterpart involves an expansion of the Hilbert space. To remain within the physical Hilbert space, we have to impose the following local single-site occupancy constraint, $f_{\bm{i} \alpha}^\dagger f^{\phantom\dagger}_{\bm{i} \alpha} = 1$, where the summation over repeating indices is implied. Consequently, the fermionic band structure of spinons is consistently constrained to operate at half-filling. Furthermore, Eq.~(\ref{eq: Spin_to_parton}) has an SU$(2)$ gauge redundancy because of particle-hole symmetry in the fermion representation. When applying mean-field decoupling to handle spin interactions, it becomes necessary to incorporate a SU$(2)$ gauge field defined on the lattice links. The nature of the saddle points in the mean-field theory determines whether they lead to a breakdown of SU$(2)$ symmetry into either U$(1)$ or $\mathbb{Z}_2$, depending on their specific structural characteristics \cite{Wen_ManyBodyQFT_Book, Wen_QuantumOrder_PSG_PRB2002}.

\begin{figure*}[tb]
    \includegraphics[width=0.9\linewidth]{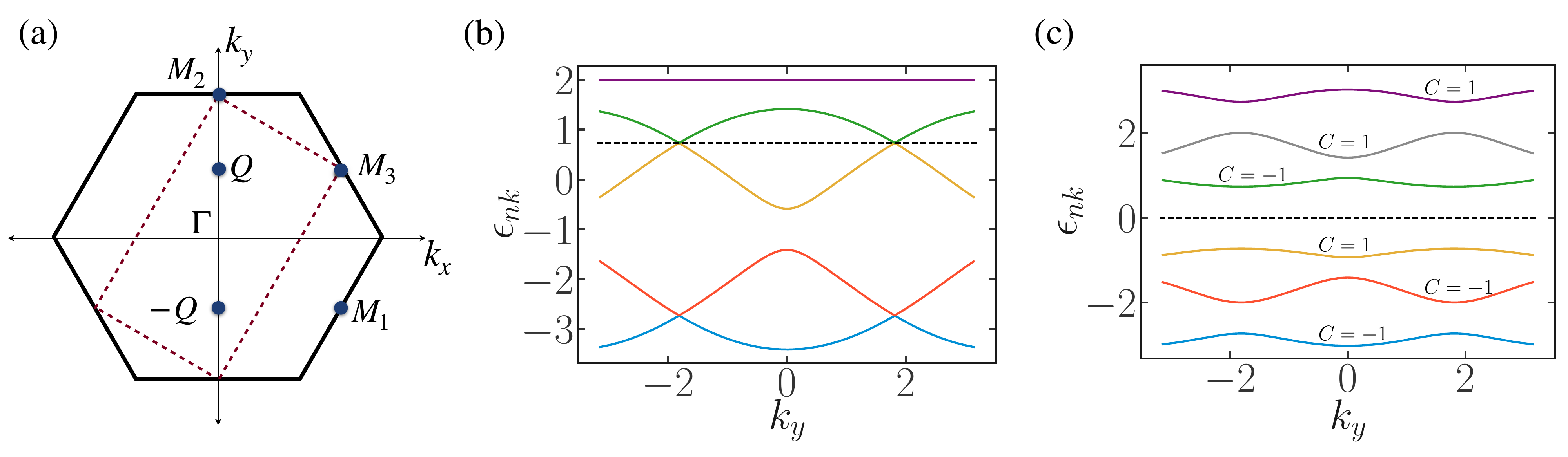}
    \caption{Brillouin zone and band structure of the flux configuration $\left[ \phi, \pi - 2\phi \right] = \left[3 \theta, \pi - 6 \theta\right]$. (\textbf{a}) To incorporate the flux configuration $\left[3 \theta, \pi - 6 \theta\right]$, we employ the doubled unit cell, with the reduced Brillouin zone outlined by red dashed lines. (\textbf{b})  The mean-field band structure of the DSL with the $[0, \pi]$ flux configuration is illustrated, revealing two Dirac points. Each spinon band is doubly degenerate, and the plots are generated along the $k_y$ line in the Brillouin zone, where the two Dirac points are situated. (\textbf{c}) Band structure of the $[\pi/2,0]$ state within our mean-field approximation. The bands have a finite gap with nonzero Chern Numbers, which are indicated on the bands.}
    \label{fig:Kagome BZ and band}
\end{figure*}

In the case where $J_\chi = 0$, we consider the ground state of the kagome Heisenberg antiferromagnet as the U$(1)$ DSL. We initiate our analysis from the U$(1)$ DSL ansatz on the kagome lattice, commonly referred to as the `$\pi$-flux' state \cite{Hastings_KagomeDSL_PRB2000, Ran_KagomeDSL_PRL2007}. Employing the conventional parton mean-field decoupling in the particle-hole channel for the Heisenberg interactions as described in Eq.~(\ref{eq: H_Heisenberg_kagome}) \cite{Wen_ManyBodyQFT_Book}, we derive
\begin{equation}
\label{eq: H1_MF_kagome}
    H_{1,\text{MF}} = - \frac{J_1}{2} \sum_{\langle\boldsymbol{i}\boldsymbol{j}\rangle, \alpha}  \left[ ( f^\dagger_{\boldsymbol{i}\alpha} f^{\phantom\dagger}_{\boldsymbol{j}\alpha}\, \chi_{\boldsymbol{ji}} + \text{H.c.} ) - \lvert \chi_{\boldsymbol{ji}} \rvert^2 \right],
\end{equation}
where $ \chi_{\boldsymbol{ij}}^{\phantom{*}} = \sum_{\alpha}  \langle f^\dagger_{\boldsymbol{i}\alpha} f^{\phantom\dagger}_{\boldsymbol{j}\alpha} \rangle = \chi_{\boldsymbol{ji}}^*$. In the case of the kagome lattice, these mean-field states are identified by the fluxes passing through the triangles and hexagons. Realizing the U$(1)$ DSL on the kagome lattice involves considering the hopping pattern depicted in Fig.~\ref{fig:Kagome lattice}, featuring zero flux through the triangular plaquettes and $\pi$ flux through the kagome hexagons \cite{Hastings_KagomeDSL_PRB2000, Ran_KagomeDSL_PRL2007}. It is important to observe that the adoption of a six-site unit cell is intended to accommodate the underlying $\pi$-flux per hexagon.  The $\pi$-flux ansatz leads to a magnetic Brillouin zone that is halved as shown in the Fig.~\ref{fig:Kagome BZ and band}a, featuring two Dirac cones at momenta $\pm \bm{Q}$, where $\bm{Q} =(0, \pi/\sqrt{3})$, for each spinon species as shown in Fig.~\ref{fig:Kagome BZ and band}b.

Next, we discuss the impact of a scalar spin chirality term on the U$(1)$ DSL ansatz defined on the kagome lattice. Using the parton representation in Eq.~(\ref{eq: Spin_to_parton}), we decouple the three spin interaction in Eq.~(\ref{eq: H_chi_kagome}) into fermion bilinears by utilizing the same mean-field parameters (see Appendix \ref{app:MF_CSL}),
\begin{align}
\label{eq: Hchi_MF_kagome}
    H_{\chi,\text{MF}} = {} & \frac{ 3 i  J_{\chi}}{16}  \sum_{\bm{ijk},\alpha}  \Big{(}  \chi^{}_{\bm{ik}}\chi^{}_{\bm{kj}}f^{\dagger}_{\bm{j}\alpha}f^{\phantom\dagger}_{\bm{i}\alpha} +  \chi^{}_{\bm{ik}}\chi^{}_{\bm{ji}}f^{\dagger}_{\bm{k}\alpha}f^{\phantom\dagger}_{\bm{j}\alpha}\nonumber \\
    & + \chi^{}_{\bm{kj}}\chi^{}_{\bm{ji}}f^{\dagger}_{\bm{i}\alpha}f^{\phantom\dagger}_{\bm{k}\alpha} - \chi^{}_{\bm{ik}}\chi^{}_{\bm{kj}}\chi^{}_{\bm{ji}} - \text{H.c.} \Big{)}.
\end{align}   
Adding \eqref{eq: H1_MF_kagome} and \eqref{eq: Hchi_MF_kagome}, the full mean-field spin Hamiltonian can be written as $H_{\text{MF}} = H_{1,\text{MF}} +H_{\chi,\text{MF}}$. Without going into the self-consistent mean-field analysis, for the sake of generality, we opt to treat $\chi_{\bm{ij}}$ as free (bounded) parameters and assume $\chi_{\bm{ij}} = \abs{\chi_{\bm{ij}}} e^{i\phi_{\bm{ij}}}$, where $\abs{\chi_{\bm{ij}}} = \chi$ is the amplitude and $\phi_{\bm{ij}}$ are the bond-dependent phases. Observing that $\chi^*_{\bm{ij}} = \chi^{\phantom{*}}_{\bm{ji}}$, it implies that $\chi^{\phantom{*}}_{\bm{ij}}$ is a directed field, requiring an orientation choice for each link. When moving from $\bm{i}$ to $\bm{j}$, $\chi^{\phantom{*}}_{\bm{ij}}$ is considered in the direction indicated by the arrow, and $\chi^*_{\bm{ij}}$ is taken opposite to the direction of the arrow, as depicted in Fig.~\ref{fig:Kagome lattice}. Now we can express the full mean-field spin Hamiltonian as follows
\begin{equation}
\label{eq: Hspin_MF_compact_kagome}
     H_{\rm{MF}} = - t \sum_{\langle \bm{ij} \rangle,\alpha} e^{i\theta_{\bm{ij}}} f^{\dagger}_{\bm{i}\alpha}f^{\phantom\dagger}_{\bm{j}\alpha}   +  \text{H.c.},
\end{equation}
where $t$ and $\theta_{\bm{ij}}$ can be computed from the values of $J_1$, $J_\chi$, $\chi$ and $\phi_{\bm{ij}}$ (see Appendix \ref{app:MF_CSL}).

Now total flux enclosed within a single triangular plaquette is given by $\theta_{\bm{ij}}+\theta_{\bm{jk}}+\theta_{\bm{ki}} = \phi$. We start from the U$(1)$ DSL, and CSL state emerges when an additional phase, represented by $\theta$, is incorporated into the directed links on the kagome lattice. Here, we consider a configuration where the up and down triangles share the same fluxes (i.e., $\phi = 3 \theta$), and the hexagon's flux is $\pi - 6 \theta$. This configuration can be denoted as $\left[ \phi, \pi - 2\phi \right] = \left[3 \theta, \pi - 6 \theta\right]$, as discussed in Refs.~\cite{Ran_KagomeDSL_PRL2007, Sheng_VMC_CSL_kagome_PRB2015}. To achieve this, a magnetic unit cell with six sites is required, deviating from the standard three-site unit cell commonly used for the kagome lattice, as also employed for the U$(1)$ DSL.

Following the procedure in Appendix \ref{app:MF_CSL}, we diagonalize the Hamiltonian in Eq.~(\ref{eq: Hspin_MF_compact_kagome}) for the flux configuration $\left[ \phi, \pi - 2\phi \right]$. When $\phi$ is set to zero, the ansatz simplifies to the well-known U$(1)$ DSL state, resulting in a spectrum featuring a pair of Dirac cones centered at half-filling for each spinon species, as illustrated in Fig.~\ref{fig:Kagome BZ and band}a. So time-reversal symmetry is preserved for $\phi = 0$. Moving forward, we examine flux phases characterized by a non-zero flux within triangles. The flux configuration  $\left[ \phi, \pi - 2\phi \right]$ with $\phi \neq 0$ breaks both time-reversal and parity symmetry while maintaining the combined symmetry of the two. The introduction of any nonzero value for $\phi$ leads to the emergence of a direct gap at each Dirac cone as shown in Fig.~\ref{fig:Kagome BZ and band}c. The lower three bands become fully occupied, while the upper three bands remain empty, maintaining the half-filling condition. We find that the filled bands have non-zero Chern numbers with values of $C = \{-1, -1, 1\}$ starting from the topmost occupied band, while the empty bands have corresponding Chern numbers of opposite sign. Taking into account the spin degeneracy, the net Chern number of the occupied bands amounts to $-2$, corresponding to a CSL phase. Consequently, the resulting fully gapped bands exhibit topologically nontrivial characteristics.

\subsection{The continuum field theory and gauge fluctuation}
Having obtained a mean-field description of the U$(1)$ DSL and CSL phases, we now derive the continuum low-energy effective field theory and consider the effect of gauge fluctuations. 

Here we briefly discuss the effective theory of the U$(1)$ DSL on the kagome lattice. Given the $\pi$-flux ansatz on the kagome lattice, featuring gapless Dirac cones at the Fermi level, our focus is narrowed to the two bands that intersect the Fermi level. Consequently, in the continuum limit, we can characterize the low-energy spinons near the Dirac points using the relativistic Lagrangian for massless Dirac fermions  \cite{Ran_KagomeDSL_PRL2007, Hermele_KagomeDSL_Properties_PRB2008},
\begin{equation}
\label{eq: L1_MF_kagome}
{\cal L}^{}_{{1, \rm MF}} = \bar{\psi}^{\phantom{\dagger}}_{l} \left(i \gamma^{\mu} \partial_{\mu} \right) \psi^{\phantom{\dagger}}_{l},
\end{equation}
where $\psi^{\phantom{\dagger}}_{l}$ is a two-component spinor whose flavor index $l$ is summed from $1$ to $N_f$. The gamma matrices are taken to be $\gamma^{\mu} = (\tau^z, i \tau^y, -i \tau^x)$, where the $\tau^a$'s are Pauli matrices acting on spinor (sublattice) indices and $\bar{\psi}^{\phantom{\dagger}}_{l} \equiv \psi^\dagger_l \tau^z$. The dispersion of the U$(1)$ DSL on the kagome lattice features two Dirac cones, resulting in a total of $N_f = 4$ when accounting for spin.

In characterizing the phase fluctuation of the mean-field ansatz, we represent $\chi_{\boldsymbol{ij}}$ as $\chi_{\boldsymbol{ij}} = \Bar{\chi}_{\boldsymbol{ij}} e^{-i a_{\boldsymbol{ij}}}$. Including the phase fluctuation, the mean-field Hamiltonian in Eq.~(\ref{eq: H1_MF_kagome}) becomes 
\begin{equation}
\label{eq: H_Heisenberg_U(1)}
    H_{1, \mathrm{U}(1)} = - \frac{J_1}{2} \sum_{\langle\boldsymbol{i}\boldsymbol{j}\rangle, \alpha} \left( \Bar{\chi}^{\phantom\dagger}_{\boldsymbol{ji}}\,  e^{- i a^{}_{\boldsymbol{ji}}}\, f^\dagger_{\boldsymbol{i}\alpha} f^{\phantom\dagger}_{\boldsymbol{j}\alpha} + \text{H.c.}\right).
\end{equation}
The Hamiltonian in Eq.~(\ref{eq: H_Heisenberg_U(1)}) is invariant under the gauge transformation
\begin{equation}
    f^{\phantom\dagger}_{\boldsymbol{i}} \rightarrow f^{\phantom\dagger}_{\boldsymbol{i}} e^{i \theta_{\bm{i}}}, \qquad a^{}_{\boldsymbol{ij}} \rightarrow a^{}_{\boldsymbol{ij}} + \theta^{}_{\bm{i}} - \theta^{}_{\bm{j}}.
\end{equation}
In this context, $a^{}_{\boldsymbol{ij}}$ serves as the spatial component of U$(1)$ gauge field. The single-particle occupancy constraint can be enforced by including a Lagrange multiplier $a_0(\bm{i},t)$, which corresponds to the time component of the gauge field $a_\mu$. Namely, we have transformed the quantum spin model into a scenario where spinons are strongly coupled to the dynamical U$(1)$ gauge field.  

Including minimal coupling of the U$(1)$ gauge field to the massless Dirac fermions, the Lagrangian in Eq.~(\ref{eq: L1_MF_kagome}) becomes \cite{Ran_KagomeDSL_PRL2007, Hermele_KagomeDSL_Properties_PRB2008},
\begin{equation}
    {\cal L}_{1}  = \bar{\psi}^{\phantom{\dagger}}_{l} i \gamma^{\mu} \big(\partial_{\mu} - i a_\mu \big)\psi^{\phantom{\dagger}}_{l}.
\end{equation}

As explained in Sec. \ref{subsec:parton_MFT_kagome}, the addition of an extra phase  $\theta$ to the U$(1)$ DSL ansatz results in the opening of a direct gap at each Dirac cone. Consequently, this modification yields a CSL with a net Chern number of $-2$. For $\phi \neq 0$, the Dirac fermions acquire ``mass'' (a ``gap'') by breaking time-reversal symmetry. So in the DSL perspective, obtaining CSL involves introducing a mass term, $\bar{\psi}_{\alpha} \psi_{\alpha}$, with the same sign for all $N_f = 4$ Dirac cones. This leads to us to a continuum low-energy effective theory of the CSL phase which involves four species of Dirac fermions with mass $m$ interacting with the U$(1)$ gauge field $a_\mu$. However, we must not overlook the contribution of the two occupied  ``heavy'' fermionic bands located significantly away from the Fermi level. These bands exhibit a non-zero Chern number, specifically $C=\{-1,-1\}$ per spin species. Integrating out the heavy fermions generates a Chern-Simons term for the U(1) gauge field $a_\mu$ with level $\mathtt{k} = -4$, accounting for spin degeneracy. In this manner, we obtain the Lagrangian of the U$(1)$ Dirac Chern-Simons theory in Minkowski space-time
\begin{equation}
\label{eq:L_Psi_CS_kagome}
{\cal L}_{\psi, a} = \bar{\psi}^{\phantom{\dagger}}_{l} i \gamma^{\mu} \big(\partial_{\mu} - i a_\mu \big)\psi^{\phantom{\dagger}}_{l} - m\ \bar{\psi}^{\phantom{\dagger}}_{l} \psi^{\phantom{\dagger}}_{l} + \mathtt{k} \ \text{CS}[a_\mu] \text{,}
\end{equation}
where the last term represents the U$(1)$ Chern-Simons term with level $\mathtt{k} =-4$. In the presence of a finite mass term $m$, a gap forms, and each light Dirac fermion induces a Chern-Simons response proportional to ${\rm{sgn}}(m)/2$. Upon integrating out the light fermions also, we obtain a U$(1)_{\hat{\mathtt{k}}}$ Chern-Simons theory,
\begin{equation}
    \mathcal{L}_a  = \frac{ \hat{\mathtt{k}}}{4 \pi}\,  \epsilon^{\mu\nu\rho} \, a_\mu \partial_\nu a_\rho,
\end{equation}
where the integer $\hat{\mathtt{k}}$ is defined by
\begin{equation}
\label{eq:kappa_hat}
    \hat{\mathtt{k}} = \mathtt{k} +\frac{ N_f}{2}\,  {\rm{sgn}}(m).
\end{equation}
As the Chern number of the band just below the Fermi level is $C=+1$, the sign of the mass term of the light Dirac fermion is positive i.e., $m > 0$, and we get a U$(1)_{-2}$ Chern-Simons theory, which is the effective field theory of the $\nu = 1/2$ bosonic Laughlin state \cite{Kalmeyer-Laughlin_SL-Heisenberg_PRB1989, Wen_CSL-Superconductivity_PRB1989}. So, the low energy physics of the CSL is captured by the following Lagrangian:
\begin{equation}
\label{eq:CS_KagomeCSL}
    \mathcal{L}_a  = - \frac{2}{4 \pi} \epsilon^{\mu\nu\rho} a_\mu \partial_\nu a_\rho.
\end{equation}
The Chern-Simons term induces a gap in the gauge particles, allowing us to ignore their fluctuations and the instanton effect in the low-energy description. So, the U$(1)$ gauge field can only help the spinons interact over short distances, making these fractionalized particles (spinons) the basic building blocks of the spin liquid ground state. Hence, the chiral gapped spin liquid remains robust against gauge fluctuations and retains its physical significance.

We highlight certain characteristics of the CSL state governed by the effective theory in Eq.~(\ref{eq:CS_KagomeCSL}). It shows a twofold degeneracy in its ground state on a torus. The Chern-Simons term induces a change in the spinons' statistics, transforming them into ``semions'' with a statistical angle of $\pi/2$ and a fractional charge of $1/2$. The two abelian anyons (trivial and semion) imply a total quantum dimension $\mathcal{D}=\sqrt{2},$ and a nontrivial topological entanglement entropy $\gamma = \ln \mathcal{D}=\frac{1}{2}\ln 2.$ In a system with an open boundary, the CSL state has a chiral edge state characterized by the U$(1)$ Kac-Moody theory at level $2$. Therefore, the chiral state that emerges is consistent with the one previously identified by Bauer and co-workers in Ref.~\cite{Bauer_KagomeCSL_Nature2014}.

\section{Edge and bulk perspective of Hall conductivities}\label{sec:bulk_edge}
We start by reviewing the edge perspective and the bulk perspective of the charge Hall and the thermal Hall conductivity, which are expected to produce equivalent results \cite{Cooper-Halperin_Thermoelectric_response_PRB1997}.  
First, the edge perspective states that the  Hall currents are carried by chiral edge modes. More precisely, for charge Hall conductivity $\sigma_{xy}$, each chiral edge mode is expected to contribute one quantum $e^2/h$, whereas for thermal Hall conductivity the counting is given by the chiral central charge $c_-$:  
\begin{equation}
    \frac{\kappa_{xy}}{T} = c_{-} \frac{\pi k_B^2}{6 \hbar}\,.
    \label{eq:c-}
\end{equation}
Second, the bulk perspective states that the thermal Hall current exists in the bulk, which can be computed using the bulk wavefunction or equivalently the Kubo formula. 
At first glance, the equivalence between the bulk and the edge perspective is not obvious, as they predict different microscopic current density profiles in the system. 

We try to reconcile the difference by first focusing on the charge response of the integer Quantum Hall effect (IQHE). In IQHE, the edge current picture is obtained by applying a different \emph{chemical potential} $\mu_L\neq\mu_R$ to the two sides of the system \cite{SMGirvin2019}. In the bulk picture, an \emph{electric field} is directly applied across the whole system, and by solving the Schrodinger equation with a bulk electric field, it is shown that there is Hall current in the bulk. This example demonstrates that the Einstein relation (equivalence between chemical potential and electrical potential) is broken in the presence of time-reversal symmetry breaking. However, the difference of microscopic current profile between these two setups is the curl of the magnetization, which does not contribute to transport \cite{Cooper-Halperin_Thermoelectric_response_PRB1997}, and therefore both of them yield the same Hall conductivity $\sigma_{xy}$. To summarize, the edge and the bulk perspectives of the charge Hall effect arise from different protocols of driving transport and the breaking of the Einstein relation. 

Next we move to thermal Hall transport. In practice, thermal transport is driven by applying a temperature gradient to the system, but as Luttinger pointed out, it can also be achieved by applying gravity to the system \cite{Luttinger_ThermalTransport_PRB1964}. Similar to the discussion in the previous paragraph, we expect the bulk and the edge perspectives correspond to driving the system with a gravitational field and a temperature gradient, respectively. When time-reversal symmetry is broken, these two perspectives will produce different microscopic energy current profiles because of the presence of energy magnetization. However, unlike the charge Hall case where the two perspectives produce the same Hall conductivity, in the thermal transport case the connection between the two perspectives is further complicated by  the fact that energy magnetization also contributes to the thermal Hall conductivity\cite{Smrcka-Streda_tranport_1977,Cooper-Halperin_Thermoelectric_response_PRB1997}, which manifests itself as unphysical divergences of Kubo formula computations at low temperatures\cite{Katsura-Lee_kxy_Magnet_PRL2010,  Murakami_kxy_Magnon_PRL2011}. This divergence is associated with microscopic circulating currents entering the Kubo formula\cite{Katsura-Lee_kxy_Magnet_PRL2010,  Murakami_kxy_Magnon_PRL2011}, which needs to be subtracted using the procedure developed in \cite{Qin_EnergyMagnetization_PRL2011}. After this substraction, the bulk result is expected to be equivalent to be edge result, as we will demonstrate later.

\section{Thermal Hall effect in the CSL phase}
\label{sec:kxy_parton_kagome}

In this section, we address the thermal Hall effect deep inside the CSL phase. The result of the total $\kappa_{xy}$ contains contributions from both the fermionic spinons and the emergent gauge fields
\begin{equation}
    \kappa_{xy}=\kappa_{xy}^\text{spinon}+\kappa_{xy}^\text{gauge}\,,
\end{equation} and we separately address them below:

\textbf{Spinon contribution: }We employ the framework introduced in Ref.~\cite{Qin_EnergyMagnetization_PRL2011} to calculate $\kappa_{xy}$ in the CSL phase. Transforming to momentum space, the mean-field Hamiltonian in Eq.~(\ref{eq: Hspin_MF_compact_kagome}) can now be formulated as 
\begin{equation}
    H_{\text{MF}} = t\, \sum_{\bm{k}}\, f^\dagger_{\bm{k} s}\, \mathcal{H}(\bm{k})^{\phantom\dagger}_{s s'}\, f^{\phantom\dagger}_{\bm{k} s'},
\end{equation}
where $s$ and $s'$ are the sublattice indices. The diagonalization of $  \mathcal{H}(\bm{k})$ for the $[\phi, \pi - 2\phi]$  flux configuration results in the dispersions for the six bands shown in Fig.~\ref{fig:Kagome BZ and band}c. Denoting the band energies for each band $n  = 1, \cdots, 6$ by $\epsilon_{n\bm{k}}$ and Berry curvature by $\Omega_{n\bm{k}}$, the thermal Hall conductivity for the fermionic system with a nonzero chemical potential $\mu$ is given by \cite{Qin_EnergyMagnetization_PRL2011}
\begin{equation}
    \kappa^{\text{spinon}}_{xy} = -\frac{k^2_{\mathrm{B}}}{ \hbar T} \int d \epsilon\, (\epsilon - \mu)^2 \sigma_{xy} (\epsilon)\, \frac{\partial f(\epsilon, \mu, T)}{\partial \epsilon}
    \label{eq:kxy_def}
\end{equation}
where  $f(\epsilon, \mu, T)$ is the Fermi function and 
\begin{equation}
    \sigma_{xy} (\epsilon) = - \int_{\epsilon_{n\bm{k}} < \epsilon  } \frac{d^2 k}{(2\pi)^2} \,  \Omega_{n \mathbf{k}}
\end{equation}
is the $\hbar/e^2$ times zero temperature anomalous Hall coefficient for a system with the chemical potential $\epsilon$. By imposing the constraint of single-site spinon occupancy, the chemical potential $\mu$ is determined in a manner that ensures the system consistently operates at half-filling. The associated Chern number is
\begin{equation}
C_n = \frac{1}{2\pi} \int d^2 k \,\, \Omega_{n {\bf k}}.
\end{equation}
In the scenario of an isolated band, distinct from all others due to an energy gap, the Chern number is well-defined and takes integer values. 
As the temperature approaches zero, Eq.~(\ref{eq:kxy_def}) converges to the Wiedemann-Franz law and gives 
\begin{equation}
    \frac{\kappa^{\text{spinon}}_{xy}}{T} = - 2 \times \frac{\pi k_B^2}{6 \hbar} \sum_{n \, \in \, {\rm filled~bands}} \hspace*{-0.3cm}C_n,
    \label{eq:kxy_zeroT}
\end{equation}
where a factor of $2$ is multiplied to account for the spin degeneracy. This implies that $\kappa_{xy}/T$ is quantized in integer units of $(\pi k_B^2 / 6 \hbar)$ at absolute zero temperature.

\begin{figure}[tb]
    \includegraphics[width=0.9\linewidth]{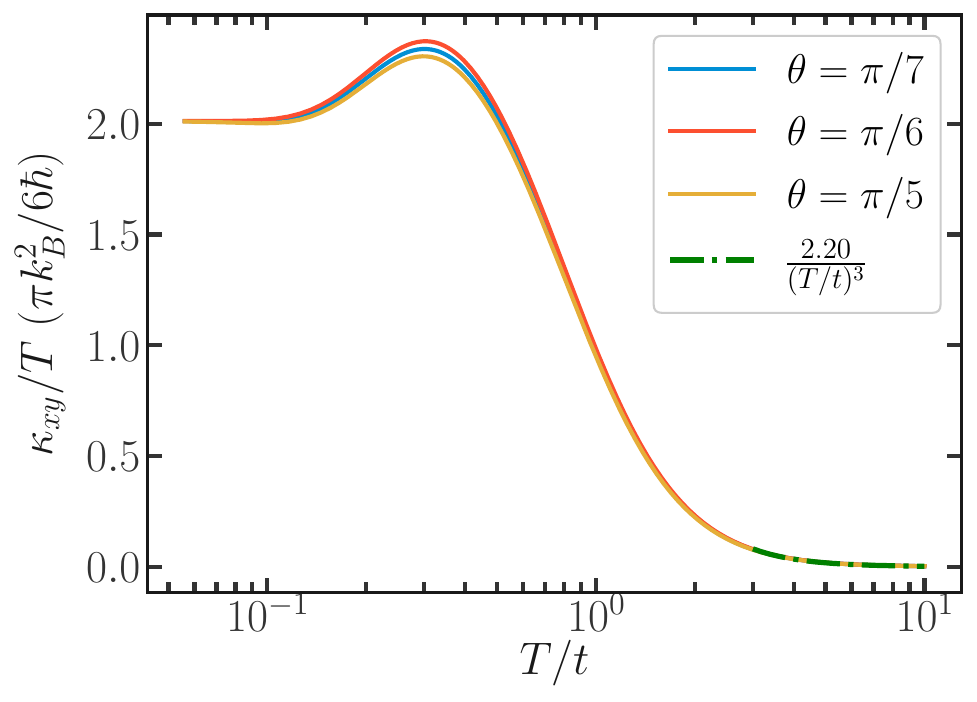}
    \caption{The temperature dependence of free fermion contribution to the thermal Hall conductivity of the CSL phase with semion topological order using Eq. (\ref{eq:kxy_def}) and the tight-binding model. The data are computed for different values of parameter $\theta$ introduced after Eq. (\ref{eq: Hspin_MF_compact_kagome}). The unit of $\kappa_{xy}/T$ here is $(\pi k_B^2 / 6 \hbar)$. $\kappa_{xy}/T$ shows quantization to 2 at zero temperature, experiences a rapid rise, and gradually decreases after reaching a maximum at temperatures comparable to the spectral gap. The non-monotonic temperature dependence with a peak exceeding the low-temperature quantized value is on account of the peculiar arrangement of Chern numbers of the chiral bands. At high temperatures, $\kappa_{xy}/T\sim T^{-3}.$ Quantum gauge fluctuations considered later qualitatively modify the temperature dependence.}
    \label{fig:Kagome Thermal Hall parton}
\end{figure}

In Fig.~\ref{fig:Kagome Thermal Hall parton}, we plot the numerically calculated temperature dependence of thermal Hall conductivity within the CSL phase in the mean-field parton approximation. In the CSL phase featuring semion topological order, the thermal Hall conductivity is quantized to $2$ in the limit of zero temperature, and it exhibits a rapid increase as the temperature rises. The initial increase with temperature is on account of the peculiar distribution of the Chern numbers of the lowest bands, $C = \{-1,-1,1,-1\},$ where the first three bands are occupied. As the temperature is increased, the first unoccupied band with Chern number $-1$ also starts contributing, which increases the thermal Hall conductivity with respect to the zero temperature limit. At temperatures significantly surpassing the maximum energy of the fermion bands, all the bands are nearly uniformly populated, dictated by the Fermi-Dirac distribution function. At high temperatures, we observe $\kappa_{xy}/T \sim T^{-3},$  similar to that predicted in Ref.~\cite{FCZhang_Universal_kxy} for noninteracting fermionic systems with a bounded spectrum; however unlike Ref.~\cite{FCZhang_Universal_kxy}, we do not find any intermediate temperature regime where $\kappa_{xy}/T$ decreases exponentially, i.e., $e^{-T/T_0}.$ For other distributions of the Chern numbers, the thermal Hall conductivity could have a monotonously decreasing temperature dependence. Experimentally, a non-monotonic temperature dependence of the thermal Hall response is seen \cite{Watanabe_KagomeExp_volborthite_PNAS2016, Doki_KagomeExp_CaK_PRL2018, Akazawa_KagomeExp_CaK_PRX2020}; however, quantization has not yet been observed.

\textbf{Gauge field contribution: }After determining the thermal Hall conductivity within the mean-field theory, we now investigate how it is influenced by the gauge field. We will calculate $\kappa_{xy}/T$ at zero-temperature by examining the edge theory of the Lagrangian in Eq.~\eqref{eq:L_Psi_CS_kagome}. In Section~\ref{sec:kxy_DCS}, we will explore the effect of the gauge field on $\kappa_{xy}$ from the perspective of the bulk theory using $1/N_f$ expansion. From the edge perspective, the total thermal Hall conductivity is given by Eq.\eqref{eq:c-} in terms of the chiral central charge $c_-$.
 For a non-interacting system, $c_{-}$ is equal to the Chern number of the system. This is consistent with the fact that $\kappa_{xy}/T$ calculated from the parton mean-field theory is quantized to $2$ in the limit of zero temperature. For the CSL state, spinons are now coupled to the U$(1)$ gauge field. We can derive the edge theory from the Chern-Simons theory in Eq.~(\ref{eq:CS_KagomeCSL}), described by a chiral compactified U$(1)$ boson CFT, also at level $-2$. The chiral central charge for this theory is given by $-1$. Consequently, in the semion topological order phase, quantum gauge fluctuations are anticipated to modify the prefactor from $2$ to the precisely quantized value of $1$ \cite{Samajdar_Emhamcement_kxy_Nature2019, Guo_GaugeThermalHall_PRB2020}. In the analysis below, we obtain the complete temperature dependence of $\kappa_{xy}/T$ taking into account the effect of $U(1)$ gauge fluctuations.

\begin{figure*}[tb]
    \includegraphics[width=0.9\linewidth]{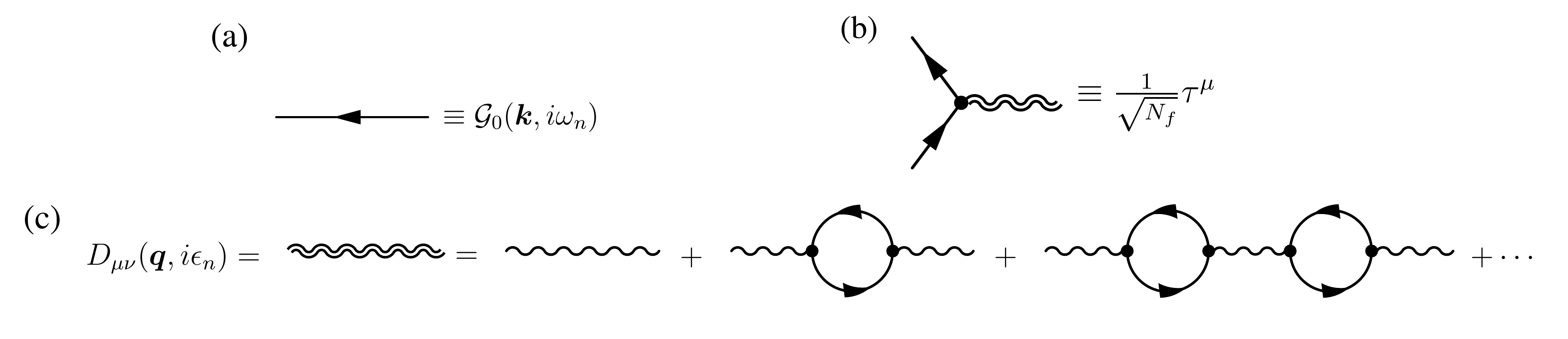}
    \caption{The propagators and vertices in the large-$N_f$ formalism. (\textbf{a}) The bare fermion propagator; (\textbf{b}) the large-$N_f$ vertices for the gauge interactions with the bare fermion; (\textbf{c}) diagrammatic representation of the renormalized gauge boson propagator in the large-$N_f$ limit. The single wiggly lines represent the bare Maxwell kinetic term, while the fermion bubbles correspond to vacuum polarization functions. The double wavy line represents the renormalized gauge boson propagator.}
    \label{fig:Feynman Rules and vertices}
\end{figure*}

\section{Thermal Hall response of the continuum field theory}
\label{sec:kxy_DCS}
The arguments based on chiral central charge are adequate for obtaining the quantized value of the low temperature thermal Hall effect. However, the finite temperature behavior is also of interest in many situations including experiments. For this purpose, we present a large-$N_f$ linear response treatment that allows a systematic accounting of the gauge fluctuation effects and can be used to understand the finite temperature behavior of the thermal Hall response. Another advantage is that this being a continuum field theoretical treatment is typically simpler to implement than, say, a numerical lattice calculation (see e.g. Ref.~\cite{Aman_KitaevThermal_PRB2023}). In a couple of recent insightful works \cite{Samajdar_Emhamcement_kxy_Nature2019, Guo_GaugeThermalHall_PRB2020}, the thermal Hall response of the SU$(2)$ Dirac Chern-Simons theory was studied. This theory describes the quantum phase transition between the N\'eel state and another featuring coexisting N\'eel and semion topological order. Here we are interested in the U$(1)$ case and examine the thermal Hall response of the low-energy continuum field theory governing the CSL phase described by the U$(1)$ Dirac Chern-Simons theory, as outlined in Eq.~(\ref{eq:L_Psi_CS_kagome}), in the $N_f \to \infty$ limit. The result of our computation is summarized in Table.~\ref{tab:summary}. 

In the limit $\abs{m}/T \to \infty$, the exact values of $\kappa_{xy}/T$ can be determined through a mapping to the CFT on the edge, utilizing the bulk-boundary correspondence discussed in Sec.~\ref{sec:bulk_edge}
\begin{align}
\label{eq:kxy_DCS_exact}
  \frac{\kappa^{\phantom{0}}_{xy}}{T} = - \frac{\pi k_B^2}{6 \hbar} \, \mathrm{sgn}( \hat{\mathtt{k}} ) \left(|\hat{\mathtt{k}}| - 1\right); \qquad \frac{\abs{m}}{T} \rightarrow \infty,
\end{align}
where  $\hat{\mathtt{k}}$ is defined in Eq.~(\ref{eq:kappa_hat}). In the abelian case, we find that the large-$N_f$ treatment up to $\mathcal{O}(1)$ gives the exact low-temperature value of the thermal Hall conductivity, and matches with the behavior obtained from the Maxwell-Chern-Simons effective theory. In contrast, in the diagrammatic expansion of the thermal Hall response in the SU$(2)$ case, nontrivial contributions appear at all orders in $1/N_f.$ 

\begin{table}[h]
    \centering
    \caption{Different contributions to $\kappa_{xy}$ in the large-$N_f$ expansion.}
    \label{tab:summary}
    
    \begin{tabular}{>{\centering\arraybackslash}p{3.2cm}|>{\centering\arraybackslash}p{2.1cm}>{\centering\arraybackslash}p{2.1cm}}
        \hhline{=|==}
        \multicolumn{3}{c}{$\kappa_{xy} = \kappa_{xy}^\text{spinon} + \kappa_{xy}^\text{gauge}$} \\
        \hline
        \textbf{Term} & \boldmath{$T = 0$} & \boldmath{$T > 0$} \\
        \hline
        $\kappa_{xy}^\text{spinon}$ ($\mathcal{O}(N_f)$) & Eq.~\eqref{eq:kxy_mM_O(1)_T0} & Eq.~\eqref{eq:kxy_Dirac_LargeN} \\
        $\kappa_{xy}^\text{gauge}$ ($\mathcal{O}(1)$)     & Eq.~\eqref{eq:kxy_AL_T=0} & Eq.~\eqref{eq:M_Q_tilde_finiteT} \\
        \hhline{=|==}
    \end{tabular}
    
\end{table}

\subsection{The U$(1)$ Dirac Chern-Simons theory}

\subsubsection{The Lagrangian and Feynman rules}

The interacting Lagrangian in Eq.~(\ref{eq:L_Psi_CS_kagome}) is still written
\begin{align}
\label{eq:L_Psi_CS_LargeN_kagome}
     \mathcal{L}_{ N_f,\psi,a} = {} & \sum_{l=1}^{N_f} \Bar{\psi}^{\phantom{\dagger}}_{l} i \gamma^\mu \left(\partial^{\phantom\dagger}_\mu  - \frac{i}{\sqrt{N_f}} a^{\phantom\dagger}_\mu \right) \psi^{\phantom{\dagger}}_{l} -  m\, \Bar{\psi}^{\phantom{\dagger}}_{l} \psi^{\phantom{\dagger}}_{l} \nonumber \\
     & + \mathtt{k} \ \text{CS}[a_\mu],
\end{align}
but the summations over fermion flavors span $N_f \gg 1$ values rather than just four, and the original fermion-boson interaction undergoes rescaling by $1/\sqrt{N_f}$. At finite temperature, Lorentz invariance is broken due to the introduction of a fixed length in the temporal direction. Since most calculations in this paper are performed at $T \neq 0$, we choose an equivalent form of the Euclidean action, deviating from the conventional relativistically invariant notation in Eq.~(\ref{eq:L_Psi_CS_LargeN_kagome}):
\begin{align}
\label{eq:L_Psi_CS_LargeN_nonrel}
     \mathcal{S}[\psi,a] = {} & \sum_{l=1}^{N_f} \int d^2 \bm{r} \,  d \tau \, \psi_l^{\dagger} \left[ \left(\partial^{\phantom\dagger}_\tau - \frac{i}{\sqrt{N_f}} a^{\phantom\dagger}_0\right) \right. \nonumber \\
  & {} \left.  -i \bm{\tau} \cdot \left(\bm{\partial} - \frac{i}{\sqrt{N_f}} \bm{a} \right)
     \right] \psi^{\phantom\dagger}_l   +  m\, \psi_l^{\dagger} \, \tau^z\, \psi_l^{\phantom\dagger} \nonumber \\
  & {} + \frac{i \mathtt{k}}{4 \pi} \epsilon^{\mu\nu\rho} \, a_\mu \partial_\nu a_\rho.
\end{align}

The Feynman rules can be derived from the action in Eq.~(\ref{eq:L_Psi_CS_LargeN_nonrel}). In the large-$N_f$ framework, each fermion loop is now accompanied by an additional factor of $N_f$ due to the summation over flavors, and simultaneously, each interaction vertex is scaled by a factor of $1/\sqrt{N_f}$. In the large-$N_f$ limit of fermion flavors, the Chern-Simons level $\mathtt{k}$ is assumed to be of the same order as $N_f$. Since all loop contributions to the fermion propagator are suppressed by $1/N_f$, we will use the free Matsubara propagator, which is given by
\begin{equation}
    \mathcal{G}_0 (\bm{k}, i\omega_n) = - \frac{1}{i\omega_n + \bm{\tau} \cdot \bm{k} + m\tau^z } .
\end{equation}
 The bare fermion-gauge boson vertex is given by
\begin{equation}
    \Gamma_{a\psi^{\dagger}\psi}^\mu = \frac{1}{\sqrt{N_f}} \tau^\mu,
\end{equation}
where $\tau^\mu =\{i \mathds{1}, \bm{\tau} \} = \{i \mathds{1}, \tau^x, \tau^y\}$. 

Throughout this paper, Greek symbols $\mu,\nu,\alpha$, etc., will represent values $0,x$ and $y$ ($2+1$ D space-time), while lowercase letters $i,j$ denote values $x$ and $y$ (two-dimensional space). We consistently use the following notation:
\begin{align}
    \bm{k} & = (k_x, k_y), \;\; k_{\mu} = (\omega_n, k_x, k_y), \;\; \abs{\bm{k}} = \sqrt{k_x^2 + k_y^2}, \nonumber \\
    k &= \sqrt{\omega_n^2 + k_x^2 + k_y^2}.
 \end{align}

\subsubsection{Renormalized gauge boson propagator at large $N_f$}

The one-particle irreducible self-energy of the gauge field consists of a closed fermion loop, which is $ \mathcal{O}(1)$ in this expansion. Summing the full geometric series of these terms yields the renormalized gauge propagator. The diagrammatic expansion is illustrated in Fig.~\ref{fig:Feynman Rules and vertices}b. So, to derive the $N_f = \infty$ renormalized gauge propagator, we first obtain the effective action to leading order in $N_f$ by perturbatively integrating out the fermions. We will interpret Eq.~\eqref{eq:L_Psi_CS_LargeN_nonrel} as an effective theory, where the Chern-Simons term is generated by integrating out the heavy fermionic band situated far from the Fermi level with non-zero Chern numbers. To acquire the U$(1)$ Chern-Simons term at level $\mathtt{k}$, it is necessary to have $2\lvert \mathtt{k} \rvert$ flavors of heavy Dirac fermions with mass $M$, satisfying ${\rm sgn}(M) = {\rm sgn}(\mathtt{k})$. In our approach, the bare Maxwell kinetic term, originating from heavy fermion integration, scales with the inverse of $M$ \cite{Guo_GaugeThermalHall_PRB2020}. Throughout this paper, we will allow $\abs{m}/T$ to vary arbitrarily, while assuming $\abs{M} \gtrsim
 T$, so that the behavior of the heavy Dirac fermion is described by the effective Maxwell-Chern-Simons theory. It is important to note that the Maxwell-Chern-Simons term generated by the heavy Dirac fermion contributes in the same order as the polarization bubble of the light Dirac fermion. The one-loop corrected effective Euclidean action of the gauge field, at leading order in $1/N_f$, is given by 
\begin{align}
\label{eq:effective_action}
     \mathcal{S}[a] & =  \frac{T}{2} \sum_{i\epsilon_n} \int \frac{d^2\bm{q}}{(2\pi)^2} \,  a_\mu(-q) \Pi^{\mu\nu}(q) a_{\nu}(q), \nonumber \\
      \Pi^{\mu\nu}(q) & =\Pi^{\mu\nu}_{f}(q) + \Pi^{\mu\nu}_{\text{MCS}}(q),
\end{align}
where the polarization bubble of the light Dirac fermion reads,
\begin{align}
\label{eq:Pi_munu_f}
    \Pi^{\mu\nu}_{f}(\bm{q}, i \epsilon_n) = {} & N_f T \sum_{i\omega_n} \int  \frac{d^2 \bm{k}}{(2\pi)^2} \nonumber \\
     & {} \times \Tr \left[\mathcal{G}_0 (\bm{k}, i\omega_n) \tau^\mu \mathcal{G}_0 (\bm{k}+\bm{q}, i\omega_n+ i \epsilon_n) \tau^\nu \right],
\end{align}
and $\Pi^{\mu\nu}_{\text{MCS}}$ is the momentum space kernel of the Maxwell-Chern-Simons term (see Appendix~\ref{app:effective_action}).

The details of the computation of the fermion polarization functions $\Pi^{\mu\nu}_{f}(\bm{q}, i \epsilon_n)$ at finite temperature are shown in Appendix~\ref{app:effective_action}. These expressions are valid for all $\abs{m}/T$. As Lorentz invariance is broken at $T \neq 0$, covariant gauge fixing, as generally used at $T = 0$, is not convenient, and we opt for working in the Coulomb gauge $\grad \cdot \bm{a} = 0$. In the $\abs{m}/T \rightarrow \infty$ limit, after imposing the Coulomb gauge condition $q_i a_i = 0$, the renormalized propagator for the gauge boson can now be written as (see Appendix~\ref{app:effective_action}) 

\begin{eqnarray}
\label{eq:D_gauge}
    D_{00} (\bm{q}, i \epsilon_n) & =& \frac{q^2}{g\bm{q}^2 \left(q^2 + m_t^2\right)}, \label{eq:D_00} \nonumber \\
    D_{0i} (\bm{q}, i \epsilon_n) & =& - D_{i0} (\bm{q}, i \epsilon_n)  = - \frac{m_t \epsilon_{ij} q_j}{g \bm{q}^2 \left(q^2 + m_t^2\right)}, \label{eq:D_0i_i0} \nonumber \\
    D_{ij} (\bm{q}, i \epsilon_n) & =& \frac{1}{ g \left(q^2 + m_t^2\right)} \left(\delta_{ij} - \frac{q_i q_j}{\bm{q}^2} \right),  \label{eq:D_ij}
\end{eqnarray}
where 
\begin{align}
\label{eq:g_mt_def}
    g & = \frac{1}{12 \pi \abs{m}} + \frac{1}{12 \pi \abs{M}},\nonumber\\
     m_t & = \frac{1}{2 \pi g N_f}  \left[\mathtt{k} + \frac{N_f}{2} {\rm{sgn}}(m) \right] = \frac{ \hat{\mathtt{k}}}{2 \pi g N_f}.
\end{align}

Because of the Chern-Simons term, the gauge fields acquire the topological mass $m_t$, evident from the pole in their propagators (note that $g$ has dimension of mass, while $\mathtt{k}$ is dimensionless). In general, it is important to note that in the Coulomb gauge, the temporal and spatial components of the gauge field decouple, leaving only the nonzero elements of the propagator given by $D_{00}$ and $D_{ij}$ \cite{RibhuKaul_QuantumCriticalityU(1)gauge_PRB2008}. However, since we are dealing with massive fermions and a Chern-Simons term, in our case, $D_{0i}$ and $D_{0i}$ are also non-zero.

At finite-$T$, we approximate $\Pi^{\mu\nu}$ by its small $|\bm{q}|, |\epsilon_n|$ dependence (which is only valid when both are smaller than $|m|$), and rewrite the general structure of $\Pi^{\mu\nu}$ using the Ward identity \cite{Dorey_QED3Superconductivity_1991kp,vafek2003thesis}
\begin{align}
\label{eq:Pi_munu_T_def}
    \Pi^{\mu\nu} (\bm{q},i \epsilon_n; T) = N_f g_a A^{\mu\nu} (q) + N_f g_b B^{\mu\nu}(q) + \frac{\Tilde{\mathtt{k}}}{2\pi} C^{\mu\nu} (q),
\end{align}
where
\begin{align}
\label{eq:ABC_munu}
    A_{\mu\nu} & = q^2 \left(\delta_{\mu 0} - \frac{q_\mu \epsilon_n}{q^2}\right) \frac{q^2}{\bm{q}^2} \left(\delta_{0\nu} - \frac{\epsilon_n q_\nu}{q^2}\right), \nonumber\\
    B_{\mu\nu} & = \bm{q}^2 \delta_{\mu i} \left(\delta_{ij} - \frac{q_i q_j}{\bm{q}^2}\right) \delta_{j\nu}, \nonumber\\
    C_{\mu\nu} & = \epsilon_{\mu\nu\rho} q_\rho.
\end{align}
$A^{\mu\nu}$ and $B^{\mu\nu}$ are the transverse projectors corresponding to $\bm{E}^2$ and $\bm{B}^2$, respectively, and $C^{\mu\nu}$ represents the odd-parity part of $\Pi^{\mu\nu}_f$. It is straightforward to see that
\begin{align}
 A_{\mu\nu} + B_{\mu\nu} = q^2 \delta_{\mu\nu}- q_\mu q_\nu.
\end{align}
The detailed expressions for $g_a$, $g_b$, and $\Tilde{\mathtt{k}}$ for general values of $\abs{m}/T$ are provided in the Appendix~\ref{app:effective_action} in Eqs.~\eqref{eq:g_mt_ainf} and \eqref{eq:g_mt_a0}. Inverting Eq.~\eqref{eq:Pi_munu_T_def}, we get the gauge field propagator (in the Coulomb gauge)
\begin{eqnarray}
\label{eq:D_gauge_T}
    D_{00} (\bm{q}, i \epsilon_n) & =&  \frac{\epsilon_n^2+(g_b/g_a) \bm{q}^2 }{g_a \bm{q}^2 \left[\epsilon_n^2+(g_b/g_a) \bm{q}^2  + \Tilde{m}_t^2\right]}, \label{eq:D_00_T} \nonumber \\
    D_{0i} (\bm{q}, i \epsilon_n) & =& - D_{i0} (\bm{q}, i \epsilon_n)  =  \frac{ - \Tilde{m}_t \epsilon_{ij} q_j}{g_a \bm{q}^2  \left[\epsilon_n^2+(g_b/g_a) \bm{q}^2  + \Tilde{m}_t^2\right]}, \label{eq:D_0i_i0_T} \nonumber \\
    D_{ij} (\bm{q}, i \epsilon_n) & =&  \frac{1}{ g_a \left[\epsilon_n^2+(g_b/g_a) \bm{q}^2  + \Tilde{m}_t^2\right]} \left(\delta_{ij} - \frac{q_i q_j}{\bm{q}^2} \right), \label{eq:D_ij_T}
\end{eqnarray}
where
\begin{equation}
\label{eq:mt_tilde}
    \Tilde{m}_t = \frac{ \Tilde{\mathtt{k}}}{2\pi g_a N_f}.
\end{equation}
We use the long wavelength limits ($\bm{q}=0,i\epsilon_n\to 0$) for the couplings $g_{a(b)}.$ We do not consider the static limit ($i\epsilon_n=0,\bm{q}\to 0$) for $g_{a(b)}$ in our calculations because in this case at low temperatures and $\bm{q}\rightarrow 0$, $1/g_a \sim |\bm{q}|^2,$ (see Appendix~\ref{app:effective_action}) and consequently, the topological mass of the gauge field acquires the momentum dependence $\tilde{m}_t \sim |\bm{q}|^{2},$ which is inconsistent with the physical expectation of a finite boson mass in the topological phase.

It is easily checked that $\tilde{m}_t$ interpolates between $m_t \sim 6\hat{\mathtt{k}}|m|/N_f$ at low temperatures ($T/|m|\ll 1$) to $6\mathtt{k} |M|/N_f$ at high temperatures $|m| \ll T \lesssim |M|.$ 
\subsubsection{Heat-current operator and thermal Hall conductivity}

To compute the thermal Hall conductivity, we require two operators \cite{Qin_EnergyMagnetization_PRL2011}: the heat current, given by
\begin{equation}
    \bm{J}^Q (\bm{r}) \equiv \bm{J}^E(\bm{r}) - \mu\, \bm{J}^N (\bm{r}),
\end{equation}
and the heat density
\begin{equation}
    K (\bm{r}) \equiv h(\bm{r}) - \mu\, n(\bm{r}),
\end{equation}
where $\mu, \bm{J}^E(\bm{r}), \bm{J}^N(\bm{r}), h(\bm{r})$ and $n(\bm{r})$ are chemical potential, energy current, particle current, local energy density, and local number density operators, respectively. Therefore, at $\mu = 0$, it is sufficient to compute the energy current and energy density operators, which can be derived from the stress-energy tensor.

The gauge invariant stress-energy tensor of the theory (\ref{eq:L_Psi_CS_LargeN_kagome}) in Minkwoski signature is given by \cite{Guo_GaugeThermalHall_PRB2020}
\begin{align}
\label{eq:T_real_Minkwoski}
    T^\mu_{\phantom{\mu}\nu} = {} & \frac{i}{2}  \left[ \bar{\psi}^{\phantom{\dagger}}_{l} \gamma^\mu \left( \partial_\nu \psi^{\phantom{\dagger}}_{l} \right) - \left( \partial_\nu \bar{\psi}^{\phantom{\dagger}}_{l}\right) \gamma^\mu \psi^{\phantom{\dagger}}_{l} \right] \nonumber \\
    & + \frac{1}{\sqrt{N_f}} \bar{\psi}^{\phantom{\dagger}}_{l} \gamma^\mu a_\nu \psi^{\phantom{\dagger}}_{l}  - \eta^\mu_{\phantom{\mu}\nu} \, \mathcal{L}_{ N_f,\psi,a},
\end{align}
where repeated fermion flavor indices $l$ are summed over.

Since the stress-energy tensor in Eq.~(\ref{eq:T_real_Minkwoski}) depends on both fermions and gauge fields, we express it as follows:
\begin{align}
    T^\mu_{\phantom{\mu}\nu} = T^\mu_{\phantom{\mu}\nu,f} + T^\mu_{\phantom{\mu}\nu,g},
\end{align}
where the fermionic contribution is given by

\begin{align}
\label{eq:T_real_f}
   T^\mu_{\phantom{\mu}\nu, f}  = {} &  \frac{i}{2} \left[ \bar{\psi}^{\phantom{\dagger}}_{l} \gamma^\mu \left( \partial_\nu \psi^{\phantom{\dagger}}_{l} \right) - \left( \partial_\nu \bar{\psi}^{\phantom{\dagger}}_{l}\right) \gamma^\mu \psi^{\phantom{\dagger}}_{l} \right] \nonumber\\
   & - \eta^\mu_{\phantom{\mu}\nu} \, \bigg{[} \frac{i}{2} \bar{\psi}^{\phantom{\dagger}}_{l} \gamma^\alpha \left( \partial_\alpha \psi^{\phantom{\dagger}}_{l} \right)  - \frac{i}{2} \left( \partial_\alpha \bar{\psi}^{\phantom{\dagger}}_{l}\right) \gamma^\alpha \psi^{\phantom{\dagger}}_{l} \nonumber\\
   & - m \, \bar{\psi}^{\phantom{\dagger}}_{l} \psi^{\phantom{\dagger}}_{l} \bigg{]},
\end{align}
and the gauge bosonic contribution is
\begin{align}
\label{eq:T_real_g}
    T^\mu_{\phantom{\mu}\nu, g} & =  \frac{1}{\sqrt{N_f}} \bar{\psi}^{\phantom{\dagger}}_{l}  \left( \gamma^\mu a_\nu  - \eta^\mu_{\phantom{\mu}\nu} \gamma^\alpha a_\alpha \right) \psi^{\phantom{\dagger}}_{l} \nonumber \\
    & =  \frac{1}{\sqrt{N_f}} \psi^{\dagger}_{l}  \left( \gamma^0 \gamma^\mu a_\nu  - \eta^\mu_{\phantom{\mu}\nu} \gamma^0 \gamma^\alpha a_\alpha \right) \psi^{\phantom{\dagger}}_{l}.
\end{align}

\begin{figure}[tb]
    \includegraphics[width=\linewidth]{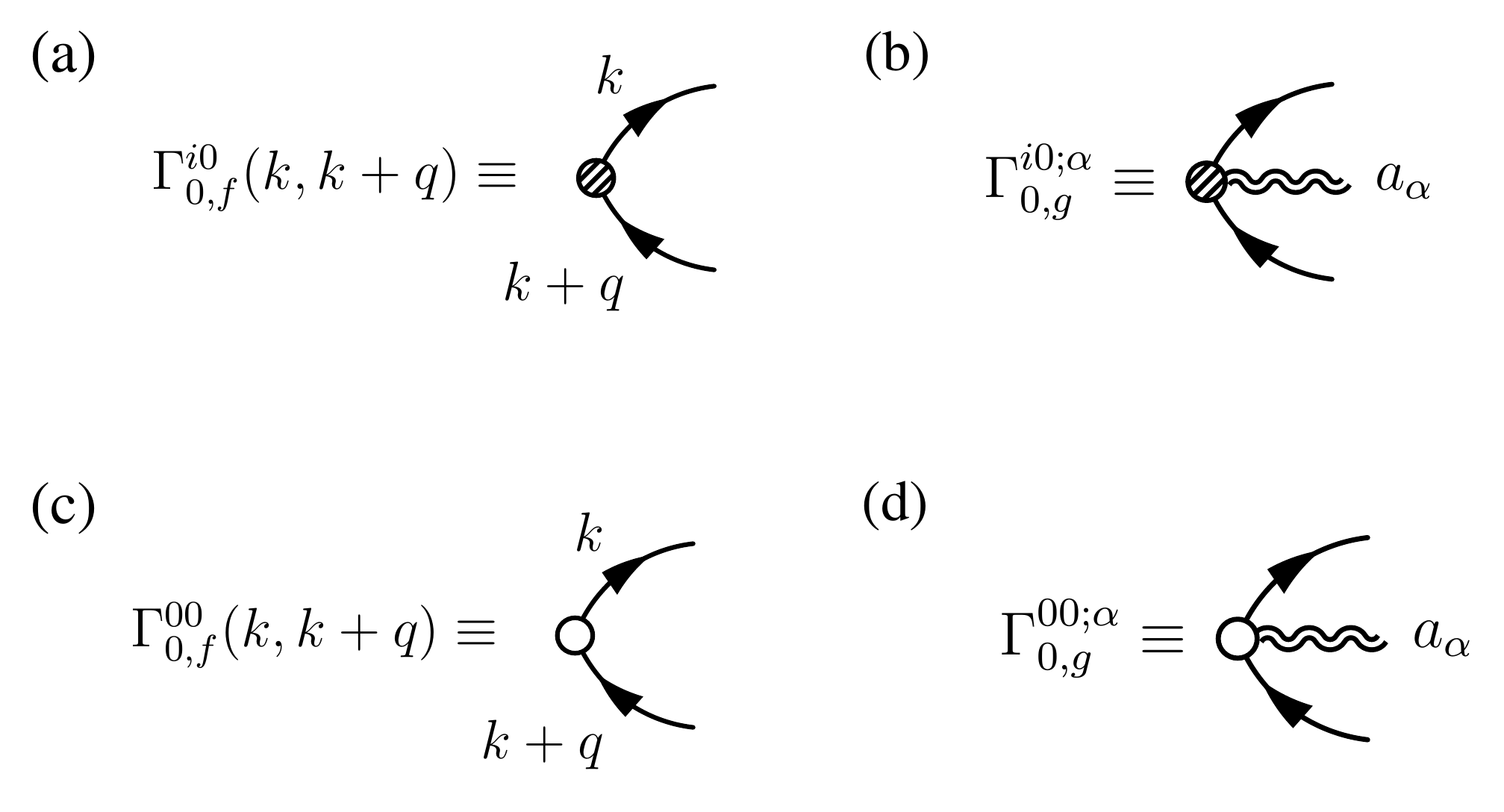}
    \caption{(\textbf{a})-(\textbf{d}) The stress tensor vertices in Eqs.~\eqref{eq:vertex_i0_f}, \eqref{eq:vertex_i0_g}, \eqref{eq:vertex_00_f}, and \eqref{eq:vertex_00_g}, respectively.}
    \label{fig:Thermal vertices}
\end{figure}

Taking the space-time Fourier transform of $T^\mu_{\phantom{\mu}\nu, f}$ in Eq.~(\ref{eq:T_real_f}) yields
\begin{align}
\label{eq:T_momentum_f}
    T^\mu_{\phantom{\mu}\nu, f} (q) = {} & \int \frac{d^3 k}{(2\pi)^3} \, \bigg{\{} \bar{\psi}^{\phantom{\dagger}}_{l}(k) \left[ \gamma^\mu \left(k+\frac{q}{2}\right)_\nu \right] \psi^{\phantom{\dagger}}_{l}(k+q) \nonumber\\
    & - \eta^\mu_{\phantom{\mu}\nu} \, \bar{\psi}^{\phantom{\dagger}}_{l}(k) \left[ \gamma^\alpha \left(k+\frac{q}{2}\right)_\alpha - m \right] \psi^{\phantom{\dagger}}_{l}(k+q) \bigg{\}}.
\end{align}


Switching to the Matsubara frequency domain from Eq.~(\ref{eq:T_momentum_f}), we obtain the fermionic part of the energy-current 
\begin{align}
    T^{i0}_{f}(\bm{q}, i \epsilon_n) = {} & - T \sum_{i\omega_n} \int  \frac{d^2 \bm{k}}{(2\pi)^2} \, \left( i \omega_n + \frac{i \epsilon_n}{2} \right) \nonumber \\
     & \times \psi^\dagger_l (\bm{k}, i\omega_n) \, \tau^i \, \psi^{\phantom{\dagger}}_{l} (\bm{k}+\bm{q}, i\omega_n+ i \epsilon_n),
\end{align}
where $\omega_n$ and $\epsilon_n$ are fermionic and bosonic Matsubara frequencies, respectively. This defines the fermionic part of the heat/energy-current vertex, illustrated in Fig.~\ref{fig:Thermal vertices}a,
\begin{align}
\label{eq:vertex_i0_f}
    \Gamma^{i0}_f (k, k+q) = - \left( i \omega_n + \frac{i \epsilon_n}{2} \right) \tau^i,
\end{align}
and we define the gauge bosonic part of the heat/energy-current vertex from Eq.~(\ref{eq:T_real_g}), as illustrated in Fig.~\ref{fig:Thermal vertices}b, as follows:
\begin{align}
\label{eq:vertex_i0_g}
    \Gamma^{i0; \alpha}_g = - \frac{1}{\sqrt{N_f}} i \tau^i \delta^{0\alpha}.
\end{align}
From Eq.~(\ref{eq:T_momentum_f}), we derive the fermionic part of the energy-density as follows:
\begin{align}
    T^{00}_{f}(\bm{q}, i \epsilon_n) = {} &  T \sum_{i\omega_n} \int  \frac{d^2 \bm{k}}{(2\pi)^2} \, \psi^\dagger_l (\bm{k}, i\omega_n) \bigg{[} \bm{\tau} \cdot \left( \bm{k}+\frac{\bm{q}}{2}\right) \nonumber \\
     & + m \tau^z \bigg{]}  \psi^{\phantom{\dagger}}_{l} (\bm{k}+\bm{q}, i\omega_n+ i \epsilon_n).
\end{align}
Similarly, the fermionic part of the energy-density vertex is defined, as shown in Fig.~\ref{fig:Thermal vertices}c, as follows;
\begin{align}
\label{eq:vertex_00_f}
    \Gamma^{00}_f (k, k+q) = \bm{\tau} \cdot \left( \bm{k}+\frac{\bm{q}}{2}\right) + m\, \tau^z,
\end{align}
and the gauge bosonic part is defined, as depicted in Fig.~\ref{fig:Thermal vertices}d, as follows:
\begin{align}
\label{eq:vertex_00_g}
    \Gamma^{00; \alpha}_g =  \frac{1}{\sqrt{N_f}} \left(-\tau^0 \delta^{0\alpha} + \tau^\alpha \right).
\end{align}

We are ultimately concerned with the stress tensor-stress tensor correlation functions, which are defined as
\begin{align}
\label{eq:Pi_TT_def}
    \Pi^{\mu \nu; \rho \lambda}(\bm{q},i \epsilon_n) = - \frac{T}{V} \left \langle   T^{\mu\nu}(\bm{q},i \epsilon_n) \, T^{\rho\lambda}(-\bm{q},- i \epsilon_n)  \right \rangle. 
\end{align}

Given our focus on the thermal Hall effect, we limit our analysis to the energy-density and energy-current sectors by setting $\nu = \lambda = 0$ in Eq.~(\ref{eq:Pi_TT_def}). Subsequently, we aim to isolate the component that is antisymmetric in $\mu$ and $\rho$ which should give the thermal Hall effect, achieved as follows:
\begin{align}
\label{eq:Pi_TT_AS_def}
    \Pi^{\mu 0; \rho 0}_{\text{AS}} = \frac{1}{2} \left( \Pi^{\mu 0; \rho 0} - \Pi^{\rho 0; \mu 0} \right).
\end{align}

As discussed in the Sec.~\ref{sec:kxy_parton_kagome}, the thermal Hall coefficient is composed of two elements: the standard linear response contribution $\kappa^{\text{Kubo}}_{xy}$ and the contribution arising from the energy magnetization $M_{Q}$ \cite{Qin_EnergyMagnetization_PRL2011}:
\begin{align}
    \kappa^{\phantom{0}}_{xy} = \kappa^{\text{Kubo}}_{xy} + \frac{2 M_{Q}}{T}.
\end{align}

The value of $\kappa^{\text{Kubo}}_{xy}$ is derived through the conventional Kubo formula and is determined by  the real-frequency retarded heat current-current correlation function $\Pi^{x 0; y 0}_{\text{AS},R}$, which is related to the imaginary-time correlation function as follows:
\begin{align}
    \kappa_{\text{Kubo}}^{xy} & = \frac{1}{T} \, \lim_{\epsilon \to 0} \, \frac{i}{\epsilon} \, \Pi^{x 0; y 0}_{\text{AS},R}( \epsilon + i 0^+) \nonumber \\
    & = \frac{1}{T} \, \lim_{\epsilon \to 0} \, \frac{i}{\epsilon} \, \Pi^{x 0; y 0}_{\text{AS}}(-i \epsilon_n \to \epsilon + i 0^+).
\end{align}
The determination of the energy magnetization proceeds as follows \cite{Qin_EnergyMagnetization_PRL2011}:
\begin{subequations}
\label{eq:EnergyMag_def}
    \begin{align}
       & 2 M_{Q}^{\phantom\dagger} - T \frac{\partial M_{\mathrm{Q}}^{\phantom\dagger}}{\partial T} = \Tilde{M}_{Q}, \label{eq:EnergyMag_def_diffeq}\\
       & \Tilde{M}_{Q} = \frac{1}{2 i} \left( \partial_{q_x} \Pi^{0 0; y 0}_{\text{AS}} - \partial_{q_y} \Pi^{0 0; x 0}_{\text{AS}} \right) \bigg\vert_{\bm{q} \rightarrow 0}. \label{eq:EnergyMag_def_curl} 
    \end{align}
\end{subequations}

\begin{figure}[tb]
    \includegraphics[width=0.9\linewidth]{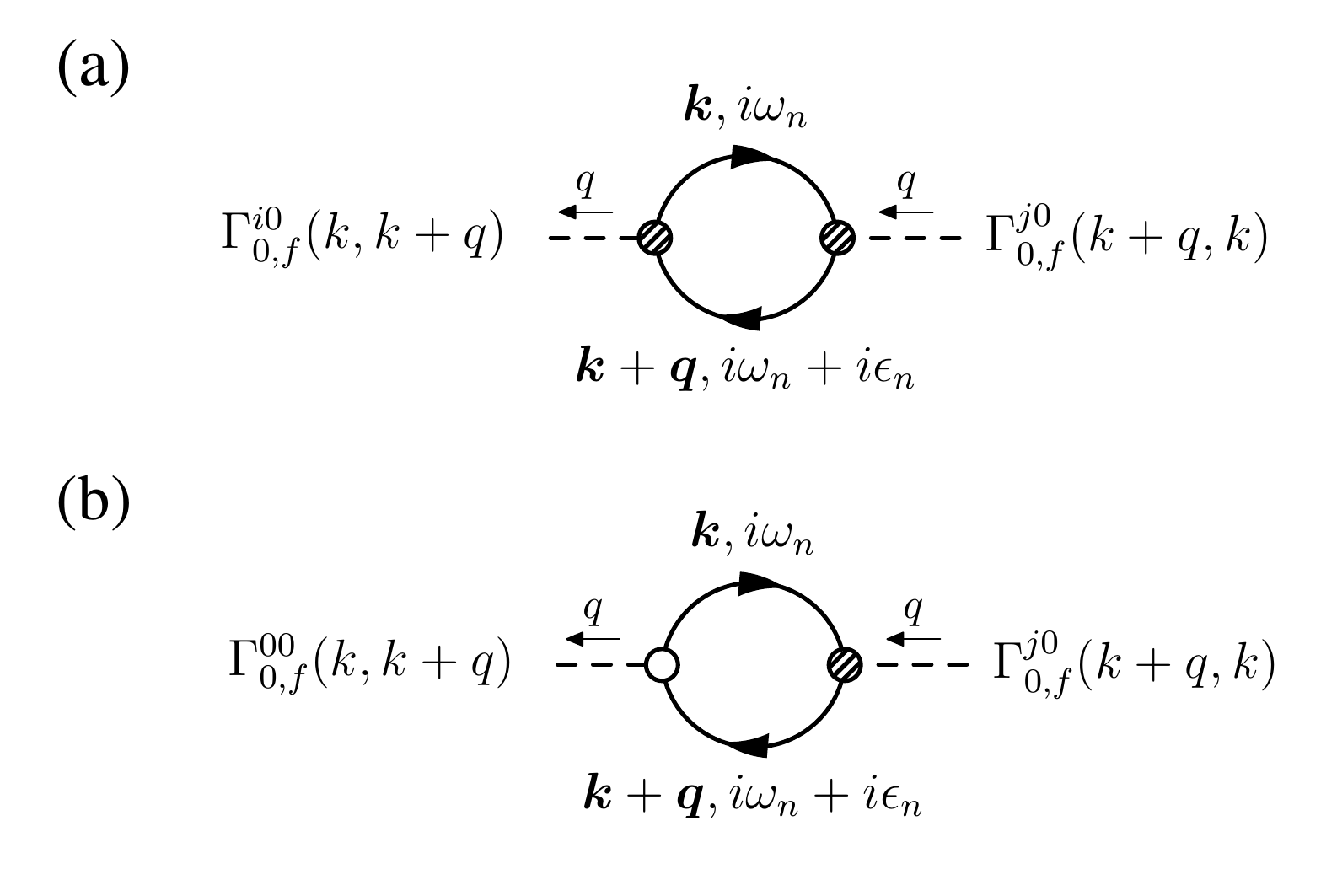}
    \caption{Feynman diagrams contributing to the thermal Hall conductivity at $\mathcal{O}(N_f).$ The bare fermion bubbles required to be evaluated for the (\textbf{a}) Kubo and (\textbf{b}) internal magnetization contribution to $\kappa_{xy}$ at $\mathcal{O}(N_f)$. The vertices are the heat current vertex $\Gamma^{i0}_f $ (shaded circle) and energy density vertex $\Gamma^{00}_f $ (empty circle).}
    \label{fig:O(N) contributions free}
\end{figure}

\subsection{Transport contribution to $\kappa^{\phantom{0}}_{xy}$ at $\mathcal{O}(N_f)$: the free Dirac fermion}

 At leading order for large $N_f$, we disregard gauge fluctuations, and the thermal Hall conductivity is mainly influenced by a single one-loop free Dirac fermion bubble at $N_f = \infty$, as depicted in the diagram shown in Figs.~\ref{fig:O(N) contributions free}a and~\ref{fig:O(N) contributions free}b, where the contribution is of $\mathcal{O}(N_f)$. 

The thermal Hall conductivity of the free Dirac fermion comprises two contributions: one originates from the light Dirac fermion with mass $m,$ and the other arises from the heavy fermion that generates the Chern-Simons term in Eq.~(\ref{eq:L_Psi_CS_LargeN_kagome}). Upon combining the contributions of the light and heavy Dirac fermions, the resulting thermal Hall conductivity is given by \cite{Guo_GaugeThermalHall_PRB2020}
\begin{equation}
\label{eq:kxy_mM_LargeN}
    \kappa^{(0)}_{xy} = N_f\, \kappa^D_{xy} (m) + 2 \lvert \mathtt{k} \rvert \, \kappa^D_{xy} (M),
\end{equation}
where $2|\mathtt{k}|=2N_f$ is the number of heavy fermion bands, and $\kappa^D_{xy}$ is the thermal Hall conductivity of a single free Dirac fermion given by \cite{Guo_GaugeThermalHall_PRB2020}
\begin{align}
\label{eq:kxy_Dirac_LargeN}
    \kappa^D_{xy} (m) = {} & - \frac{\pi \, \text{sgn}(m)}{12}\,  T - \frac{m}{2 \pi } \bigg[\frac{2 \text{Li}_2\left(-e^{-\lvert m \rvert/T}\right)}{\lvert m \rvert /T}  \nonumber \\
  &  -  \ln \left(e^{-\lvert m \rvert /T}+1 \right) \bigg].
\end{align}
In the limit $\abs{m}/T \to \infty$ and $\abs{M}/T \to \infty$, Eq.~(\ref{eq:kxy_mM_LargeN}) simplifies to
\begin{align}
\label{eq:kxy_mM_O(1)_T0}
    \frac{\kappa^{(0)}_{xy}}{T} =  - \frac{\pi}{6} \, \mathrm{sgn}( \hat{\mathtt{k}} ) \, |\hat{\mathtt{k}}|,
\end{align}
which is exactly the first term in Eq.~(\ref{eq:kxy_DCS_exact}).

\begin{figure}[tb]
    \includegraphics[width=0.85\linewidth]{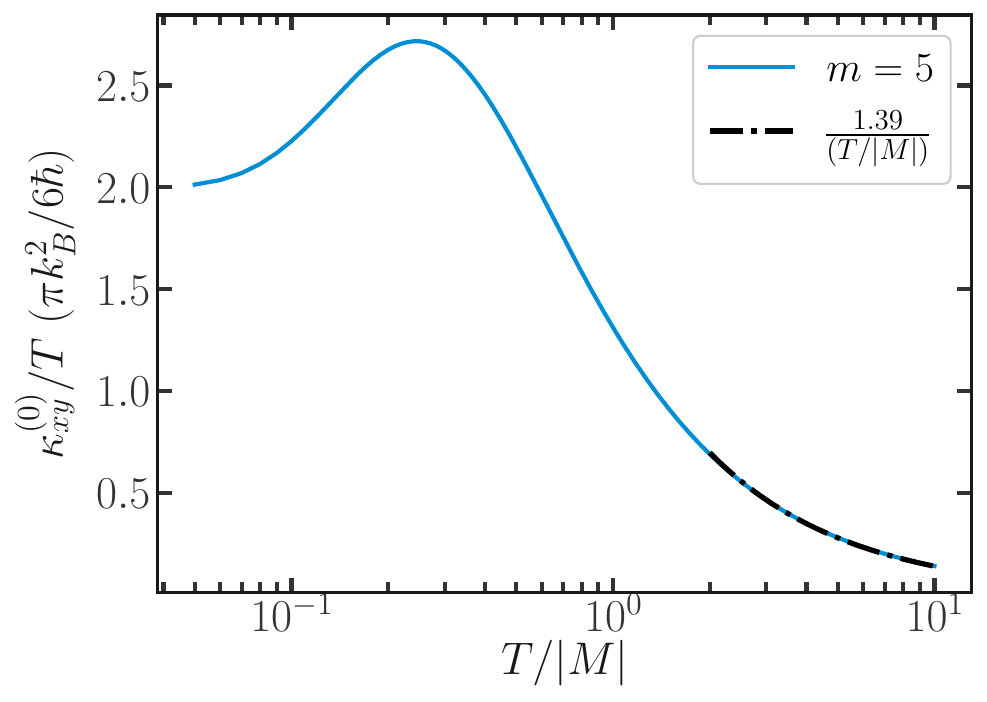}
    \caption{The bare fermion contribution to the thermal Hall conductivity as obtained in Eq.~(\ref{eq:kxy_mM_LargeN}) in the continuum theory. We plot $\kappa_{xy}/T$ for different light fermion mass $m$ and heavy fermion mass $M = -15$. The results are in qualitative agreement with the parton mean-field calculation on the lattice as shown in Fig.~\ref{fig:Kagome Thermal Hall parton}. Quantum corrections from $\mathcal{O}(1)$ processes change the behavior qualitatively, renormalizing the quantized value and finite temperature behavior (see Fig.~\ref{fig:Kagome kxy DCS full zero and finite T}).}
    \label{fig:Kagome Thermal Hall DCS}
\end{figure}

Fig.~\ref{fig:Kagome Thermal Hall DCS} illustrates a plot of Eq.~(\ref{eq:kxy_Dirac_LargeN}) for the case of $N_f = \lvert \mathtt{k} \rvert = 4$. In the topological phase, the thermal Hall conductivity is precisely quantized to $2$ in the zero-temperature limit, consistent with mean-field theory. However, it exhibits a rapid increase with temperature, followed by a gradual decrease after reaching a maximum in the finite-temperature regime. Therefore, the continuum field theory yields a temperature-dependent $ \kappa_{xy}/T$ that exhibits qualitative agreement with the corresponding outcomes obtained from the lattice model as shown in Fig.~\ref{fig:Kagome Thermal Hall parton}. The unbounded excitation spectrum of the continuum theory results in a different asymptotic behavior compared to the $\kappa_{xy}^{(0)}/T \sim T^{-3}$ law we obtained for free fermions with bounded spectra. Here, using Eq. \ref{eq:kxy_mM_LargeN} and Eq. (\ref{eq:kxy_Dirac_LargeN}) we obtain the asymptote
\begin{align}
\label{eq:free_asymptote}
    \frac{\kappa_{xy}^{(0)}}{T} & \sim -\frac{N_f\ln 2}{2\pi (T/|M|)}\left(\frac{|m|}{|M|}\text{sgn}(m) + 2 \text{ sgn}(M)\right). 
\end{align}
The difference between the free and bound spectrum asymptotic behavior arises because for the system with the bound spectrum, the specific heat vanishes at high temperatures in accordance with a power-law.

\subsection{Transport contribution to $\kappa^{\phantom{0}}_{xy}$ at $\mathcal{O}(N_f)$: U$(1)$ gauge field fluctuations}

\begin{figure}[tb]
    \includegraphics[width=0.9\linewidth]{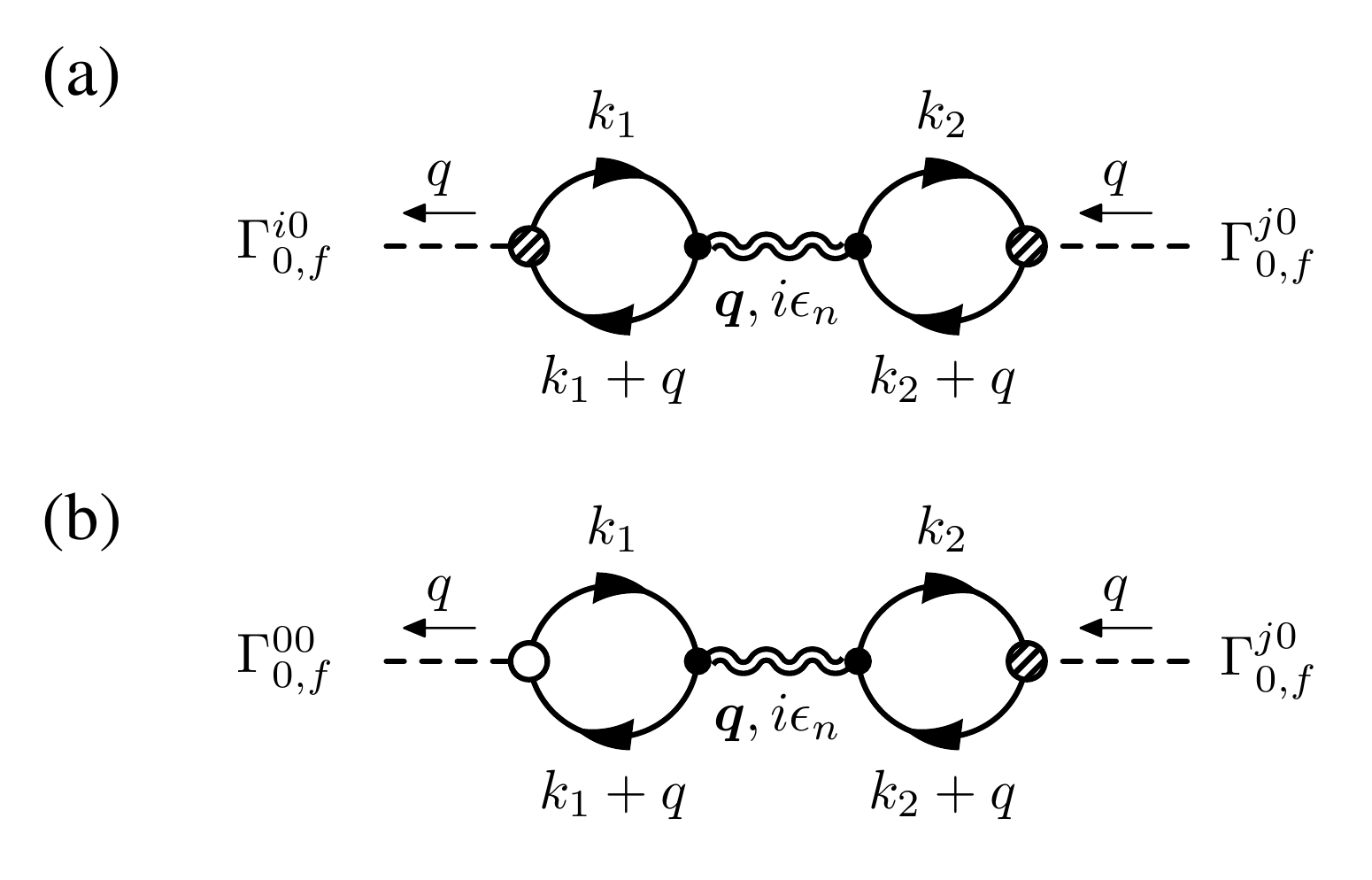}
    \caption{The primary Feynman graphs for the (\textbf{a}) Kubo and (\textbf{b}) energy magnetization contribution to $\kappa_{xy}$ at $\mathcal{O}(N_f)$ that potentially plays a leading role in the correction to the gauge field in the limit $N_f \rightarrow \infty$. The double wiggly represents the renormalized gauge boson propagator.}
    \label{fig:O(N) contributions gauge}
\end{figure}

Upon investigating the influence of gauge fluctuations within the $1/N_f$ expansion, it has been demonstrated in Ref.~\cite{Guo_GaugeThermalHall_PRB2020} that certain Feynman graphs, as shown in Figs.~\ref{fig:O(N) contributions gauge}a and ~\ref{fig:O(N) contributions gauge}b, have the potential to contribute to the thermal Hall conductivity at $\mathcal{O}(N_f).$ However, owing to the non-Abelian nature of the  SU$(2)$ gauge theory, it has been shown in the Ref.\cite{Guo_GaugeThermalHall_PRB2020} that both diagrams are proportional to $\text{Tr}(\sigma^a)$, where $\sigma^a$ represents Pauli matrices in color space. Consequently, these contributions are identically zero. Therefore, in the limit  $N_f \to \infty$, there are no modifications to the thermal Hall conductivity arising from SU$(2)$ gauge-field fluctuations within the SU$(2)$ Dirac Chern-Simons theory \cite{Guo_GaugeThermalHall_PRB2020}. 

However, since our U$(1)$ gauge field is Abelian, lacking Pauli matrices in color space, it remains uncertain a priori whether there are any corrections to the thermal Hall conductivity due to U$(1)$ gauge-field fluctuations. The determination of potential corrections necessitates the calculation of contributions from both the Kubo formula and the energy magnetization component. In this section, we investigate these graphs and confirm their disappearance, eliminating the need for any corrections. We will first focus on these diagrams in the zero-$T$ limit ($\abs{m}/T \rightarrow \infty$), and also assume that the ratio of light fermion and heavy fermion masses, $|m|/|M|\ll 1.$

We write the Matsubara correlators, illustrated in Figs.~\ref{fig:O(N) contributions gauge}a and ~\ref{fig:O(N) contributions gauge}b, as follows:

\begin{align}
 \label{eq:Pi_TT_(1)}
 \Pi^{\mu 0; \rho 0}_{(1)}(\bm{q},i \epsilon_n) = - N_f \mathcal{B}_1 (\bm{q}, i \epsilon_n)  D_{\alpha \beta}(\bm{q}, i \epsilon_n) \mathcal{B}_2 (\bm{q}, i \epsilon_n),
\end{align}
where
\begin{align}
   \mathcal{B}_1 (\bm{q}, i \epsilon_n) = {} & - T \sum_{i\omega_{1n}} \int  \frac{d^2 \bm{k}_1}{(2\pi)^2} \, \mathrm{Tr} \left[ \Gamma^{ \mu 0}_f (k_1, k_1+q) \right. \nonumber \\
  &\left. \mathcal{G}_0(\bm{k}_1+\bm{q}, i \omega_{1n} +i \epsilon_n)\,  \tau^\alpha \, \mathcal{G}_0(\bm{k}_1, i \omega_{1n}) \right], \nonumber\\
  \mathcal{B}_2 (\bm{q}, i \epsilon_n) = {} & - T \sum_{i\omega_{2n}} \int  \frac{d^2 \bm{k}_2}{(2\pi)^2} \,
  \mathrm{Tr} \left[ \Gamma^{\rho 0}_f (k_2+q, k_2) \right. \nonumber \\
  &\left. \mathcal{G}_0(\bm{k}_2, i \omega_{2n}) \, \tau^\beta \, \mathcal{G}_0(\bm{k}_2+\bm{q}, i \omega_{2n} +i \epsilon_n) \right].
\end{align}
Here $D_{\alpha \beta}(\bm{q}, i \epsilon_n)$ is the renormalized gauge boson propagator, the complete expressions of which can be found in Eq.~\eqref{eq:D_gauge}. Note that the frequency and momentum of the gauge propagator coincide with the external frequency and momentum of the thermal vertices. By employing Eq.~\eqref{eq:Pi_TT_AS_def}, we can isolate the part which is antisymmetric in $\mu\rho$ as
\begin{align}
\label{eq:Pi_TT_AS_(1)}
    \Pi^{\mu 0; \rho 0}_{\text{AS},(1)} (\bm{q}, i \epsilon_n) = \frac{1}{2} \left[ \Pi^{\mu 0; \rho 0}_{(1)} (\bm{q}, i \epsilon_n) - \Pi^{\rho 0; \mu 0}_{(1)} (\bm{q}, i \epsilon_n)\right].
\end{align}

We first calculate the transport contribution from the Kubo formula by setting $\mu = x$ and $\rho = y$ in Eq.~\eqref{eq:Pi_TT_AS_(1)}. Additionally, for acquiring the DC response, it is essential to perform an analytical continuation to real frequencies, where $i \epsilon_n$ transforms into $\epsilon + i 0^+$. Subsequently, the limit $\epsilon \rightarrow 0$ should be taken after $\bm{q} \rightarrow 0$. Consequently, our attention is directed towards the gauge boson propagator $D_{ij}(\bm{q} = 0, i \epsilon_n)$. As $\bm{q} \to 0$, $D_{\alpha\beta}$ in Eq.~\eqref{eq:D_gauge} experiences an infrared divergence (IR) due to the presence of a $1/\bm{q}^2$ factor in their expressions. Thus, the $\bm{q} = 0$ limit requires careful handling due to these subtleties. This IR divergence can be cured by replacing $\bm{q}^2$ with $\bm{q}^2 + \alpha^2$ and then setting $\alpha = 0$ after the integral has been evaluated. After taking the limit $\bm{q} \rightarrow 0$ in Eq.~\eqref{eq:Pi_TT_(1)}, we find that the denominator of $\Pi^{x 0; y 0}_{\text{AS},(1)}$ becomes even in $k_{1x}, k_{1y}, k_{2x}$ and $k_{2y}$, while the numerator consists terms proportional $k_{1x}k_{2y}$ and $k_{1y}k_{2x}$. Since these terms are odd in $k_{1x}$, $k_{1y}$, $k_{2x}$, and $k_{2y}$, they vanish upon integration over momenta and can be safely dropped. Therefore, the diagram in Fig.~\ref{fig:O(N) contributions gauge}a does not contribute to $\kappa^{\text{Kubo}}_{xy}$.

To compute the energy magnetization, we first need to evaluate $\Tilde{M}^{Q}_{(1)}$ in Eq.~\eqref{eq:EnergyMag_def}, which is expressed as follows:
\begin{align}
\label{eq:MQ_(1)}
    \Tilde{M}^{Q}_{(1)} = \frac{1}{2 i} \left( \partial_{q_x} \Pi^{0 0; y 0}_{\text{AS},(1)} - \partial_{q_y} \Pi^{0 0; x 0}_{\text{AS},(1)} \right) \bigg\vert_{\bm{q} \rightarrow 0}.
\end{align}
Before proceeding with the calculation of correlators to evaluate the energy magnetization, it is important to make a few preliminary remarks. First, $\Pi^{0 0; i 0}_{\text{AS},(1)}$ is evaluated by letting $\epsilon \rightarrow 0$ and then $\bm{q} \rightarrow 0$. So, we evaluate the summations over $\omega_{1n}$ and $\omega_{2n}$ at $\epsilon_n = 0$. Moreover, focusing on the limit $\bm{q} \rightarrow 0$, we can expand the full integrand in powers of $q_x$ and $q_y$. Subsequently, we retain only the terms of $\mathcal{O}(q_x) \left(\mathcal{O}(q_y)\right)$ in the correlator $\Pi^{0 0; y 0}_{\text{AS},(1)} \left(\Pi^{0 0; x 0}_{\text{AS},(1)}\right)$  that remain relevant after taking the partial derivative with respect to $q_x(q_y)$. This implies that we can directly substitute $\bm{q} = 0$ into the denominator of $\Pi^{0 0; i 0}_{\text{AS},(1)}$ which makes the denominator even in $\omega_{1n},k_{1x}, k_{1y},\omega_{2n}, k_{2x}$ and $k_{2y}$. Now after performing the partial derivative, we find that the numerator of Eq.~\eqref{eq:MQ_(1)} consists of terms that are odd in $\omega_{1n},k_{1x}, k_{1y},\omega_{2n}, k_{2x}$ and $k_{2y}$ and hence, cancel out when integration over all internal momenta and summing over all internal frequencies. Consequently, the diagram illustrating the contribution from magnetization in Fig.~\ref{fig:O(N) contributions gauge}b also goes to zero.

We also checked similarly using the finite-$T$ gauge propagator in Eq.~\eqref{eq:D_gauge_T} that  at $\mathcal{O}(N_f)$, there are no modifications to the thermal Hall conductivity arising from U$(1)$ gauge field fluctuations shown in Figs.~\ref{fig:O(N) contributions gauge}a and ~\ref{fig:O(N) contributions gauge}b.

\subsection{Transport contribution to $\kappa^{\phantom{0}}_{xy}$ at $\mathcal{O}(1)$: self-energy and vertex corrections}
\label{sec:kxy_DCS_O(1)}

\begin{figure}[tb]
    \includegraphics[width=0.9\linewidth]{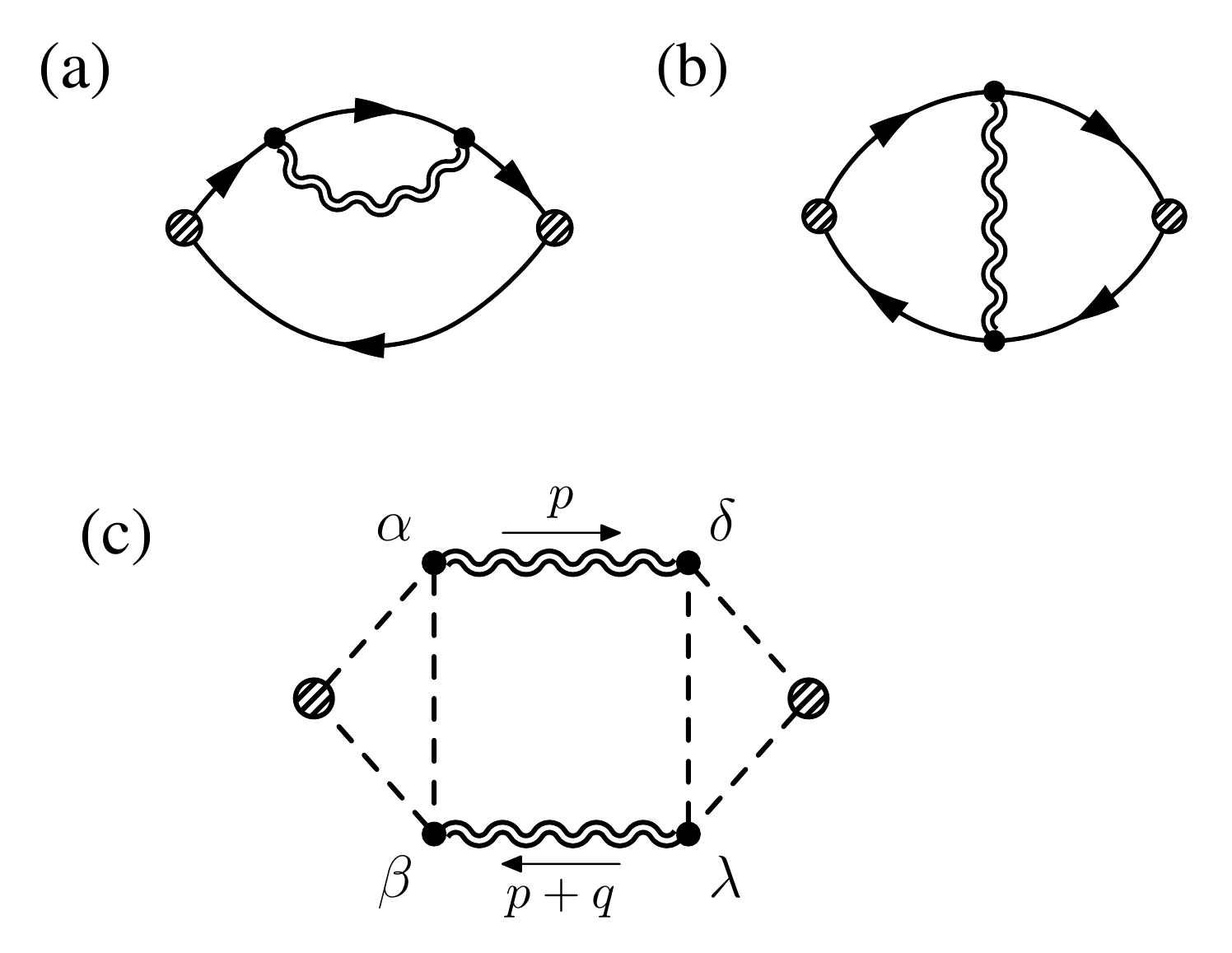}
    \caption{Three different types of diagrams, which contribute to $\kappa_{xy}$ at $\mathcal{O}(N_f)$: (\textbf{a}) the density of states, (\textbf{b}) the Maki-Thomson, and (\textbf{c}) the Aslamazov-Larkin diagrams. Furthermore, the energy magnetization correction involves evaluating similar diagrams, where the left thermal vertex is replaced with an energy-density vertex. The dashed lines in the triangles represent processes described in Fig.~\ref{fig:Sum of four AL diagrams}}
    \label{fig:O(1) contributions}
\end{figure}

We will now examine the contributions to $\kappa_{xy}$ from the theory in Eq.~\eqref{eq:L_Psi_CS_LargeN_nonrel} at $\mathcal{O}(1),$ which happen to be the leading ones involving the gauge bosons. The Feynman diagrams contributing to this order are shown in Fig.~\ref{fig:O(1) contributions}. The three corrections are respectively (a) the density of states (DoS), (b) the Maki-Thompson (MT) and (c) the Aslamazov-Larkin (AL) corrections. We discuss each of these below. 

\subsubsection{The density of states (DoS) and Maki-Thompson (MT) diagrams}

In the earlier work related to the SU$(2)$ case \cite{Guo_GaugeThermalHall_PRB2020}, it was argued that the DoS and MT \cite{Maki_diagram_1968, Thompson_diagram_1970} diagrams listed in Figs.~\ref{fig:O(1) contributions}a and \ref{fig:O(1) contributions}b, respectively, are not important in the $\abs{m}/T \rightarrow \infty$ limit. The same reasons are applicable here. The DoS diagram is essentially a self-energy correction to the fermion propagator, yielding a fermion mass correction $\delta m \propto 1/N_f$ in the zero momentum limit since the gauge boson mode is gapped. The DoS diagram can also generate fermion anomalous dimension at higher momentum. However, it turns out that this is exactly cancelled out by contributions from the MT diagram (a vertex correction) in the following manner. Using a gravitational Ward identity \cite{Grav_Ward_Brout_1996}, it can be shown that the anomalous dimension from the vertex correction cancels that from the self-energy, so the net effect is again a fermion mass renormalization $\delta m $ consistent with the self-energy calculation. This finite renormalization $\delta m $ which is an $\mathcal{O}(1/N_f)$ correction, can be ignored in the $\abs{m}/T \rightarrow \infty$ limit. Since the corrections are all $\mathcal{O}(1/N_f),$ we posit that the DoS and MT corrections are not important even at finite temperatures.

\subsubsection{The Aslamazov-Larkin diagram}
Thus, the crucial Feynman diagram is the Aslamazov-Larkin (AL) correction \cite{Aslamazov_Larkin}, shown in Fig.~\ref{fig:O(1) contributions}c , which consists of two dashed triangle diagrams and two renormalized gauge boson propagators. Based on the gauge invariant stress-energy tensor in Eq.~\eqref{eq:T_real_Minkwoski}, there are four types of diagrams that contribute to the dashed triangle diagram, as shown in Fig.~\ref{fig:Sum of four AL diagrams}.

The first two diagrams are fermion triangles, consisting of three fermion propagators with the stress-energy vertex given in Eqs.~\eqref{eq:vertex_i0_f} and \eqref{eq:vertex_00_f}, integrated over the fermion momentum and summed over the internal fermionic frequency. These diagrams have the form:

\begin{widetext}

\begin{align}
\Gamma^{\mu 0; \alpha \beta}_{(\text{AL}),1}(p, p+q) = & -  N_f \left(\frac{1}{\sqrt{N_f}} \right)^2 \, T \sum_{i\omega_{n}} \int  \frac{d^2 \bm{k}}{(2\pi)^2} \, \nonumber \\
  & \times \mathrm{Tr} \left[ \Gamma^{ \mu 0}_f (k, k+q) \, \mathcal{G}_0(\bm{k}+\bm{q}, i \omega_{n} +i \epsilon_n)\,  \tau^\beta \, \mathcal{G}_0(\bm{k}-\bm{p}, i \omega_{n} - i \Omega_n)\, \tau^\alpha\, \mathcal{G}_0(\bm{k}, i \omega_{n}) \right], \label{eq:Gamma_AL_1}\\  
\Gamma^{\mu 0; \alpha \beta}_{(\text{AL}),2}(p, p+q) = & -  N_f \left(\frac{1}{\sqrt{N_f}} \right)^2 \, T \sum_{i\omega_{n}} \int  \frac{d^2 \bm{k}}{(2\pi)^2} \, \nonumber \\
  & \times \mathrm{Tr} \left[ \Gamma^{ \mu 0}_f (k, k+q) \, \mathcal{G}_0(\bm{k}+\bm{q}, i \omega_{n} +i \epsilon_n)\,  \tau^\alpha \, \mathcal{G}_0(\bm{k}+\bm{p}+\bm{q}, i \omega_{n}+i \Omega_n + i\epsilon_n)\, \tau^\beta\, \mathcal{G}_0(\bm{k}, i \epsilon_{n}) \right]. \label{eq:Gamma_AL_2}
\end{align}

The trace integral of $\Gamma^{\mu 0; \alpha \beta}_{(\text{AL}),2}$ shares the same structure as $\Gamma^{\mu 0; \alpha \beta}_{(\text{AL}),1}$, but with different momentum dependence ($p, \alpha \leftrightarrow p+q, \beta$).

\begin{figure}[tb]
    \includegraphics[width=0.95\linewidth]{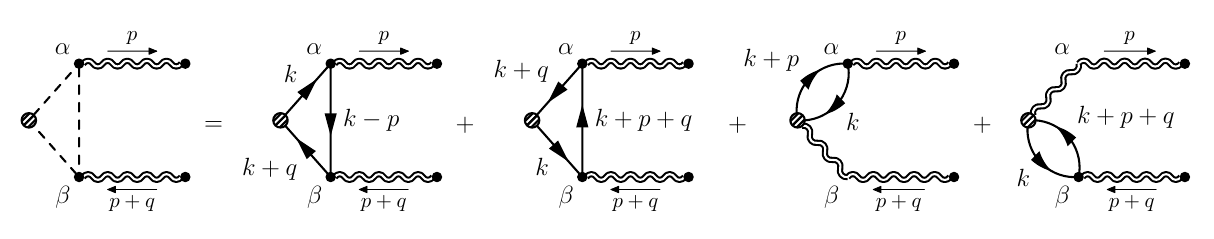}
    \caption{The dashed triangle diagram in Fig.~\ref{fig:O(1) contributions}c is written as a sum of four diagrams. The first two diagrams in the series of graphs contribute when the stress tensor vertex in \eqref{eq:vertex_i0_f} is used, while the last two diagrams contribute when the stress tensor vertex in \eqref{eq:vertex_i0_g} is used. All four diagrams are necessary to maintain gauge invariance.  }
    \label{fig:Sum of four AL diagrams}
\end{figure}

Similarly, the last two diagrams are fermion bubbles, each consisting of two fermion propagators with the stress-energy vertex given in Eqs.~\eqref{eq:vertex_i0_g} and \eqref{eq:vertex_00_g}. Their expressions can be obtained in the form
\begin{align}
\Gamma^{\mu 0; \alpha \beta}_{(\text{AL}),3}(p, p+q) & =  -  N_f \left(\frac{1}{\sqrt{N_f}} \right) \, T \sum_{i\omega_{n}} \int  \frac{d^2 \bm{k}}{(2\pi)^2} \, \mathrm{Tr} \left[ \Gamma^{ \mu 0; \beta}_g \, \mathcal{G}_0(\bm{k}, i \omega_{n})\,  \tau^\alpha \, \mathcal{G}_0(\bm{k}+\bm{p}, i \omega_{n}+i \Omega_n) \right],\label{eq:Gamma_AL_3}\\
\Gamma^{\mu 0; \alpha \beta}_{(\text{AL}),4}(p, p+q) & = -  N_f \left(\frac{1}{\sqrt{N_f}} \right) \, T \sum_{i\omega_{n}} \int  \frac{d^2 \bm{k}}{(2\pi)^2} \, \mathrm{Tr} \left[ \Gamma^{ \mu 0; \alpha}_g  \, \mathcal{G}_0(\bm{k}+\bm{p}+\bm{q}, i \omega_{n}+i \Omega_n + i\epsilon_n)\, \tau^\beta\, \mathcal{G}_0(\bm{k}, i \epsilon_{n}) \right]. \label{eq:Gamma_AL_4}
\end{align}

By combining \eqref{eq:Gamma_AL_1}, \eqref{eq:Gamma_AL_2}, \eqref{eq:Gamma_AL_3} and \eqref{eq:Gamma_AL_4}, we can write the final expression of the dashed triangle diagram as
\begin{align}
\label{eq:Gamma_AL_sum}
    \Gamma^{\mu 0; \alpha \beta}_{f,(\text{AL})} =\Gamma^{\mu 0; \alpha \beta}_{(\text{AL}),1} +\Gamma^{\mu 0; \alpha \beta}_{(\text{AL}),2} + \Gamma^{\mu 0; \alpha \beta}_{(\text{AL}),3} +\Gamma^{\mu 0; \alpha \beta}_{(\text{AL}),4}.
\end{align}

After extensive calculations, the details of which can be found in Appendix~\ref{app:AL_diagram_calculation}, we compute $\Gamma^{\mu 0; \alpha \beta}_{(\text{AL})}$ for a general value of $\abs{m}/T$. However let us first consider the zero-temperature limit $\abs{m}/T \rightarrow \infty$, where we obtain (see Appendix~\ref{app:AL_diagram_calculation}):
\begin{align}
    \Gamma^{0 0; \alpha \beta}_{f,(\text{AL})} (p, p+q) & = \frac{1}{12 \pi \abs{m}} \Lambda^{0 0; \alpha \beta}, \label{eq:Gamma_AL_00_zeroT}\\
     \Gamma^{j 0; \alpha \beta}_{f,(\text{AL})} (p, p+q) & = \frac{i}{12 \pi \abs{m}} \Lambda^{j 0; \alpha \beta}. \label{eq:Gamma_AL_i0_zeroT}
\end{align}
where
\begin{align}
    \Lambda^{\mu 0; \alpha \beta} (p, p+q) = {} & \left[ p_\mu (\Omega_n + \epsilon_n ) +  \Omega_n ( p + q)_\mu- \delta_{\mu 0} \bm{p} \cdot (\bm{p} + \bm{q}) \right] \delta_{\alpha \beta}  + p \cdot (p+q) \left( \delta_{\mu \alpha} \delta_{0 \beta} + \delta_{\mu \beta} \delta_{0 \alpha} \right) \nonumber\\
    & - (p+q)_\alpha \left[ \delta_{\beta \mu} \Omega_n + \delta_{\beta 0} p_\mu  \right] - p_\beta \left[ \delta_{\alpha \mu} (\Omega_n + \epsilon_n ) + \delta_{\alpha 0} (p + q)_\mu\right] + \delta_{\mu 0} (p+q)_\alpha p_\beta .
\end{align}
This matches precisely with the stress-tensor vertex of the Maxwell-Chern-Simons theory \cite{Guo_GaugeThermalHall_PRB2020}. However, we cannot completely disregard the stress-tensor vertex of the Maxwell-Chern-Simons theory, which arises from the integration process of the heavy Dirac fermion. In this way, we arrive at
\begin{align}
    \Gamma^{0 0; \alpha \beta}_{(\text{AL})} (p, p+q) & = g \Lambda^{0 0; \alpha \beta}, \label{eq:Gamma_00_zeroT}\\
    \Gamma^{j 0; \alpha \beta}_{(\text{AL})} (p, p+q) & = i g \Lambda^{j 0; \alpha \beta}, \label{eq:Gamma_i0_zeroT}
\end{align}
where $g$ is defined in Eq.~\eqref{eq:g_mt_def}.

Given the effective vertex function, we proceed to calculate the corresponding AL current-current correlation function, as shown in Fig.~~\ref{fig:O(1) contributions}c, which takes the form
\begin{align}
\label{eq:Pi_AL_def}
    \Pi^{\mu 0; \rho 0}_{(\text{AL})}(\bm{q},i \epsilon_n) = - T \sum_{i\Omega_{n}} \int  \frac{d^2 \bm{p}}{(2\pi)^2}  \, \frac{1}{2} \, \Gamma^{\mu 0; \alpha \beta}_{(\text{AL})} (p, p+q) \,  D_{\beta \lambda}(p+q) \, \Gamma^{\rho 0; \lambda \delta}_{(\text{AL})} (p+q, p) \,  D_{\delta \alpha}(p).
\end{align}
A symmetry factor of $1/2$ has been included here. The above expression is valid for arbitrary $|m|/T,$ provided the temperature-dependent expressions for $\Gamma_{(\text{AL})}^{\mu 0;\alpha\beta}$ and $D_{\beta\lambda}$ etc. obtained in Appendix~\ref{app:AL_diagram_calculation} and Appendix~\ref{app:effective_action} respectively are used. For simplicity, we first proceed to calculate in the regime $|m|/T \rightarrow \infty,$ where further analytic simplification is possible.

We focus on the antisymmetric component in $\mu \rho$, which is responsible for the thermal Hall effect:
\begin{align}
\label{eq:Pi_TT_AS_(AL)}
    \Pi^{\mu 0; \rho 0}_{\text{AS},(\text{AL})} (\bm{q}, i \epsilon_n) & = \frac{1}{2} \left[ \Pi^{\mu 0; \rho 0}_{(\text{AL})} (\bm{q}, i \epsilon_n) - \Pi^{\rho 0; \mu 0}_{(\text{AL})} (\bm{q}, i \epsilon_n)\right] \nonumber \\
    & = \frac{T}{2} \sum_{i\Omega_{n}} \int  \frac{d^2 \bm{p}}{(2\pi)^2} \, \frac{m_t\, \epsilon^{\mu\rho\sigma} \mathcal{N}_\sigma (p,q)}{\left( \Omega_n^2 + \bm{p}^2 + m_t^2 \right) \left( (\Omega_n+\epsilon_n)^2 + \abs{\bm{p}+\bm{q}}^2 + m_t^2 \right)},
\end{align}
where $\mathcal{N}_\sigma (p,q)$ represent three distinct polynomials in terms of $p$ and $q$:
\begin{align}
   & \mathcal{N}_0  = (q_x^2 + q_y^2) \Omega_n  + (p_x q_x + p_y q_y) (\epsilon_n + 2 \Omega_n), \nonumber \\
    &  \mathcal{N}_x = - i \left[ 2 p_x^2 q_x + p_y q_y (q_x + 2 p_x) - \Omega_n q_x (\epsilon_n + 2 \Omega_n) + p_x( \epsilon_n^2 + 2 q_x^2 + q_y^2 + 2 \epsilon_n \Omega_n) \right] , \nonumber \\
    &  \mathcal{N}_y = - i \left[ 2 p_y^2 q_y + p_x q_x (q_y + 2 p_y) - \Omega_n q_y (\epsilon_n + 2 \Omega_n) + p_y( \epsilon_n^2 +  q_x^2 + 2 q_y^2 + 2 \epsilon_n \Omega_n) \right].
\end{align}
We note that $\Pi^{\mu 0; \rho 0}_{\text{AS},(\text{AL})}$ satisfies the Ward identity, which, however, tells us that
\begin{align}
    q_\mu \Pi^{\mu 0; \rho 0}_{\text{AS},(\text{AL})} = 0, \;\; q_\rho \Pi^{\mu 0; \rho 0}_{\text{AS},(\text{AL})} = 0.
\end{align}

This suggests that $\Pi^{\mu 0; \rho 0}_{\text{AS},(\text{AL})}$ is proportional to $\epsilon^{\mu\rho\sigma} q_\sigma$. It is convenient to extract the tensor structure in the following way \cite{Guo_GaugeThermalHall_PRB2020}:
\begin{align}
\label{eq:Pi_AL_Ward}
    \Pi^{\mu 0; \rho 0}_{\text{AS},(\text{AL})} (\bm{q}, i \epsilon_n)= m_t \, \epsilon^{\mu\rho\sigma} q_\sigma \, \mathcal{P}(\bm{q}, i \epsilon_n),
\end{align}
and 
\begin{align}
\label{eq:P_cal}
    \mathcal{P}(\bm{q}, i \epsilon_n) = \frac{1}{2 (\epsilon_n^2 + \bm{q}^2)}\, T \sum_{i\Omega_{n}} \int  \frac{d^2 \bm{p}}{(2\pi)^2} \, \frac{\epsilon_n \mathcal{N}_0 + q_x \mathcal{N}_x + q_y \mathcal{N}_y}{\left( \Omega_n^2 + \varepsilon_{\bm{p}}^2  \right) \left( (\Omega_n+\epsilon_n)^2 + \varepsilon_{\bm{p}+\bm{q}}^2  \right)},
\end{align}
where $\varepsilon_{\bm{p}}^2 = \abs{\bm{p}}^2 + m_t^2$.

After performing the sum over $\Omega_n$ in Eq.~\eqref{eq:P_cal} by contour integration, we obtain $\mathcal{P}(\bm{q}, i \epsilon_n) = \mathcal{P}_0(\bm{q}, i \epsilon_n) + \mathcal{P}_\beta(\bm{q}, i \epsilon_n)$. Here, $\mathcal{P}_0(\bm{q}, i \epsilon_n)$ is the zero-temperature contribution, and the finite-temperature part in which we are interested reads
\begin{align}
\label{eq:P_cal_beta}
    \mathcal{P}_\beta (\bm{q}, i \epsilon_n) =  \frac{1}{2 (\epsilon_n^2 + \bm{q}^2)}\, \int  \frac{d^2 \bm{p}}{(2\pi)^2} \, \frac{1}{\left( (\varepsilon_{\bm{p}} + i \epsilon_n)^2 - \varepsilon_{\bm{p}+\bm{q}}^2 \right) \left( (\varepsilon_{\bm{p}} - i \epsilon_n)^2 - \varepsilon_{\bm{p}+\bm{q}}^2 \right)} \, \left[ \frac{n_B(\varepsilon_{\bm{p})}}{\varepsilon_{\bm{p}}}\, \mathcal{M}_1 + \frac{n_B(\varepsilon_{\bm{p}+\bm{q})}}{\varepsilon_{\bm{p}+\bm{q}}}\, \mathcal{M}_2   \right],
\end{align}

where
\begin{align}
    \mathcal{M}_1 = {} & (1+i) \left\{ (1+i) \bm{q}^2 \varepsilon_{\bm{p}}^4 - \bm{p} \cdot \bm{q} \left[ (1+i) ( \bm{p}\cdot \bm{q} + \bm{q}^2 ) + i \epsilon_n^2  \right] (\varepsilon_{\bm{p}+\bm{q}}^2 + \epsilon_n^2)  \right. \nonumber \\
  &{} \left. + (1+i) \varepsilon_{\bm{p}}^2 \left[ - \bm{q}^2 \varepsilon_{\bm{p}+\bm{q}}^2 + \bm{p}\cdot \bm{q} ( \bm{p} \cdot \bm{q} + \bm{q}^2 ) \right] - i \varepsilon_{\bm{p}}^2 \left[ 3 \bm{p} \cdot \bm{q} + (1+i)\bm{q}^2  \right] \epsilon_n^2 \right\}, \nonumber\\
   \mathcal{M}_2 = {} & (1+i) \left\{ (1+i) \bm{q}^2 \varepsilon_{\bm{p}+\bm{q}}^4 - ( \bm{p} \cdot \bm{q} + \bm{q}^2 ) \left[ (1+i) \bm{p} \cdot \bm{q}  - i \epsilon_n^2 \right] (\varepsilon_{\bm{p}}^2 + \epsilon_n^2)  \right. \nonumber \\
  &{} \left. + (1+i) \varepsilon_{\bm{p}+\bm{q}}^2  \left[ - \bm{q}^2 \varepsilon_{\bm{p}}^2 + \bm{p}\cdot \bm{q} ( \bm{p} \cdot \bm{q} + \bm{q}^2 ) \right] + \varepsilon_{\bm{p}+\bm{q}}^2  \left[ 3 \bm{p} \cdot \bm{q} + (1+2i)\bm{q}^2  \right] \epsilon_n^2 \right\}.
\end{align}
\end{widetext}

It is important to note that $\mathcal{P}_0(\bm{q}, i \epsilon_n)$ can be obtained from $\mathcal{P}_\beta(\bm{q}, i \epsilon_n)$ by substituting the Bose-Einstein distribution function $n_B$ with $1/2$ in Eq.~\eqref{eq:P_cal_beta}. Hereafter, we have omitted the zero-temperature part.

To compute the thermal Hall conductivity, we begin by examining the Kubo term  $\kappa^{\text{Kubo}}_{xy}$, which is evaluated in the $\epsilon \rightarrow 0$ limit after taking $\bm{q} \rightarrow 0$.  By examining Eq.~\eqref{eq:P_cal_beta}, we observe that $\mathcal{P}_\beta (\bm{q}=0, i \epsilon_n) = 0$, leading us to conclude that $\kappa^{\text{Kubo}}_{xy} = 0$. A similar conclusion can be reached by directly taking the Kubo or long-wavelength limit $\bm{q} \rightarrow  0$ in Eq.~\eqref{eq:Pi_TT_AS_(AL)}, where it is evident that $ \Pi^{x 0; y0}_{\text{AS},(\text{AL})} (\bm{q}=0, i \epsilon_n) = 0$.

The second contribution to the thermal Hall conductivity arises from the energy magnetization, which can be determined by solving the differential equation
\begin{align}
\label{eq:EnergyMag_def_AL}
     2 M^{Q}_{(\text{AL})}  - {} & T \frac{\partial M^{\mathrm{Q}}_{(\text{AL})}}{\partial T}  = \Tilde{M}^{Q}_{(\text{AL})}; \nonumber\\
    \Tilde{M}^{Q}_{(\text{AL})} = {} & \frac{1}{2 i} \bigg( \partial_{q_x} \Pi^{0 0; y 0}_{\text{AS},(\text{AL})}  - \partial_{q_y} \Pi^{0 0; x 0}_{\text{AS},(\text{AL})} \bigg) \bigg\vert_{\bm{q} \rightarrow 0} \nonumber \\
      = {} & i m_t  \mathcal{P}_\beta (\bm{q} \rightarrow 0, \epsilon_n = 0),
\end{align}
where the correlators on the RHS are evaluated in the \textit{static} limit, i.e., $\bm{q} \rightarrow 0$ after $\epsilon_n \rightarrow 0$. 

In the static limit, we can explicitly perform the momentum integration in Eq.~\eqref{eq:P_cal_beta} to obtain
\begin{align}
    \mathcal{P}_\beta (\bm{q} \rightarrow 0, \epsilon_n = 0) = &  -\frac{i}{4 \pi} \, \abs{m_t} \, n_B(\abs{m_t})  \nonumber \\
    &  + \frac{i}{4 \pi}\,  T \ln \left( 1- e^{-\abs{m_t}/T} \right).
\end{align}

We can now solve the differential equation in Eq.~\eqref{eq:EnergyMag_def_AL} analytically, which yields
\begin{align}
    M^{Q}_{(\text{AL})} = {} & c_1 \, T^2 + \frac{1}{4 \pi} \bigg{[}  m_t T \ln \left( 1- e^{-\abs{m_t}/T} \right) \nonumber \\
    & - 2\, T^2 \, \text{sgn}(m_t /T) \, \text{Li}_2 \left(e^{-\abs{m_t}/T} \right) \bigg{]},
\end{align}
where $c_1$ is an arbitrary constant. Thus the physical thermal Hall conductivity is given by 
\begin{align}
\label{eq:kxy_AL_c1}
  \frac{\kappa_{(\text{AL})}^{xy}}{T}  =  \frac{2 M^{Q}_{(\text{AL})}}{T^2}  = {} & 2 \, c_1  + \frac{1}{2 \pi} \bigg{[}  \frac{m_t}{T} \ln \left( 1- e^{-\abs{m_t}/T} \right) \nonumber \\
    &  - 2  \, \text{sgn}(m_t /T) \, \text{Li}_2 \left(e^{-\abs{m_t}/T} \right) \bigg{]},
\end{align}
which reproduces results of $\kappa_{xy}/T$ of pure Maxwell-Chern-simons theroy in Ref.~\cite{Guo_GaugeThermalHall_PRB2020}. The constant $c_1$ can be determined by inspecting the behavior of $\kappa_{(\text{AL})}^{xy}/T$ in the $\abs{m_t}/T \rightarrow 0$ limit. We observe that $\kappa_{(\text{AL})}^{xy}/T$ in Eq.~\eqref{eq:kxy_AL_c1} depends solely on the dimensionless variable $\abs{m_t}/T$ alone. In the high-temperature limit $T \gg m_t$,  the system is expected to be unaffected by the energy gap ($\sim m_t$) \cite{Guo_GaugeThermalHall_PRB2020}, hence $\kappa_{(\text{AL})}^{xy}/T$ should approach zero. Thus, in the $T \rightarrow \infty$ limit, Eq.~\eqref{eq:kxy_AL_c1} simplifies to
\begin{align}
    2\, c_1 - \text{sgn}(m_t) \frac{\pi}{6} = 0,
\end{align}
which implies that $c_1 = \text{sgn} (m_t)\, \pi/12$. Consequently, when $T \rightarrow 0$ with $m_t \neq 0$, we get
\begin{align}
\label{eq:kxy_AL_T=0}
    \frac{\kappa_{(\text{AL})}^{xy}}{T} = \frac{\pi}{6}\, \text{sgn}(m_t) = \frac{\pi}{6} \, \text{sgn}(\hat{\mathtt{k}}); \qquad \frac{\abs{m_t}}{T} \rightarrow \infty,
\end{align}
which yields the second term in Eq.~\eqref{eq:kxy_DCS_exact}. For temperatures exceeding $m_t,$ $\kappa_{(\text{AL})}^{xy}/T$ in Eq. \ref{eq:kxy_AL_c1} has the asymptotic form $\kappa_{(\text{AL})}^{xy}/T \sim \text{sgn}(\hat{\mathtt{k}})T^{-1}\ln(T/|m_t|),$ and the logarithmic factor crucially distinguishes this from the high temperature contribution from the noninteracting fermions.  The above analysis does not take into account the temperature dependence of the triangle vertices, nor that of the parameters $g$ and $m_t$ in the gauge field propagators. The temperature dependence of the triangle vertices is a feature that does not exist in the pure Maxwell-Chern-Simons counterpart, and reflects the fact that the fermion masses are finite. In the rest of the analysis below, we study the AL corrections up to higher temperatures, $T\lesssim |M|$, where we expect that a description in terms of the effective Maxwell-Chern-Simons theory for the heavy Dirac fermion is expected to apply.

\begin{figure}[tb]
    \includegraphics[width=0.90\linewidth]{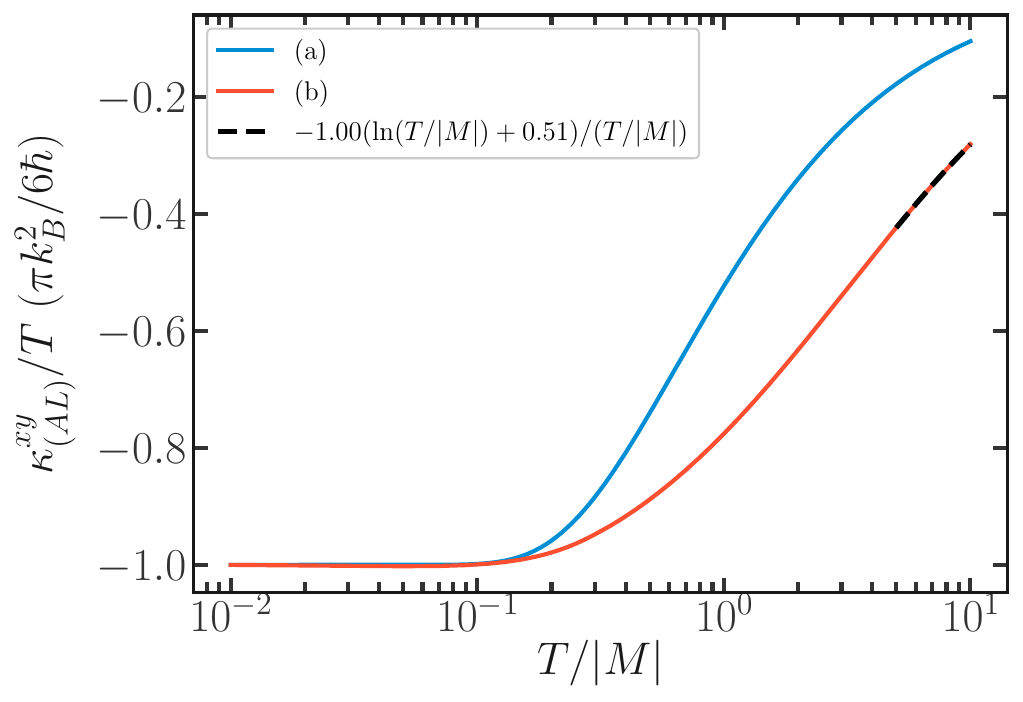}
    \caption{The thermal Hall conductivity from the AL diagrams. (\textbf{a}) $\kappa_{xy}/T$ in Eq.~\eqref{eq:kxy_AL_c1} is calculated under the assumption that the zero-temperature limit ($\abs{m}/T \rightarrow \infty$) of the dashed triangle vertices and gauge propagator have been taken into account. (\textbf{b}) $\kappa_{xy}/T$ is computed by numerically solving the differential equation in \eqref{eq:EnergyMag_def_AL}, using the temperature-dependent dashed triangle vertices and gauge propagator. Here, we choose light and heavy fermions of masses $m = 5$ and $M = -15$, respectively.}
    \label{fig:Kagome kxy DCS AL zero and finite T}
\end{figure}

\begin{figure}[tb]
    \includegraphics[width=0.88\linewidth]{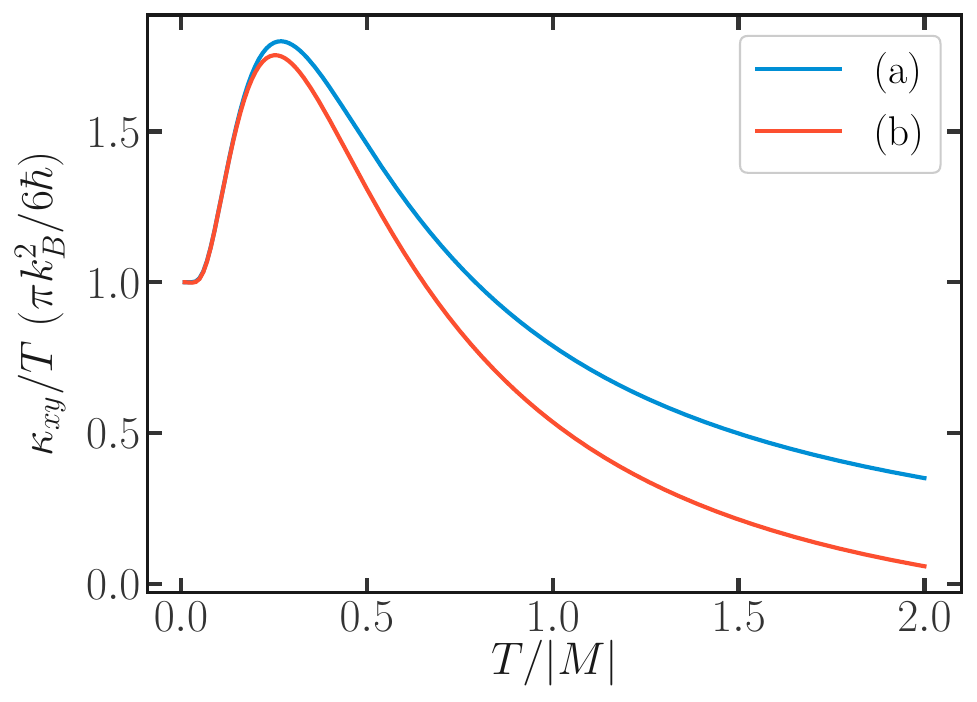}
    \caption{(\textbf{a}) The full thermal Hall conductivity is computed by adding Eqs.~\eqref{eq:kxy_mM_LargeN} and \eqref{eq:kxy_AL_c1}, where we choose $m=5$ and $M = -15.$ This bulk calculation is consistent with $\kappa_{xy}/T|_{T\rightarrow 0}$ in Eq.~\eqref{eq:kxy_DCS_exact} obtained earlier in this paper from gravitational anomaly arguments. (\textbf{b}) $\kappa_{xy}/T$ is computed by using the temperature-dependent dashed triangle vertices and gauge propagator.} 
    \label{fig:Kagome kxy DCS full zero and finite T}
\end{figure}

The AL corrections modify the free fermion results. Since $\text{sgn}(\hat{\mathtt{k}})=-1$ in our case, the AL corrections \emph{decrease} the quantized low-temperature value of $\kappa_{xy}/T$ from $\pi/3$ to $\pi/6.$ The AL corrections have a monotonous temperature dependence, smoothly decreasing from the quantized value towards zero with increasing temperature, as shown in Fig.~\ref{fig:Kagome kxy DCS AL zero and finite T}. Thus the overall temperature dependence of $\kappa_{xy}/T$, illustrated in Fig.~\ref{fig:Kagome kxy DCS full zero and finite T}, should show a non-monotonic profile with a somewhat sharper intermediate temperature peak than the free fermion plot of Fig. \ref{fig:Kagome Thermal Hall DCS}.

Now we analyze the higher temperature behavior. 
We begin by focusing on the Kubo conductivity. Using the finite-$T$ gauge field propagators from Eq. \ref{eq:D_gauge_T} and the triangle vertices calculated in the long-wavelength limit ($\epsilon_n\rightarrow0, \abs{\bm{q}}=0$), we find $\kappa^{\text{Kubo}}_{xy} = 0$ (see Appendix~\ref{app:AL_diagram_calculation}).

Next, we focus on the transport contribution from the energy magnetization. At finite temperatures, the dashed triangle diagrams in the static limit ($\epsilon_n =0, \abs{\bm{q}} \rightarrow 0$) have the form (see Appendix~\ref{app:AL_diagram_calculation}):
\begin{align}
\label{eq:Gamma_mu0_mag_def}
    \Gamma^{0 0; \alpha \beta}_{(\text{AL}),M} (\bm{p},i\Omega_n; T) = {} & A_{0;1}^M P_1^{\alpha\beta} (p) + A_{0;2}^M P_2^{\alpha\beta}(p) \nonumber\\ 
    & {} + A_{0;3}^M P_3^{\alpha\beta}(p), \nonumber\\
    \Gamma^{x 0; \alpha \beta}_{(\text{AL}),M} (\bm{p},i\Omega_n; T) = {} & A_{x;1}^M P_1^{\alpha\beta} (p) + A_{x;2}^M P_2^{\alpha\beta}(p) \nonumber\\
    & {} + A_{x;3}^M P_3^{\alpha\beta}(p)  + B_x^M \mathcal{X}^{\alpha\beta}(p), \nonumber\\
    \Gamma^{y 0; \alpha \beta}_{(\text{AL}),M} (\bm{p},i\Omega_n; T) = {} & A_{y;1}^M P_1^{\alpha\beta} (p) + A_{y;2}^M P_2^{\alpha\beta}(p) \nonumber\\\
    & {} + A_{y;3}^M P_3^{\alpha\beta}(p) + B_y^M \mathcal{Y}^{\alpha\beta}(p)
\end{align}

where the projectors are defined as
\begin{align}
\label{eq:stress_projector}
    P_1^{\alpha\beta} (p) &  = p^2 \left(\delta_{\alpha 0} - \frac{p_\alpha \Omega_n}{p^2}\right) \frac{p^2}{\bm{p}^2} \left(\delta_{0\beta} - \frac{\Omega_n p_\beta}{p^2}\right), \nonumber\\
    P_2^{\alpha\beta} (p) & = \bm{p}^2 \delta_{\alpha i} \left(\delta_{ij} - \frac{p_i p_j}{\bm{p}^2}\right) \delta_{j\beta}, \nonumber\\
    P_3^{\alpha\beta} (p) & = \epsilon^{\alpha\beta\rho} p_\rho, \nonumber\\
    \mathcal{X}^{\alpha\beta}(p) & = \begin{pmatrix}
        0 & p_y^2 & -p_x p_y \\
        p_y^2 & 0 & - p_y \Omega_n \\
        -p_x p_y &  - p_y \Omega_n & 2p_x\Omega_n
    \end{pmatrix}, \nonumber\\
    \mathcal{Y}^{\alpha\beta}(p) & = \begin{pmatrix}
       0 & -p_x p_y &  p_x^2 \\
        -p_x p_y & 2p_y\Omega_n & - p_x \Omega_n \\
        -p_x^2 &  - p_x \Omega_n & 0
    \end{pmatrix}.
\end{align}

As we are focusing on the case where $q \rightarrow 0$, we drop $(\bm{q}, i\epsilon_n)$ from the arguments and use a superscript $M$ to indicate that we are working in the static limit. After extensive calculations, we determine the coefficients of the above projectors for general values of $\abs{m}/T$, as follows (see Appendix~\ref{app:AL_diagram_calculation}):

\begin{align}
\label{eq:stress_coeff_mag}
    A_{0;1}^M (p) & = g_a (p),\,\,\,
    A_{0;2}^M (p) = - g_b (p),\nonumber\\
    A_{0;x}^M (p) & = i g_a (p),\,\,\,
    A_{0;y}^M (p)  = i g_a (p).
\end{align}
where the terms $g_{a(b)}$ also appear in the expression for the gauge propagator at finite temperature in Eq.~\eqref{eq:D_gauge_T}, with their detailed forms provided in Appendix~\ref{app:effective_action}. Sending $\epsilon_n =0, \abs{\bm{q}} \rightarrow 0$ in the static limit, we obtain
\begin{align}
   \Tilde{M}_Q = T \sum_{i\Omega_n} \int \frac{d^2\bm{p}}{(2\pi)^2} \frac{\Tilde{m}_t}{2 g_a} \left[ \frac{g_b \bm{p}^2 - 2 g_a \Omega_n^2}{\left( \Omega_n^2 + (g_b/g_a) \bm{p}^2 + \Tilde{m}_t^2 \right)^2}  \right],
\end{align}
where $\tilde{M}_Q$ is defined in Eq. (\ref{eq:EnergyMag_def_AL}).
By taking the values of $g_a$, $g_b$, and $\Tilde{m}_t$ in the ($\Omega_n\rightarrow0, \abs{\bm{p}}=0$) limit and performing the Matsubara summation and momentum integration analytically, we obtain the following result (see Appendix~\ref{app:AL_diagram_calculation}):
\begin{align}
\label{eq:M_Q_tilde_finiteT}
    \Tilde{M}_Q = {} & \frac{g_a \Tilde{m}_t}{2 g_b} \left\{ - \frac{T}{2\pi} \ln{\left[2\sinh{\left(\frac{\abs{\Tilde{m}_t}}{2 T}\right)} \right]} \right. \nonumber \\
  &{} \left. + \frac{\abs{\Tilde{m}_t}}{4\pi} \coth{\left(\frac{\abs{\Tilde{m}_t}}{2 T}\right)}  \right\}.
\end{align}
After the RHS is evaluated, we now solve the differential equation numerically. Fig.~\ref{fig:Kagome kxy DCS AL zero and finite T} shows the result of the two different approximations (i.e. with or without taking the temperature dependence of the gauge propagators as well as the triangle diagrams associated with massive light Dirac fermion excitations into account). When these temperature dependences are taken into account, the validity of the analysis is now extended to a significantly wider temperature range $0< T/|M|\lesssim 1.$ At even higher temperatures, the excitations of the heavy fermions will result in further finite temperature corrections to the polarizability, triangle diagrams, level and topological mass, similar to those encountered in the vicinity of $T/|m|\sim 1.$ Since $|M|$ is the only energy scale in this high temperature regime, we would get $\kappa_{(\text{AL})}^{xy}/T \sim \text{sgn}(M)T^{-1}\ln(T/|M|),$ for $T/|M|\gg 1.$ This behavior is in agreement with our numerical results in Fig.~\ref{fig:Kagome kxy DCS AL zero and finite T}. In Fig.~\ref{fig:Kagome kxy DCS full zero and finite T} we show the temperature dependence of the total thermal Hall conductivity taking the bare fermion and gauge fluctuations arising from the AL diagrams into account. 

We conclude with some comments on the temperature range of validity of the large-$N_f$ approximation. From Eq.~\ref{eq:free_asymptote}, we note that at high temperatures, $\kappa_{xy}/T \sim  \frac{N_f\ln 2}{2\pi (T/|M|)},$ and the $\mathcal{O}(1)$ AL perturbative corrections become comparable to the bare fermion contribution at a temperature $T=T_\text{high} \sim |M|e^{N_f \ln 2} \gg |M|.$

\section{Conclusion and Outlook}

In this study, we investigated the thermal Hall conductivity of the Heisenberg model on the kagome lattice in the presence of a scalar spin chirality interaction. We utilized a mean-field formulation with fermionic spinons to derive a low-energy continuum action for the CSL state featuring semionic topological order.

We took two different approaches. First, we introduced the fermionic spinon to do conventional mean-field decoupling of the spin model and subsequently, we numerically calculated the temperature dependence of the thermal Hall conductivity using the formula given in Ref.~\cite{Qin_EnergyMagnetization_PRL2011} for the non-interacting system. Second, we discussed how fluctuations in the gauge field could change the quantized values at zero temperature from $2$ to exactly $1$, based on gravitational anomaly arguments. The limitation of the gravitational anomaly arguments is that it is valid only in the zero temperature limit. In the second approach, we retained only the relevant low-energy spinon degrees of freedom and investigated the linear response thermal Hall conductivity in the resulting continuum Dirac Chern-Simons theory. The results from the two approaches were in quantitative agreement as $T\rightarrow 0.$

We also performed a large-$N_f$ perturbation expansion for systematic accounting of U$(1)$ gauge fluctuation effects in the thermal Hall response, with results summarized in Table.~\ref{tab:summary}. The leading $O(N_f)$ contributions to $\kappa_{xy}$ containing gauge boson propagators were shown to vanish identicaly. The physical origin of such vanishing is qualitatively different from the recently studied SU$(2)$ Dirac Chern-Simons theories in Ref.~\cite{Guo_GaugeThermalHall_PRB2020} where the vanishing of such contributions is on account of the tracelessness of the Pauli matrices in SU$(2)$ color space. The leading nontrivial gauge contributions are $\mathcal{O}(1)$ and are associated with Aslamazov-Larkin (AL) like diagrams containing two gauge propagators and two fermion loops. 

We obtained the full temperature dependence of the dashed triangle vertices and the gauge propagators, in a temperature-dependent Maxwell-Chern-Simons approximate form for the gauge field propagator. 
At low temperatures, the vertices associated with AL fermionic loops become essentially local and the contribution becomes completely equivalent to that obtained from the U$(1)$ Maxwell-Chern-Simons gauge theory \cite{Guo_GaugeThermalHall_PRB2020}. The AL corrections introduce a modification to the free fermion result, specifically by lowering the quantized value of $\kappa_{xy}/T$ at low temperatures. This correction halves the original value from $\pi/3$ to $\pi/6.$ However at nonzero temperatures, the temperature dependence of the vertices must be taken into account, for the light Dirac fermions and the gauge bosons have comparable masses. The temperature dependence of the triangle vertices and that of the gauge propagators largely compensate each other, saving the temperature dependence of the topological mass $m_t$ that now enters the calculations in this regime. We provide explicit expressions for the finite-temperature thermal Hall response by calculating the Aslamazov-Larkin diagram at finite temperature. This computation is performed within the framework of a temperature-dependent Maxwell-Chern-Simons action for the gauge field. At high temperatures, the AL diagrams make a logarithmic correction to the result for noninteracting fermions.

We conclude with some remarks about experimental data on the thermal Hall response of Kagome antiferromagnets. While quantized thermal Hall response has not been reported, a non-monotonic temperature dependence has been observed \cite{Watanabe_KagomeExp_volborthite_PNAS2016, Doki_KagomeExp_CaK_PRL2018, Akazawa_KagomeExp_CaK_PRX2020}, reminiscent of our parton theory. Recently, thermal transport measurements were performed on disordered herbertsmithite \cite{Barthelemy_heat_herbertsmithite_PRB2023}. Although the thermal Hall response was not significant, the field dependence of thermal conductivity was used to extract the defect contribution. The impurity contribution was not consistent with spin-$1/2$ defects. If the CSL ground state has semionic topological order as we argued here, then vacancy defects are likely to localize fractional moments. It would be interesting to see the behaviour of impurity magnetization in real systems as a function of a magnetic field. An important future problem is studying the thermal Hall response in the U$(1)$ spin liquid with vacancies and gauge fluctuations \cite{Kolezhuk_2006}.

\subsection*{Acknowledgements}
 A.M. thanks Subhro Bhattacharjee, Jagannath Das, Unmesh Ghorai, and Souparna Nath for useful discussions. A.M. and V.T. acknowledge the support of the Department of Atomic Energy, Government of India, under Project Identification No. RTI 4002. H.G. was supported by the Bethe-Wilkins-KIC postdoctoral fellowship at Cornell University. S. S. was supported by the U.S. National Science Foundation grant No. DMR-2245246 and by the Simons Collaboration on Ultra-Quantum Matter which is a grant from the Simons Foundation (651440, S.S.).

\appendix
\section{Mean-field Hamiltonian of the CSL state}
\label{app:MF_CSL}
In this appendix, we begin by diagonalizing the mean-field Hamiltonian in Eq.~(\ref{eq: Hspin_MF_compact_kagome}) in momentum for the flux configuration $\left[ \phi, \pi - 2\phi \right] = \left[3 \theta, \pi - 6 \theta\right]$, followed by a discussion of its band structure. To start, we demonstrate the mean-field decoupling of the three-spin interaction, expressed for a single triangle as:
    \begin{align}
    \label{eq:H_ijk}
    H_{\bm{ijk}}  = {} & J_{\chi} \bm{S}_{\bm{i}} \cdot (\bm{S}_{\bm{j}} \times \bm{S}_{\bm{k}}) \nonumber \\
         = {} & \frac{i J_{\chi}}{2} \left[ \left(S^+_{\bm i} S^-_{\bm j} - S^-_{\bm i} S^+_{\bm j} \right)S^z_{\bm{k}} + \left(S^+_{\bm{k}} S^-_{\bm{i}} - S^-_{\bm{k}} S^+_{\bm{i}} \right)S^z_{\bm{j}}\right. \nonumber \\
  & {} \left.+ \left(S^+_{\bm{j}} S^-_{\bm{k}} - S^-_{\bm{j}} S^+_{\bm{k}} \right)S^z_{\bm{i}}\right].
\end{align}

Using the parton representation in Eq.~(\ref{eq: Spin_to_parton}), we rewrite the first term in Eq.~(\ref{eq:H_ijk}) as a sum of products of six-fermion interaction terms

\begin{align}
    J_{\chi}\frac{i}{2} & \left(S^+_{\bm{i}} S^-_{\bm{j}} - S^-_{\bm{i}} S^+_{\bm{j}} \right)S^z_{\bm{k}} \nonumber \\ 
         = {} & J_{\chi}\frac{i}{4} \left(f^{\dagger}_{\bm{i}\uparrow} f^{\phantom\dagger}_{\bm{k}\uparrow} f^{\dagger}_{\bm{k}\uparrow} f^{\phantom\dagger}_{\bm{j}\uparrow} f^{\dagger}_{\bm{j}\downarrow} f^{\phantom\dagger}_{\bm{i}\downarrow}   -   f^{\dagger}_{\bm{i}\downarrow} f^{\phantom\dagger}_{\bm{j}\downarrow} f^{\dagger}_{\bm{j}\uparrow} f^{\phantom\dagger}_{\bm{k}\uparrow} f^{\dagger}_{\bm{k}\uparrow} f^{\phantom\dagger}_{\bm{i}\uparrow} \right. \nonumber \\
  &\left.  -   f^{\dagger}_{\bm{i}\uparrow} f^{\phantom\dagger}_{\bm{j}\uparrow} f^{\dagger}_{\bm{j}\downarrow} f^{\phantom\dagger}_{\bm{k}\downarrow} f^{\dagger}_{\bm{k}\downarrow} f^{\phantom\dagger}_{\bm{i}\downarrow}   +   f^{\dagger}_{\bm{i}\downarrow} f^{\phantom\dagger}_{\bm{k}\downarrow} f^{\dagger}_{\bm{k}\downarrow} f^{\phantom\dagger}_{\bm{j}\downarrow} f^{\dagger}_{\bm{j}\uparrow} f^{\phantom\dagger}_{\bm{i}\uparrow}\right).
\end{align}

To achieve the mean-field decoupling of the given Hamiltonian, we replace the link operators with their expectation values and expand up to first-order in the fluctuations, resulting in the following expression
\begin{align}
\label{eq:MF_Hijk_1stTerm}
    & \frac{i J_{\chi}}{16} \sum_\alpha\left( \chi_{\bm{ik}}\chi_{\bm{kj}}f^{\dagger}_{\bm{j}\alpha}f^{\phantom\dagger}_{\bm{i}\alpha} +  \chi_{\bm{ik}}\chi_{\bm{ji}}f^{\dagger}_{\bm{k}\alpha}f^{\phantom\dagger}_{\bm{j}\alpha} \right. \nonumber \\
  &\left.  + \chi_{\bm{kj}}\chi_{\bm{ji}}f^{\dagger}_{\bm{i}\alpha}f^{\phantom\dagger}_{\bm{k}\alpha} - \chi_{\bm{ik}}\chi_{\bm{kj}}\chi_{\bm{ji}} - \text{H.c.}\right).
\end{align}

Similarly, we perform the mean-field decoupling of the other two terms in Eq.~(\ref{eq:H_ijk}) which gives the same expression as in Eq.~(\ref{eq:MF_Hijk_1stTerm}). Ultimately, this leads to Eq.~(\ref{eq: Hchi_MF_kagome}). Combining Eqs.~(\ref{eq: H1_MF_kagome}) and (\ref{eq: Hchi_MF_kagome}), the full mean-field spin Hamiltonian can be expressed as 
\begin{align}
\label{eq:Hspin_MF}
    H_{\text{MF}} = & - \frac{J_1}{2} \sum_{\langle\boldsymbol{i}\boldsymbol{j}\rangle, \alpha}  \left[ ( f^\dagger_{\boldsymbol{i}\alpha} f^{\phantom\dagger}_{\boldsymbol{j}\alpha}\, \chi_{\boldsymbol{ji}} + \text{H.c.} ) - \lvert \chi_{\boldsymbol{ji}} \rvert^2 \right] \nonumber \\
    & + \frac{ 3 i  J_{\chi}}{16}  \sum_{\bm{ijk},\alpha}  \left(  \chi^{}_{\bm{ik}}\chi^{}_{\bm{kj}}f^{\dagger}_{\bm{j}\alpha}f^{\phantom\dagger}_{\bm{i}\alpha} +  \chi^{}_{\bm{ik}}\chi^{}_{\bm{ji}}f^{\dagger}_{\bm{k}\alpha}f^{\phantom\dagger}_{\bm{j}\alpha} \right. \nonumber \\
  &\left. + \chi^{}_{\bm{kj}}\chi^{}_{\bm{ji}}f^{\dagger}_{\bm{i}\alpha}f^{\phantom\dagger}_{\bm{k}\alpha} - \chi^{}_{\bm{ik}}\chi^{}_{\bm{kj}}\chi^{}_{\bm{ji}} - \text{H.c.} \right).
\end{align}

 Without going into the self-consistent mean-field analysis, we assume $\chi_{\bm{ij}} = |\chi_{\bm{ij}}| e^{i\phi_{\bm{ij}}}$ and the orientation of the bond is shown in Fig.~\ref{fig:Kagome lattice}. By combining the hopping phases $\phi_{\bm{ij}}$'s, the Hamiltonian in Eq.~(\ref{eq:Hspin_MF}) can be expressed in a concise form as:
\begin{equation}
     H_{\rm{MF}} \approx - t \sum_{\langle \bm{ij} \rangle,\alpha} e^{i\theta_{\bm{ij}}} f^{\dagger}_{\bm{i}\alpha}f^{\phantom\dagger}_{\bm{j}\alpha}   +  \text{H.c.},
\end{equation}
where $t$ is the spinon hopping amplitude which is given by \cite{Banerjee_CSL_2023}
    \begin{align}
        & t \cos \theta_{\bm{ij}} = \frac{J_1 \chi}{2} \cos \phi_{\bm{ji}} + \frac{ 3 i  J_{\chi} \chi^2}{16} \sin \left(\phi_{\bm{ik}}+\phi_{\bm{kj}}\right), \\
        & t \sin \theta_{\bm{ij}} = \frac{J_1 \chi}{2} \sin \phi_{\bm{ji}} + \frac{ 3 i  J_{\chi} \chi^2}{16} \cos \left(\phi_{\bm{ik}}+\phi_{\bm{kj}}\right).
    \end{align}
Now the total flux enclosed within a single triangular plaquette is given by $\theta_{\bm{ij}}+\theta_{\bm{jk}}+\theta_{\bm{ki}} = \phi$ and physically diverse states can be induced by selecting $\theta_{\bm{ij}}$ in a manner that leads to varying magnetic fluxes passing through the triangles and hexagons of the kagome lattice. Currently, we refrain from computing $\theta_{\bm{ij}}$ and the hopping parameter $t$. Instead, we initiate our analysis with the $\pi$-flux ansatz, gradually increasing the flux on the triangular plaquette. This can be conceptualized as the influence of adding a scalar spin chirality term. In general, the unit cell of the kagome lattice has three sublattices and the primitive lattice vector is given by $\boldsymbol{a}_1 = \boldsymbol{x}$ and $\boldsymbol{a}_2 = (1/2) \boldsymbol{x} + (\sqrt{3}/2)\boldsymbol{y}$. But to accommodate the flux configuration $\left[ \phi, \pi - 2\phi \right] = \left[3 \theta, \pi - 6 \theta\right]$, we double the unit cell in the horizontal direction. We choose the six-site unit cell labeled by $(\boldsymbol{R}, s)$, where $\boldsymbol{R} = 2 n_1 \boldsymbol{a}_1 + n_2 \boldsymbol{a}_2$ and $s$ stands for sublattice indices. The positions of the six sites are labeled by indices  $\boldsymbol{i}(\boldsymbol{R},s) = \boldsymbol{R} + \boldsymbol{v}_i$, where 

\begin{align}
\boldsymbol{v}_i =\begin{cases}
    0 & s = 1\\
    \boldsymbol{a}_{1}/2 & s = 2\\
    \boldsymbol{a}_{2}/2 & s = 3\\ 
     \boldsymbol{a}_{1} & s = 4\\
    3 \boldsymbol{a}_{1}/2 & s = 5\\
    \boldsymbol{a}_{1} + \boldsymbol{a}_{1}/2 & s = 6
  \end{cases}
\end{align}

The mean-field Hamiltonian corresponding to the flux configuration $ \left[3 \theta, \pi - 6 \theta\right]$ may be written as

\begin{widetext}
    \begin{align}
        H_{\text{MF}} = {} &  - t\sum_{\boldsymbol{R}} \left\{f^\dagger_{\boldsymbol{R}1} \left[e^{i\theta}f^{\phantom{\dagger}}_{\boldsymbol{R}2} + e^{-i\theta}f^{\phantom{\dagger}}_{\boldsymbol{R}3}\right]  + f^\dagger_{\boldsymbol{R}2} \left[e^{i\theta}f^{\phantom{\dagger}}_{\boldsymbol{R}3} - e^{-i\theta}f^{\phantom{\dagger}}_{\boldsymbol{R}4}\right] + f^\dagger_{\boldsymbol{R}3} \left[- e^{i\theta}f^{\phantom{\dagger}}_{{\boldsymbol{R}+\boldsymbol{a}_2,1}} - e^{-i\theta}f^{\phantom{\dagger}}_{{\boldsymbol{R}-2\boldsymbol{a}_1+\boldsymbol{a}_2,5}}\right] \nonumber  \right. \nonumber \\
  & {} \left. + f^\dagger_{\boldsymbol{R}4} \left[e^{i\theta}f^{\phantom{\dagger}}_{\boldsymbol{R}5} + e^{-i\theta}f^{\phantom{\dagger}}_{\boldsymbol{R}6}\right] + f^\dagger_{\boldsymbol{R}5} \left[e^{i\theta}f^{\phantom{\dagger}}_{\boldsymbol{R}6} + e^{-i\theta}f^{\phantom{\dagger}}_{{\boldsymbol{R}+2\boldsymbol{a}_1,3}}\right] + f^\dagger_{\boldsymbol{R}6} \left[e^{i\theta}f^{\phantom{\dagger}}_{{\boldsymbol{R}+\boldsymbol{a}_2,4}} - e^{-i\theta}f^{\phantom{\dagger}}_{{\boldsymbol{R}+\boldsymbol{a}_2,2}}\right]  + \text{H.c.}\right\},
    \end{align}
where for simplicity we drop the spin index. It can be written in the momentum space as 
\begin{equation}
    H_{{\rm MF}} = t \sum_{\bm{k}} f^\dagger_{\bm{k} s} \mathcal{H}(\bm{k})_{s s'} f^{\dagger}_{\bm{k} s'},
\end{equation}
 where the $6 \times 6$ matrix $\mathcal{H}$ is given by
\begin{equation}
\label{eq:CSL MFT Hamiltoian momentum space}
- \mathcal{H}(\bm{k}) = \left( \begin{array}{cccccc}
0 & e^{i\theta} & e^{-i\theta}(1 - K_2^*) & 0 & e^{i\theta}K_1^* & 0 \\
e^{-i\theta} & 0 & e^{i\theta} & -e^{-i\theta} & 0 & -e^{i\theta}K_2^* \\
e^{i\theta}(1 - K_2) & e^{-i\theta} & 0 & 0 & -e^{-i\theta} K_1^* K_2 & 0 \\
0 & -e^{i\theta} & 0 & 0 & e^{i\theta} & e^{-i\theta}(1 + K_2^*) \\
e^{-i\theta}K_1 & 0 & - e^{i\theta} K_1 K_2^* & e^{-i\theta}& 0 & e^{i\theta} \\
0 & -e^{-i\theta}K_2 & 0 & e^{i\theta}(1 + K_2) & e^{-i\theta} & 0
\end{array} \right) \text{.}
\end{equation}
\end{widetext}

Here, we have defined $K_1 = e^{2 i \bm{k} \cdot \bm{a}_1}$ and $K_2 = e^{i \bm{k} \cdot \bm{a}_2}$ and the Fourier transform is defined by
\begin{equation}
    f_{\bm{R} s } = \frac{1}{\sqrt{N_c}} \sum_{\bm{k}} e^{i \bm{k} \cdot \bm{R}} f_{\bm{k}s}.
\end{equation}
Note that, for $\theta = 0$, the Hamiltonian in Eq.~\eqref{eq:CSL MFT Hamiltoian momentum space} reduces to the Hamiltonian corresponding to $\pi$-flux ansatz as given in Ref.~\cite{Hermele_KagomeDSL_Properties_PRB2008}. Diagonalizing it for $\theta = \pi/6$, we obtain the corresponding spinon band structure as shown in Fig.~\ref{fig:Kagome BZ and band}c.

\section{Effective action at Large-$N_f$}
\label{app:effective_action}
In this section, we detail the computation of the vacuum polarization tensor of the light Dirac fermion and the renormalized gauge boson propagator in the Coulomb gauge, which are used to calculate the gauge-field correction to the thermal Hall conductivity at both leading and next-to-leading order in the $N_f \rightarrow \infty$ limit. We will begin by computing the vacuum polarization tensor at finite temperature and then extend the calculations from Refs.~\cite{vafek2003thesis,RibhuKaul_QuantumCriticalityU(1)gauge_PRB2008, Lee_Mulligan_Conductivity_PRB2023} to include massive Dirac fermions with a Chern-Simons term in order to calculate the renormalized gauge boson propagator.

\subsection{Computation of $\Pi^{\mu\nu}_f$ }
\label{subsec:Pi_munu_f}

To begin, we calculate the fermion polarization function $\Pi^{\mu\nu}_f (\boldsymbol{q}, i \epsilon_n)$ as defined in Eq.~\eqref{eq:Pi_munu_f} at zero temperature, which can be expressed as:
\begin{align}
\label{eq:Pi_f_1}
    \Pi^{\mu\nu}_{f}(\bm{q}, i \epsilon_n) = {} &  N_f T \sum_{i\omega_n} \int  \frac{d^2 \bm{k}}{(2\pi)^2}  \frac{L^{\mu\nu}(k;q)}{\left(\omega_n^2+\bm{k}^2+m^2 \right)} \nonumber\\
    & \times \frac{1}{\left((\omega_n +\epsilon_n)^2 + (\bm{k}+\bm{q})^2 + m^2\right)},
\end{align}
where
\begin{align}
    L^{\mu\nu}(k;q) = {} & \Tr \left[ \left(i\omega_n - \bm{\tau} \cdot \bm{k} - m\tau^z\right) \tau^\mu \right. \nonumber \\
  & {} \left. \left((i\omega_n+i\epsilon_n) - \bm{\tau} \cdot (\bm{k}+\bm{q}) - m\tau^z \right) \tau^\nu \right].
\end{align}
In the zero-temperature limit, we can replace the summation over the Matsubara frequency with an integral:
\begin{align}
    T \sum_{i\omega_{n}} \rightarrow \int \frac{d \omega_n}{2 \pi}.
\end{align}
The integrals can be evaluated using standard Feynman parameter techniques. Using the Feynman parametrization formula
\begin{equation}
    \frac{1}{A\, B} = \int^1_0 \, dx\, \frac{1}{\left[x A + (1-x) B  \right]^2},
\end{equation}
Eq.~\eqref{eq:Pi_f_1} can be expressed as
\begin{align}
   \Pi^{\mu\nu}_f (\bm{q},i \epsilon_n) = N_f \int^1_0 d x \, T \sum_{i l_{n}} \int  \frac{d^2 \bm{l}}{(2\pi)^2} \, \frac{ L^{\mu\nu}}{\left[l_n^2 + \bm{l}^2 + Q^2  \right]^2},
\end{align}
where
\begin{align}
    l_n & = \omega_n + x\epsilon_n,\nonumber\\
    \bm{l} & = \bm{k} + x \bm{q}, \nonumber\\
    Q^2 & = x(1-x)\, \epsilon_n^2 + x(1-x)\, \bm{q}^2 + m^2.
\end{align}

The integral can be computed using dimensional regularization. Although closed-form solutions are possible for arbitrary momenta and mass, we expand the result to second order in momenta, yielding:
\begin{align}
\label{eq:Pi_f_zeroT}
     \Pi^{\mu\nu}_f (\bm{q},i \epsilon_n) =  \frac{N_f}{12 \pi \abs{m}} \left( q^2 \delta^{\mu\nu} - q^\mu q^\nu \right) + \frac{ N_f}{2 \pi} \frac{\text{sgn}(m)}{2} \epsilon^{\mu\nu\lambda} q_\lambda.
\end{align}

Several noteworthy aspects arise from this result. First, the massive light Dirac fermions contribute to gauge field dynamics similar to the Maxwell term $f_{\mu\nu}^2$, as indicated by the first term in Eq.~\eqref{eq:Pi_f_zeroT}. The second term represents the anticipated Chern-Simons term with level $N_f \text{sgn}(m)/2$, which renormalizes the bare Chern-Simons term $\mathtt{k}$. A similar calculation can be carried out for the heavy Dirac fermion in the $\abs{M}/T \rightarrow \infty$ limit. This yields a Maxwell kinetic term that scales with the inverse of $M$ and the bare Chern-Simons term $\mathtt{k}$. In this manner, we obtain the effective Maxwell-Chern-Simons theory with level $\mathtt{k} +  N_f \text{sgn}(m)/2.$

Using Feynman parameters at finite temperature offers no particular benefit. Instead, we first sum over the fermionic Matsubara frequencies $\omega_n$, followed by integrating over $\bm{k}$. First, we will rewrite the general structure of the fermion polarization tensor at large-$N_f$ for the light Dirac fermion:
\begin{align}
\label{eq:Pi_munu_f_T}
    \Pi^{\mu\nu}_f (\bm{q},i \epsilon_n; T) = \Pi_A A^{\mu\nu} (q) + \Pi_B B^{\mu\nu}(q) + \Pi_C C^{\mu\nu} (q),
\end{align}
where $A^{\mu\nu}$, $B^{\mu\nu}$, and $C^{\mu\nu}$ are defined in Eq.~\eqref{eq:ABC_munu}.

To proceed, we define
\begin{align}
    \Pi^{\mu\nu}_{f}(\bm{q}, i \epsilon_n; T) =  N_f \int  \frac{d^2 \bm{k}}{(2\pi)^2}  \mathcal{S}^{\mu\nu} (\bm{k},T;\bm{q},i\epsilon_n),
\end{align}
where
\begin{align}
    \mathcal{S}^{\mu\nu} (\bm{k},T;\bm{q},i\epsilon_n)  = {} & T \sum_{i\omega_{n}} \frac{L^{\mu\nu}(\bm{k},i\omega_n;\bm{q},i\epsilon_n)}{\left((i\omega_n)^2-\bm{k}^2-m^2 \right)} \nonumber\\
    &{} \times \frac{1}{\left((i\omega_n +i\epsilon_n)^2 - (\bm{k}+\bm{q})^2 - m^2\right)}.
\end{align}
\begin{widetext}
Using standard contour methods to perform the summation, we obtain
    \begin{align}
\label{eq:S_munu}
    \mathcal{S}^{\mu\nu} (\bm{k},T;\bm{q},i\epsilon_n) ={} & - \frac{n_F(E_{\bm{k}})}{2E_{\bm{k}}} \frac{L^{\mu\nu}(\bm{k},E_{\bm{k}};\bm{q},i\epsilon_n)}{E_{\bm{k}+\bm{q}}^2-E_{\bm{k}}^2 - 2 i E_{\bm{k}} \epsilon_n + \epsilon_n^2} + \frac{n_F(-E_{\bm{k}})}{2E_{\bm{k}}} \frac{L^{\mu\nu}(\bm{k},-E_{\bm{k}};\bm{q},i\epsilon_n)}{E_{\bm{k}+\bm{q}}^2-E_{\bm{k}}^2 + 2 i E_{\bm{k}} \epsilon_n + \epsilon_n^2} \nonumber\\
    & {} - \frac{n_F(E_{\bm{k}+\bm{q}})}{2E_{\bm{k}+\bm{q}}} \frac{L^{\mu\nu}(\bm{k},E_{\bm{k}+\bm{q}}-i \epsilon_n;\bm{q},i\epsilon_n)}{E_{\bm{k}+\bm{q}}^2-E_{\bm{k}}^2 + 2 i E_{\bm{k}} \epsilon_n + \epsilon_n^2} + \frac{n_F(-E_{\bm{k}+\bm{q}})}{2E_{\bm{k}+\bm{q}}} \frac{L^{\mu\nu}(\bm{k},-E_{\bm{k}+\bm{q}}-i \epsilon_n;\bm{q},i\epsilon_n)}{E_{\bm{k}+\bm{q}}^2-E_{\bm{k}}^2 - 2 i E_{\bm{k}} \epsilon_n + \epsilon_n^2}
\end{align}
From $\Pi^{ij}_f$, we shall be able to extract $\Pi_A$, $\Pi_B$, and $\Pi_C$. Since we eventually integrate over all $\bm{k}$, we can shift $\bm{k}+\bm{q} \rightarrow \bm{k}$ in the last two terms in the Eq.~\eqref{eq:S_munu} as well as reverse the direction of $\bm{k}$ to obtain
\begin{align}
    \Pi^{ij}_{f}(\bm{q}, i \epsilon_n; T)  = {} & 2 N_f \int  \frac{d^2 \bm{k}}{(2\pi)^2} \frac{1-2 n_F({E_{\bm{k}}})}{2 E_{\bm{k}}} \bigg[ \frac{2 k_i k_j + k_i q_j + k_j q_i - \delta_{ij}(\bm{k}\cdot\bm{q} + i \epsilon_n E_{\bm{k}}) + \epsilon_{ij} m \epsilon_n}{\epsilon_n^2+\bm{q}^2+ 2 i E_{\bm{k}} \epsilon_n + 2 \bm{k}\cdot\bm{q}} \nonumber \\
    & {} + \frac{2 k_i k_j + k_i q_j + k_j q_i - \delta_{ij}(\bm{k}\cdot\bm{q} - i \epsilon_n E_{\bm{k}}) + \epsilon_{ij} m \epsilon_n}{\epsilon_n^2+\bm{q}^2 - 2 i E_{\bm{k}} \epsilon_n + 2 \bm{k}\cdot\bm{q}} \bigg].    
\end{align}
We note that all $T$ dependence resides in the Fermi occupation factor $n_F$. Since its argument in the equation above is always positive definite, at $T=0$, $n_F({E_{\bm{k}}})=0$ and the result is given by 
\begin{align}
     \Pi^{\mu\nu}_f (\bm{q},i \epsilon_n; T=0)  =  \frac{N_f}{12 \pi \abs{m}} \left( q^2 \delta^{\mu\nu} - q^\mu q^\nu \right) + \frac{ N_f}{2 \pi} \frac{\text{sgn}(m)}{2} \epsilon^{\mu\nu\lambda} q_\lambda.
\end{align}
which matches the one obtained using dimensional regularization, as shown in Eq.~\eqref{eq:Pi_f_zeroT}. Therefore, at zero-$T$, we get
\begin{align}
    \Pi_A(T=0) =  \frac{N_f}{12 \pi \abs{m}},\,\,& \,\, 
    \Pi_B(T=0)  =  \frac{N_f}{12 \pi \abs{m}}, \nonumber\\  \Pi_C(T=0) & = \frac{ N_f}{2 \pi} \frac{\text{sgn}(m)}{2}.
\end{align}

Therefore at finite $T$ we can write
\begin{align}
    \Pi^{\mu\nu}_f (\bm{q},i \epsilon_n; T) = \Pi^{\mu\nu}_f (\bm{q},i \epsilon_n; T=0) + \delta\Pi^{\mu\nu}_f (\bm{q},i \epsilon_n),
\end{align}
where
\begin{align}
\label{eq:delta_Pi_ij}
    \delta \Pi^{ij}_{f}(\bm{q}, i \epsilon_n) = {} & - 2 N_f \int  \frac{d^2 \bm{k}}{(2\pi)^2} \frac{n_F({E_{\bm{k}}})}{E_{\bm{k}}} \bigg[ \frac{2 k_i k_j + k_i q_j + k_j q_i - \delta_{ij}(\bm{k}\cdot\bm{q} + i \epsilon_n E_{\bm{k}}) + \epsilon_{ij} m \epsilon_n}{\epsilon_n^2+\bm{q}^2+ 2 i E_{\bm{k}} \epsilon_n + 2 \bm{k}\cdot\bm{q}} \nonumber \\
    & {} + \frac{2 k_i k_j + k_i q_j + k_j q_i - \delta_{ij}(\bm{k}\cdot\bm{q} - i \epsilon_n E_{\bm{k}}) + \epsilon_{ij} m \epsilon_n}{\epsilon_n^2+\bm{q}^2 - 2 i E_{\bm{k}} \epsilon_n + 2 \bm{k}\cdot\bm{q}} \bigg].    
\end{align}
To perform the integration over $\bm{k}$, we decompose this into an angular integral and an integral over the magnitude of $\bm{k}$. Rewriting this in polar coordinates $\bm{k} = \abs{\bm{k}}\left(\cos \theta_{\bm{k}}, \sin{\theta_{\bm{k}}} \right)$ and $\bm{q} = \abs{\bm{q}}\left(\cos \theta_{\bm{q}}, \sin{\theta_{\bm{q}}} \right)$, and using the identities 
\begin{align}
    \int_0^{2\pi} d\theta_{\bm{k}} \frac{1}{iA+\cos{\left(\theta_{\bm{k}} - \theta_{\bm{q}}\right)}} & = \int_0^{2\pi} d\theta_{\bm{k}} \frac{1}{iA+\cos{\theta_{\bm{k}}}}, \nonumber\\
    \int_0^{2\pi} d\theta_{\bm{k}} \frac{\cos{\theta_{\bm{k}}}}{iA+\cos{\left(\theta_{\bm{k}} - \theta_{\bm{q}}\right)}} & = \int_{-\theta_{\bm{q}}}^{2\pi-\theta_{\bm{q}}} d\theta_{\bm{k}} \frac{\cos{\theta_{\bm{k}}} \cos{\theta_{\bm{q}}} - \sin{\theta_{\bm{k}}}\sin{\theta_{\bm{q}}}}{iA+\cos{\theta_{\bm{k}}}} = \cos{\theta_{\bm{q}}} \int_{0}^{2\pi} d\theta_{\bm{k}} \frac{\cos{\theta_{\bm{k}}}}{iA+\cos{\theta_{\bm{k}}}} , \nonumber\\
    \int_0^{2\pi} d\theta_{\bm{k}} \frac{\sin{\theta_{\bm{k}}}}{iA+\cos{\left(\theta_{\bm{k}} - \theta_{\bm{q}}\right)}} & =  \int_{0}^{2\pi} d\theta_{\bm{k}} \frac{\sin{\theta_{\bm{q}}} \cos{\theta_{\bm{k}}}}{iA+\cos{\theta_{\bm{k}}}},\,\, \int_0^{2\pi} d\theta_{\bm{k}} \frac{\cos{(2\theta_{\bm{k}}})}{iA+\cos{\left(\theta_{\bm{k}} - \theta_{\bm{q}}\right)}} = \cos{(2\theta_{\bm{q}})} \int_{0}^{2\pi} d\theta_{\bm{k}} \frac{\cos{(2\theta_{\bm{k}}})}{iA+\cos{\theta_{\bm{k}}}}, \nonumber\\
    \int_0^{2\pi} d\theta_{\bm{k}} \frac{\sin{(2\theta_{\bm{k}}})}{iA+\cos{\left(\theta_{\bm{k}} - \theta_{\bm{q}}\right)}} & = \sin{(2\theta_{\bm{q}})} \int_{0}^{2\pi} d\theta_{\bm{k}} \frac{\cos{(2\theta_{\bm{k}}})}{iA+\cos{\theta_{\bm{k}}}},
\end{align}
we can simplify Eq.~\eqref{eq:delta_Pi_ij} to read
\begin{align}
    \delta \Pi^{ij}_{f}(\bm{q}, i \epsilon_n) = {} & N_f \delta_{ij} \epsilon_n^2 \int_0^\infty \frac{\abs{\bm{k}} d\abs{\bm{k}}}{(2\pi)^2} \frac{n_F \left(\sqrt{\abs{\bm{k}}^2+m^2}\right)}{\sqrt{\abs{\bm{k}}^2+m^2}} \int_0^{2\pi} d\theta_{\bm{k}} \left[ \frac{\mathcal{J}_A(\abs{\bm{q}},i\epsilon_n,\abs{\bm{k}};\cos{\theta_{\bm{k}}})}{i\mathcal{K}(\abs{\bm{q}},i\epsilon_n,\abs{\bm{k}})+\cos{\theta_{\bm{k}}}} +(\epsilon_n \rightarrow -\epsilon_n) \right] \nonumber\\
    & {} + N_f \bm{q}^2 \left(\delta_{ij} - \frac{q_i q_j}{\bm{q}^2}\right) \int_0^\infty \frac{\abs{\bm{k}} d\abs{\bm{k}}}{(2\pi)^2} \frac{n_F \left(\sqrt{\abs{\bm{k}}^2+m^2}\right)}{\sqrt{\abs{\bm{k}}^2+m^2}} \int_0^{2\pi} d\theta_{\bm{k}} \left[ \frac{\mathcal{J}_B(\abs{\bm{q}},i\epsilon_n,\abs{\bm{k}};\cos{\theta_{\bm{k}}})}{i\mathcal{K}(\abs{\bm{q}},i\epsilon_n,\abs{\bm{k}})+\cos{\theta_{\bm{k}}}} +(\epsilon_n \rightarrow -\epsilon_n) \right] \nonumber\\
    & {} + N_f \epsilon_n \int_0^\infty \frac{\abs{\bm{k}} d\abs{\bm{k}}}{(2\pi)^2} \frac{n_F \left(\sqrt{\abs{\bm{k}}^2+m^2}\right)}{\sqrt{\abs{\bm{k}}^2+m^2}} \int_0^{2\pi} d\theta_{\bm{k}} \left[ \frac{-2m}{i\mathcal{K}(\abs{\bm{q}},i\epsilon_n,\abs{\bm{k}})+\cos{\theta_{\bm{k}}}} +(\epsilon_n \rightarrow -\epsilon_n) \right],
\end{align}

where
\begin{align}
\mathcal{J}_A(\abs{\bm{q}},i\epsilon_n,\abs{\bm{k}};\cos{\theta_{\bm{k}}}) & =  - \frac{1}{\abs{\bm{k}}\abs{\bm{q}}\epsilon_n^2} \left(\abs{\bm{k}}^2 -i \epsilon_n \sqrt{\abs{\bm{k}}^2+m^2}+ \abs{\bm{k}}\abs{\bm{q}} \cos{\theta_{\bm{k}}} + \abs{\bm{k}}^2 \cos{(2\theta_{\bm{k}})}\right), \nonumber\\
\mathcal{J}_B(\abs{\bm{q}},i\epsilon_n,\abs{\bm{k}};\cos{\theta_{\bm{k}}}) &=  \frac{1}{\abs{\bm{k}}\abs{\bm{q}}} \left(\frac{2\abs{\bm{k}}}{\abs{\bm{q}}} \cos{\theta_{\bm{k}}} + \frac{2\abs{\bm{k}}^2}{\abs{\bm{q}}^2} \cos{(2\theta_{\bm{k}})}\right), \nonumber\\
\mathcal{K}(\abs{\bm{q}},i\epsilon_n,\abs{\bm{k}}) & =  \frac{1}{2\abs{\bm{k}}\abs{\bm{q}}} \left(2\epsilon_n \sqrt{\abs{\bm{k}}^2+m^2}- i \epsilon_n^2 - i \abs{\bm{q}}^2 \right).
\end{align}
Once written in this form, we can read off the finite-$T$ corrections to $\Pi_A$, $\Pi_B$, and $\Pi_C$:
\begin{align}
\label{eq:delta_Pi_ABC}
    \delta\Pi_A & = N_f \int_0^\infty \frac{\abs{\bm{k}} d\abs{\bm{k}}}{(2\pi)^2} \frac{n_F \left(\sqrt{\abs{\bm{k}}^2+m^2}\right)}{\sqrt{\abs{\bm{k}}^2+m^2}} \int_0^{2\pi} d\theta_{\bm{k}} \left[ \frac{\mathcal{J}_A(\abs{\bm{q}},i\epsilon_n,\abs{\bm{k}};\cos{\theta_{\bm{k}}})}{i\mathcal{K}(\abs{\bm{q}},i\epsilon_n,\abs{\bm{k}})+\cos{\theta_{\bm{k}}}} +(\epsilon_n \rightarrow -\epsilon_n) \right], \nonumber\\
    \delta\Pi_B & =N_f \int_0^\infty \frac{\abs{\bm{k}} d\abs{\bm{k}}}{(2\pi)^2} \frac{n_F \left(\sqrt{\abs{\bm{k}}^2+m^2}\right)}{\sqrt{\abs{\bm{k}}^2+m^2}} \int_0^{2\pi} d\theta_{\bm{k}} \left[ \frac{\mathcal{J}_B(\abs{\bm{q}},i\epsilon_n,\abs{\bm{k}};\cos{\theta_{\bm{k}}})}{i\mathcal{K}(\abs{\bm{q}},i\epsilon_n,\abs{\bm{k}})+\cos{\theta_{\bm{k}}}} +(\epsilon_n \rightarrow -\epsilon_n) \right], \nonumber\\
    \delta\Pi_C & = N_f \int_0^\infty \frac{\abs{\bm{k}} d\abs{\bm{k}}}{(2\pi)^2} \frac{n_F \left(\sqrt{\abs{\bm{k}}^2+m^2}\right)}{\sqrt{\abs{\bm{k}}^2+m^2}} \int_0^{2\pi} d\theta_{\bm{k}} \left[ \frac{-2m}{i\mathcal{K}(\abs{\bm{q}},i\epsilon_n,\abs{\bm{k}})+\cos{\theta_{\bm{k}}}} +(\epsilon_n \rightarrow -\epsilon_n) \right].
\end{align}
Let us begin by calculating $\Pi_C$. The angular integrals can be computed analytically using contour integration:
\begin{align}
    \int_0^{2\pi} d\theta_{\bm{k}} \frac{1}{iA+\cos{\theta_{\bm{k}}}} = - \frac{2\pi i}{\sqrt{A^2+1}} \text{sgn}(\text{Re}(A))
\end{align}

Although the angular integral can be done analytically, the integral over $\abs{\bm{k}}$ cannot be expressed in closed form. However, it can be evaluated analytically in the limit where the frequency $\epsilon_n$ and momentum $\abs{\bm{q}}$ approach zero—or more specifically, when 
$\epsilon_n$ and $\abs{\bm{q}}$ are much smaller than the mass scale $\abs{m}$. But, the behavior depends on the ratio $a=\epsilon_n/\abs{\bm{q}}$ \cite{Aitchison1995}. In this discussion, we focus on two specific cases: the static limit $a=0$ ($\epsilon_n=0, \abs{\bm{q}}\rightarrow0$) and the long wavelength limit $a=\infty$ ($\epsilon_n\rightarrow0, \abs{\bm{q}}=0$). So, after performing the angular integration, we investigate the region $\epsilon_n = a \abs{\bm{q}}$ for small $\abs{\bm{q}}$ and obtain
\begin{align}
\label{eq:delta_Pi_C}
    \delta\Pi_C (\epsilon_n = a \abs{\bm{q}}, \abs{\bm{q}} \ll \abs{m}) = - \frac{m N_f}{2\pi} \int_{\abs{m}}^\infty dE \frac{a(1+a)^2 E n_F(E)}{(E^2(1+a^2)-m^2)^{3/2}}.
\end{align} 
We still cannot take the limit $a\rightarrow 0$ in Eq.~\eqref{eq:delta_Pi_C}, but after a partial integration we obtain
\begin{align}
    \delta\Pi_C (\epsilon_n = a \abs{\bm{q}}, \abs{\bm{q}} \ll \abs{m}) = - \frac{m N_f}{2\pi \abs{m}} n_F(\abs{m}) - \frac{m N_f}{2\pi} \int_{\abs{m}}^\infty dE \frac{d n_F(E)}{dE}\frac{a}{(E^2(1+a^2)-m^2)^{1/2}}.
\end{align}
We may now safely take the $a \rightarrow 0$ to obtain:
\begin{align}
\label{eq:delta_Pi_C_a0}
    \delta\Pi_C (\epsilon_n = a \abs{\bm{q}}, \abs{\bm{q}} \ll \abs{m}) \overset{a \rightarrow 0}{\sim}  - \frac{\text{sgn}(m) N_f}{2\pi} n_F(\abs{m}).
\end{align}
The $a\rightarrow \infty$ limit can be taken directly in Eq.~\eqref{eq:delta_Pi_C}, and the result is given by
\begin{align}
\label{eq:delta_Pi_C_ainf}
    \delta\Pi_C (\epsilon_n = a \abs{\bm{q}}, \abs{\bm{q}} \ll \abs{m}) \overset{a \rightarrow \infty}{\sim}  - \frac{N_f m}{2\pi} \int_{\abs{m}}^\infty dE \frac{n_F(E)}{E^2}.
\end{align}

We can employ a similar procedure for $\delta \Pi_A$ and $\delta \Pi_B$ in Eq.~\eqref{eq:delta_Pi_ABC}:
\begin{align}
    \delta\Pi_A (\epsilon_n = a \abs{\bm{q}}, \abs{\bm{q}} \ll \abs{m}) = {} & -\frac{N_f}{8\pi} \int_{\abs{m}}^\infty dE\,  n_F(E) \frac{a(1+a^2)^2 E \left[(1+a^2) E^4 + (-5+a^2) m^2 E^2 + 4 m^4 \right]}{(E^2(1+a^2)-m^2)^{7/2}} \nonumber\\
    & {} + \frac{N_f}{\pi \abs{\bm{q}}^2} \int_{\abs{m}}^\infty dE\,  n_F(E) \frac{-a E^3 - a^3 E^3 + 2 a E m^2 + (E^2(1+a^2)-m^2)^{3/2} }{(E^2(1+a^2)-m^2)^{3/2}}, \label{eq:delta_Pi_A} \\
    \delta\Pi_B (\epsilon_n = a \abs{\bm{q}}, \abs{\bm{q}} \ll \abs{m}) = {} & -\frac{N_f}{8\pi} \int_{\abs{m}}^\infty dE\,  n_F(E) \frac{a(1+a^2)^2 E (E^2-m^2) \left[(1+a^2) E^2 + (-1+5a^2) m^2\right]}{(E^2(1+a^2)-m^2)^{7/2}} \nonumber\\
    & {} + \frac{N_f a}{\pi \abs{\bm{q}}^2} \int_{\abs{m}}^\infty dE\,  n_F(E) \frac{(1+3a^2+2a^4)E^3-(1+3a^2)m^2 E-2a (E^2(1+a^2)-m^2)^{3/2} }{(E^2(1+a^2)-m^2)^{3/2}}. \label{eq:delta_Pi_B}
\end{align}
By applying a similar partial integration method, we obtain the following result in the $a \rightarrow 0$ limit:
\begin{align}
    \delta\Pi_A & (\epsilon_n = a \abs{\bm{q}}, \abs{\bm{q}} \ll \abs{m}) \overset{a \rightarrow 0}{\sim}  -\frac{N_f}{6\pi\abs{m}} n_F(\abs{m}) + \frac{N_f T}{\pi \abs{\bm{q}}^2} \ln{\left(1+e^{-\abs{m}/T} \right)} + \frac{N_f \abs{m}}{\pi\abs{\bm{q}}^2} n_F(\abs{m}), \label{eq:delta_Pi_A_a0}\\
     \delta\Pi_B & (\epsilon_n = a \abs{\bm{q}}, \abs{\bm{q}} \ll \abs{m}) \overset{a \rightarrow 0}{\sim}  -\frac{N_f}{6\pi\abs{m}} n_F(\abs{m}). \label{eq:delta_Pi_B_a0}
\end{align}
\end{widetext}
Before examining the $a\rightarrow\infty$ limit, we first consider the expression for $\Pi^{00}_f$ in the static limit, which is given by
\begin{align}
    \Pi^{00}_f(\bm{q}, i\epsilon_n;T) \overset{a \rightarrow 0}{\sim} {} & \frac{ N_f \abs{\bm{q}}^2}{12\pi\abs{m}} + \frac{N_f T}{\pi} \ln{\left(1+e^{-\abs{m}/T} \right)} \nonumber\\
    & {} -\frac{N_f \abs{\bm{q}}^2 }{6\pi\abs{m}} n_F(\abs{m}) + \frac{N_f \abs{m}}{\pi} n_F(\abs{m}).
\end{align}
Taking the $\abs{\bm{q}} \rightarrow 0$ limit, we get 
\begin{align}
    \Pi^{00}_f \overset{a \rightarrow 0}{\sim}   \frac{N_f \abs{m}}{\pi} n_F(\abs{m}) + \frac{N_f T}{\pi} \ln{\left(1+e^{-\abs{m}/T} \right)}.
\end{align}
which is constant and vanishes as $T \rightarrow 0$. It represents the thermal mass, which arises in the longitudinal component without breaking gauge invariance.

The $a\rightarrow \infty$ limit can be taken directly in Eqs.~\eqref{eq:delta_Pi_A} and \eqref{eq:delta_Pi_B}, and the results are given by
\begin{align}
    \delta\Pi_A (\epsilon_n = a \abs{\bm{q}}, \abs{\bm{q}} \ll \abs{m}) \overset{a \rightarrow \infty}{\sim} {} & - \frac{N_f}{8\pi} \int_{\abs{m}}^\infty dE\, n_F(E) \nonumber\\
    & {} \times \frac{E^2+m^2}{E^4}, \label{eq:delta_Pi_A_ainf}\\
    \delta\Pi_B (\epsilon_n = a \abs{\bm{q}}, \abs{\bm{q}} \ll \abs{m})  \overset{a \rightarrow \infty}{\sim} {} & - \frac{N_f}{8\pi} \int_{\abs{m}}^\infty dE\, n_F(E) \nonumber\\
    & {} \times \frac{(E^2-m^2)(E^2+5m^2)}{E^6}.\label{eq:delta_Pi_B_ainf}
\end{align}

Since we have the polarization tensor of the light Dirac fermion for a general $\abs{m}/T$, we combine Eq.~\eqref{eq:Pi_munu_f_T} with the effective Maxwell-Chern-Simons theory of the heavy Dirac fermion to obtain the complete polarization tensor, expressed as
\begin{align}
    \Pi^{\mu\nu} (\bm{q},i \epsilon_n;T) = N_f g_a A^{\mu\nu} (q) + N_f  g_b B^{\mu\nu}(q) + \frac{\Tilde{\mathtt{k}}}{2\pi} C^{\mu\nu} (q).
\end{align}
Here, in the long-wavelength limit ($\epsilon_n\rightarrow0, \abs{\bm{q}}=0$), we obtain
\begin{align}
\label{eq:g_mt_ainf}
    g_a (\bm{q},i\epsilon_n)  \overset{a \rightarrow \infty}{\sim} {} & g  - \frac{1}{8\pi} \int_{\abs{m}}^\infty dE\, n_F(E) \frac{E^2+m^2}{E^4}, \nonumber\\
    g_b (\bm{q},i\epsilon_n)  \overset{a \rightarrow \infty}{\sim} {} & g  - \frac{1}{8\pi} \int_{\abs{m}}^\infty dE\, n_F(E) \nonumber\\
   & {} \times \frac{(E^2-m^2)(E^2+5m^2)}{E^6}, \nonumber\\
    \Tilde{\mathtt{k}} (\bm{q},i\epsilon_n)  \overset{a \rightarrow \infty}{\sim} {} & \mathtt{k} + \frac{N_f}{2} {\rm{sgn}}(m) - N_f m \int_{\abs{m}}^\infty dE \frac{n_F(E)}{E^2},
\end{align}
and in the static limit ($\epsilon_n =0, \abs{\bm{q}} \rightarrow 0$), we arrive at the following result:
\begin{align}
\label{eq:g_mt_a0}
    g_a (\bm{q},i\epsilon_n)  \overset{a \rightarrow 0}{\sim} {} & g  -\frac{1}{6\pi\abs{m}} n_F(\abs{m}) + \frac{\abs{m}}{\pi\abs{\bm{q}}^2} n_F(\abs{m}) \nonumber\\
    & {} + \frac{T}{\pi \abs{\bm{q}}^2} \ln{\left(1+e^{-\abs{m}/T} \right)}, \nonumber\\
    g_b (\bm{q},i\epsilon_n)  \overset{a \rightarrow 0}{\sim} {} & g    -\frac{N_f}{6\pi\abs{m}} n_F(\abs{m}), \nonumber\\
    \Tilde{\mathtt{k}} (\bm{q},i\epsilon_n)  \overset{a \rightarrow 0}{\sim} {} & \mathtt{k} + \frac{N_f}{2} {\rm{sgn}}(m) - \text{sgn}(m) N_f n_F(\abs{m}),
\end{align}
where $g$ is defined in Eq.~\eqref{eq:g_mt_def}. 

\subsection{Effective gauge boson propagator $D_{\mu\nu}$}
With the full polarization tensor determined, we proceed to calculate the renormalized gauge boson propagator in the Coulomb gauge. We can decompose the spatial component of the gauge field ($a_i$) into longitudinal ($a_L$) and transverse ($a_T$) components \cite{Kachru_Half_filled_PRB2015}:

\begin{align}
    a_i (\bm{q},i \epsilon_n) = i \frac{q_i}{\abs{\bm{q}}} a_L (\bm{q},i \epsilon_n) + i \epsilon_{ij} \frac{q_i q_j}{\abs{\bm{q}}} a_T (\bm{q},i \epsilon_n),
\end{align}
where $a_L$ and $a_T$ are related to the Cartesian components ($a_x, a_y$) via
\begin{align}
    a_L (\bm{q},i \epsilon_n) & = - i \frac{q_i}{\abs{\bm{q}}}  a_i (\bm{q},i \epsilon_n) , \nonumber \\
      a_T (\bm{q},i \epsilon_n) & = \frac{i}{\abs{\bm{q}}} \left( q_x a_y - q_y a_x  \right).
\end{align}
In the Coulomb gauge, where $q_i a_i = 0$, the longitudinal component of $a_i$ is zero. Now, we can relate polarization tensor in $a_0$ , $a_T$ basis to the $\Pi^{\mu\nu}$ in the Cartesian basis as follows:
\begin{align}
    \Pi_{0T} (q) & =  i \frac{q_y}{\abs{\bm{q}}} \Pi_{0x} (q) - i \frac{q_x}{\abs{\bm{q}}} \Pi_{0y} (q), \nonumber \\
    \Pi_{T0} (q) & = - i \frac{q_y}{\abs{\bm{q}}} \Pi_{x0} (q) + i \frac{q_x}{\abs{\bm{q}}} \Pi_{y0} (q), \nonumber \\
    \Pi_{TT} (q) & = \left( \delta_{ij} - \frac{q_i q_j}{\bm{q}^2} \right) \Pi_{ij}.    
\end{align}
With the full polarization tensor derived in Subsection~\ref{subsec:Pi_munu_f} for both zero and finite-$T$, we can compute $\Pi_{0T}$, $\Pi_{T0}$, and $\Pi_{TT}$.

Using the full polarization tensor both at zero and finite-$T$ obtained in the Subsection~\ref{subsec:Pi_munu_f}, one can calculate $\Pi_{0T}$ , $\Pi_{T0}$ and $\Pi_{TT}$. We express the full effective action for the gauge field in Eq.~\eqref{eq:effective_action} using the $a_0$ , $a_T$ basis as
\begin{align}
\label{eq:S_eff_0T}
     \mathcal{S}[a]  =  \frac{T}{2} \sum_{i\epsilon_n} \int \frac{d^2\bm{q}}{(2\pi)^2} \, \begin{pmatrix}
         a_0 & a_T
     \end{pmatrix}_{-q} \begin{pmatrix}
         \Pi_{00} & \Pi_{0T} \\
         \Pi_{T0} & \Pi_{TT} 
     \end{pmatrix} \begin{pmatrix}
         a_0 \\
         a_T
     \end{pmatrix}_{q}.
\end{align}
Using the effective action in \eqref{eq:S_eff_0T}, we can calculate the gauge boson propagators in the $a_0$ and $a_T$ basis. However, since all Feynman diagrams involving gauge fields in our main text are expressed in the Cartesian basis, we can rewrite the effective action in the $a_0$ , $a_x$ and $a_y$ basis as follows \cite{Lee_Mulligan_Conductivity_PRB2023}:

\begin{widetext}

\begin{align}
\label{eq:S_eff_gaugefixed}
    \mathcal{S}[a]  = \frac{T}{2} \sum_{i\epsilon_n} \int \frac{d^2\bm{q}}{(2\pi)^2} \, \begin{pmatrix}
         a_0 & a_x & a_y
     \end{pmatrix}_{-q} \begin{pmatrix}
         \begin{pmatrix}
             \Pi_{00} & - \frac{i q_y}{\abs{\bm{q}}} \Pi_{0T} &  \frac{i q_x}{\abs{\bm{q}}} \Pi_{0T}\\ 
             \frac{i q_y}{\abs{\bm{q}}} \Pi_{T0} & \frac{q_y^2}{\abs{\bm{q}}^2} \Pi_{TT}   &  - \frac{q_x q_y}{\abs{\bm{q}}^2} \Pi_{TT}  \\
             - \frac{i q_x}{\abs{\bm{q}}} \Pi_{T0} & - \frac{q_x q_y}{\abs{\bm{q}}^2} \Pi_{TT} &\frac{q_x^2}{\abs{\bm{q}}^2} \Pi_{TT}
         \end{pmatrix} + \frac{1}{\xi} \begin{pmatrix}
             0 & 0 & 0 \\
             0 & q_x^2 & q_x q_y \\
             0 & q_x q_y & q_y^2
         \end{pmatrix} 
     \end{pmatrix} \begin{pmatrix}
         a_0 \\
         a_x \\
         a_y
     \end{pmatrix}_{q},
\end{align}

where we add the gauge fixing term $\frac{1}{2 \xi} (\grad \cdot \bm{a})^2 $ to the Lagrangian. To derive the gauge boson propagators in \eqref{eq:D_gauge} and \eqref{eq:D_gauge_T}, we first invert the $3 \times 3$ matrix in Eq.~\eqref{eq:S_eff_gaugefixed} and subsequently take the limit $\xi \rightarrow 0$.

\section{Calculation of the Aslamazov-Larkin diagram}
\label{app:AL_diagram_calculation}
In this Appendix, we calculate the four diagrams, both at zero and finite temperature, that contribute to the dashed triangle diagram, which is subsequently used to determine the Aslamazov-Larkin (AL) correction to the thermal Hall conductivity.
\subsection{Dashed triangle diagram in the $\abs{m}/T \rightarrow \infty$ limit}
Here, we first demonstrate the zero temperature calculation for $\Gamma^{\mu 0; \alpha \beta}_{(\text{AL}),1} (p, p+q)$ in Eq.~\eqref{eq:Gamma_AL_1}, which can be written as
\begin{align}
\label{eq:Gamma_AL_1_zeroT}
    \Gamma^{\mu 0; \alpha \beta}_{(\text{AL}),1}(p, p+q)  = {} & - N_f  \left(\frac{1}{\sqrt{N_f}} \right)^2 T \sum_{i\omega_n} \int  \frac{d^2 \bm{k}}{(2\pi)^2} \nonumber\\
    & {} \times \frac{L^{\mu 0; \alpha \beta}_{(\text{AL}),1}(k;p;q)}{\left(\omega_n^2+\bm{k}^2+m^2 \right) \left((\omega_n +\epsilon_n)^2 + (\bm{k}+\bm{q})^2 + m^2\right) \left((\omega_n -\Omega_n)^2 + (\bm{k}-\bm{p})^2 + m^2\right)},
\end{align}
where
\begin{align}
    L^{\mu 0; \alpha \beta}_{(\text{AL}),1}(k;p;q) = & \Tr \bigg[\Gamma^{ \mu 0}_f (k, k+q) \left((i\omega_n+i\epsilon_n) - \bm{\tau} \cdot (\bm{k}+\bm{q}) - m\tau^z \right) \tau^{\beta}\nonumber\\
    &  \left((i\omega_n - i\Omega_n) - \bm{\tau} \cdot (\bm{k}-\bm{p}) - m\tau^z \right) \tau^{\alpha}\left(i\omega_n - \bm{\tau} \cdot \bm{k} - m\tau^z\right) \bigg].
\end{align}

\end{widetext}
In the $\abs{m}/T \rightarrow \infty$ limit, we can replace the summation over the Matsubara frequency with an integral. The integrals can be evaluated using standard Feynman parameter techniques. Using the Feynman parametrization formula
\begin{align}
    \frac{1}{A B C} = 2 \int_0^1 dx \int_0^{1-x} dy \, \frac{1}{\left[  A x + B y + C (1-x-y) \right]^3},
\end{align}

which gives
\begin{align}
\Gamma^{\mu 0; \alpha \beta}_{(\text{AL}),1}(p, p+q) =  -\int \frac{dl_n}{2 \pi} \int  \frac{d^2 \bm{l}}{(2\pi)^2} \, \frac{L^{\mu 0; \alpha \beta}_{(\text{AL}),1}}{(l_n^2 + \bm{l}^2 + \Delta_1^2)^3},
\end{align}
where 
\begin{align}
    l_n = &\, \omega_n + x \epsilon_n - y \Omega_n,\nonumber\\
    \bm{l}  = &\, \bm{k} + x \bm{q} - y \bm{p}, \nonumber\\
    \left(\Delta_1\right)^2  = &\, x(1-x) (\epsilon_n^2 + \bm{q}^2) + y(1-y) (\Omega_n^2 + \bm{p}^2) \nonumber\\
    & + 2 xy \, \epsilon_n \Omega_n + 2 xy \, \bm{q} \cdot \bm{p} + m^2.
\end{align}
The integral is convergent and can be evaluated using dimensional regularization. We expand the result to second order in momenta and, as an example, obtain the following:
\begin{align}
    \Gamma^{0 0; 0 x}_{(\text{AL}),1}(p, p+q) = \frac{1}{24 \pi \abs{m}} \left[3 m p_y - p_x(\epsilon_n+2\Omega_n) \right].
\end{align}
Similarly,
\begin{align}
    \Gamma^{0 0; 0 x}_{(\text{AL}),2}(p, p+q) = \frac{1}{24 \pi \abs{m}} \left[3 m p_y - p_x(\epsilon_n+2\Omega_n) \right].
\end{align}
The computation of the bubble diagrams $\Gamma^{\mu 0; \alpha \beta}_{(\text{AL}),3}(p, p+q)$ and $\Gamma^{\mu 0; \alpha \beta}_{(\text{AL}),4}(p, p+q)$ follows a similar procedure to that of the fermionic vacuum polarization bubble. For example,
\begin{align}
    \Gamma^{0 0; 0 x}_{(\text{AL}),3}(p, p+q) & = -\frac{1}{4 \pi \abs{m}} m p_y + \frac{1}{24 \pi \abs{m}} p_x \Omega_n, \nonumber\\
    \Gamma^{0 0; 0 x}_{(\text{AL}),4}(p, p+q) & = 0.
\end{align}
By combining all of them, we can write the final expression as
\begin{align}
\label{eq:Gamma_AL_sum}
    \Gamma^{0 0; 0 x}_{(\text{AL})} & = \Gamma^{0 0; 0 x}_{(\text{AL}),1} +\Gamma^{0 0; 0 x}_{(\text{AL}),2} + \Gamma^{0 0; 0 x}_{(\text{AL}),3} +\Gamma^{0 0; 0 x}_{(\text{AL}),4} \nonumber\\
    & = - \frac{1}{12 \pi \abs{m}} \left[p_x(\epsilon_n+\Omega_n) \right].
\end{align}

\subsection{Transport contribution from the Kubo formula at finite-$T$}
\label{subsec:Kubo_AL}
We now provide details of the computation of the dashed triangle diagram for general values of $\abs{m}/T$ in the long-wavelength limit ($\epsilon_n\rightarrow0, \abs{\bm{q}}=0$) to obtain the DC response. To proceed, we redefine Eq.~\eqref{eq:Gamma_AL_1_zeroT} as
\begin{align}
     \Gamma^{\mu 0; \alpha \beta}_{(\text{AL}),1}(p, p+q; T)  =- \int  \frac{d^2 \bm{k}}{(2\pi)^2}  \mathcal{S}^{\mu 0; \alpha \beta}_{(\text{AL}),1} (\bm{k},T;\bm{p},i\Omega_n;\bm{q},i\epsilon_n),
\end{align}
where
\begin{widetext}
    \begin{align}
    \label{eq:S_AL_1}
    \mathcal{S}^{\mu 0; \alpha \beta}_{(\text{AL}),1} (\bm{k},T;\bm{p},i\Omega_n;\bm{q},i\epsilon_n)  = - T \sum_{i\omega_{n}}  \frac{L^{\mu 0; \alpha \beta}_{(\text{AL}),1}(\bm{k},i\omega_n;\bm{p},i\Omega_n;\bm{q},i\epsilon_n)}{\left((i\omega_n)^2-E_{\bm{k}}^2 \right) \left((i\omega_n +i\epsilon_n)^2 -E_{\bm{k}+\bm{q}}^2\right) \left((i\omega_n - i\Omega_n)^2 -E_{\bm{k}-\bm{p}}^2\right)}.
\end{align}

Using standard contour methods to perform the summation, we obtain
\begin{align}
\label{eq:S_munu_AL_1}
    \mathcal{S}^{\mu 0; \alpha \beta}_{(\text{AL}),1} &(\bm{k},T;\bm{p},i\Omega_n;\bm{q},i\epsilon_n) \nonumber\\
     ={} &  - \frac{n_F(E_{\bm{k}})}{2E_{\bm{k}}} \frac{L^{\mu 0; \alpha \beta}_{(\text{AL}),1}(\bm{k},E_{\bm{k}};\bm{p},i\Omega_n;\bm{q},i\epsilon_n)}{\left(E_{\bm{k}+\bm{q}}^2-E_{\bm{k}}^2 - 2 i E_{\bm{k}} \epsilon_n + \epsilon_n^2 \right) \left(E_{\bm{k}-\bm{p}}^2-E_{\bm{k}}^2 + 2 i E_{\bm{k}} \Omega_n + \Omega_n^2 \right)} \nonumber\\
   &+\frac{n_F(-E_{\bm{k}})}{2E_{\bm{k}}} \frac{L^{\mu 0; \alpha \beta}_{(\text{AL}),1}(\bm{k},-E_{\bm{k}};\bm{p},i\Omega_n;\bm{q},i\epsilon_n)}{\left(E_{\bm{k}+\bm{q}}^2-E_{\bm{k}}^2 + 2 i E_{\bm{k}} \epsilon_n + \epsilon_n^2 \right) \left(E_{\bm{k}-\bm{p}}^2-E_{\bm{k}}^2 - 2 i E_{\bm{k}} \Omega_n + \Omega_n^2 \right)} \nonumber\\
   &- \frac{n_F(E_{\bm{k}+\bm{q}})}{2E_{\bm{k}+\bm{q}}} \frac{L^{\mu 0; \alpha \beta}_{(\text{AL}),1}(\bm{k},E_{\bm{k}+\bm{q}};\bm{p},i\Omega_n;\bm{q},i\epsilon_n)}{\left(E_{\bm{k}+\bm{q}}^2-E_{\bm{k}}^2 - 2 i E_{\bm{k}+\bm{q}} \epsilon_n - \epsilon_n^2 \right) \left(-E_{\bm{k}-\bm{p}}^2 + E_{\bm{k}+\bm{q}}^2 - 2 i E_{\bm{k}+\bm{q}} \epsilon_n - \epsilon_n^2 - 2 i E_{\bm{k}+\bm{q}} \Omega_n - \Omega_n^2 - 2 \epsilon_n \Omega_n\right)}\nonumber\\
   &+\frac{n_F(-E_{\bm{k}+\bm{q}})}{2E_{\bm{k}+\bm{q}}} \frac{L^{\mu 0; \alpha \beta}_{(\text{AL}),1}(\bm{k},-E_{\bm{k}+\bm{q}};\bm{p},i\Omega_n;\bm{q},i\epsilon_n)}{\left(E_{\bm{k}+\bm{q}}^2-E_{\bm{k}}^2 + 2 i E_{\bm{k}+\bm{q}} \epsilon_n - \epsilon_n^2 \right) \left(-E_{\bm{k}-\bm{p}}^2 + E_{\bm{k}+\bm{q}}^2 + 2 i E_{\bm{k}+\bm{q}} \epsilon_n - \epsilon_n^2 + 2 i E_{\bm{k}+\bm{q}} \Omega_n - \Omega_n^2 - 2 \epsilon_n \Omega_n\right)}\nonumber\\
   &- \frac{n_F(E_{\bm{k}-\bm{p}})}{2E_{\bm{k}-\bm{p}}} \frac{L^{\mu 0; \alpha \beta}_{(\text{AL}),1}(\bm{k},E_{\bm{k}-\bm{p}};\bm{p},i\Omega_n;\bm{q},i\epsilon_n)}{\left(E_{\bm{k}-\bm{p}}^2-E_{\bm{k}}^2 + 2 i E_{\bm{k}-\bm{p}} \Omega_n - \Omega_n^2 \right) \left(E_{\bm{k}-\bm{p}}^2 - E_{\bm{k}+\bm{q}}^2 + 2 i E_{\bm{k}-\bm{p}} \epsilon_n - \epsilon_n^2 + 2 i E_{\bm{k}-\bm{p}} \Omega_n - \Omega_n^2 - 2 \epsilon_n \Omega_n\right)}\nonumber\\
   &+ \frac{n_F(E_{\bm{k}-\bm{p}})}{2E_{\bm{k}-\bm{p}}} \frac{L^{\mu 0; \alpha \beta}_{(\text{AL}),1}(\bm{k},-E_{\bm{k}-\bm{p}};\bm{p},i\Omega_n;\bm{q},i\epsilon_n)}{\left(E_{\bm{k}-\bm{p}}^2-E_{\bm{k}}^2 - 2 i E_{\bm{k}-\bm{p}} \Omega_n - \Omega_n^2 \right) \left(E_{\bm{k}-\bm{p}}^2 - E_{\bm{k}+\bm{q}}^2 - 2 i E_{\bm{k}-\bm{p}} \epsilon_n - \epsilon_n^2 - 2 i E_{\bm{k}-\bm{p}} \Omega_n - \Omega_n^2 - 2 \epsilon_n \Omega_n\right)}.
\end{align}
\end{widetext}
Similarly, we can also perform Matsubara summation for $\Gamma^{\mu 0; \alpha \beta}_{(\text{AL}),2}(p, p+q)$ also. The calculation of the bubble diagrams $\Gamma^{\mu 0; \alpha \beta}_{(\text{AL}),3}(p, p+q)$ and $\Gamma^{\mu 0; \alpha \beta}_{(\text{AL}),4}(p, p+q)$ at finite temperature follows the same procedure as the fermionic vacuum polarization bubble.

At $T=0$, $n_F({E_{\bm{k}}})=0$, so the result for the zero-temperature case, $\Gamma^{\mu 0; \alpha \beta}_{(\text{AL})} (\bm{p},i\Omega_n;\bm{q},i\epsilon_n; T=0)$ aligns with that obtained via dimensional regularization, as shown in Eqs.~\eqref{eq:Gamma_AL_00_zeroT} and \eqref{eq:Gamma_AL_i0_zeroT}. Therefore at finite $T$ we can write
\begin{align}
    \Gamma^{\mu 0; \alpha \beta}_{f,(\text{AL})} (\bm{p},i\Omega_n;\bm{q},i\epsilon_n; T) = & \Gamma^{\mu 0; \alpha \beta}_{(\text{AL})} (\bm{p},i\Omega_n;\bm{q},i\epsilon_n; T=0) \nonumber\\
    & + \delta\Gamma^{\mu 0; \alpha \beta}_{f,(\text{AL})} (\bm{p},i\Omega_n;\bm{q},i\epsilon_n).
\end{align}
Before carrying out the integration over $\bm{k}$, a few remarks are necessary. First, after completing the Matsubara summation, integrating over $\bm{k}$ becomes challenging with both $p$ and $q$ nonzero. However, since our main focus in this section is the transport contribution from the Kubo formula, we first take the long-wavelength limit  ($\epsilon_n\rightarrow0, \abs{\bm{q}}=0$) in Eq.~\eqref{eq:S_munu_AL_1}. As we observed earlier, the fermion polarization tensor can generally be decomposed into two independent transverse projectors. Here, we explore whether a similar decomposition into independent transverse projectors is possible for the triangle diagram. 

To proceed, we first take the long-wavelength limit  ($\epsilon_n\rightarrow0, \abs{\bm{q}}=0$) in Eq.~\eqref{eq:Gamma_AL_00_zeroT} to extract the zero-$T$ part which is given by
\begin{align}
    \Gamma^{0 0; \alpha \beta}_{f,(\text{AL})} &(\bm{p},i\Omega_n;\bm{q}=0,i\epsilon_n \rightarrow 0; T=0) \nonumber\\
    & = \frac{1}{12\pi\abs{m}} \begin{pmatrix}
        \bm{p}^2 & - p_x \Omega_n & -p_y \Omega_n \\
        - p_x \Omega_n & -p_y^2 + \Omega_n^2 & p_x p_y \\
        -p_y \Omega_n & p_x p_y & -p_x^2 + \Omega_n^2
    \end{pmatrix}.
\end{align}
Next, we notice that the  dashed triangle diagram satisfies the Ward identities for the gauge boson vertices
\begin{align}
    p_\alpha \Gamma^{\mu 0; \alpha \beta}_{(\text{AL})} (\bm{p},i\Omega_n;\bm{q},i\epsilon_n; T=0) & = 0, \nonumber\\
    (p+q)_\beta \Gamma^{\mu 0; \alpha \beta}_{(\text{AL})} (\bm{p},i\Omega_n;\bm{q},i\epsilon_n; T=0) & = 0.
\end{align}
In the $q \rightarrow 0$ limit, these identities reduce to 
\begin{align}
    p_\alpha \Gamma^{\mu 0; \alpha \beta}_{(\text{AL})} (\bm{p},i\Omega_n;\bm{q}=0,i\epsilon_n \rightarrow 0; T=0) & = 0, \nonumber\\
    p_\beta \Gamma^{\mu 0; \alpha \beta}_{(\text{AL})} (\bm{p},i\Omega_n;\bm{q}=0,i\epsilon_n \rightarrow 0; T=0) & = 0.
\end{align}
Following the Ward identities, we can decompose the tensor structure of $\Gamma^{0 0; \alpha \beta}_{f,(\text{AL})} (\bm{p},i\Omega_n;\bm{q}=0,i\epsilon_n \rightarrow 0; T=0)$ into two independent tensors
\begin{align}
    \Gamma^{0 0; \alpha \beta}_{f,(\text{AL})} (\bm{p},i\Omega_n; q \rightarrow 0; T=0) = \frac{1}{12 \pi \abs{m}} \left( P_1^{\alpha\beta} - P_2^{\alpha\beta} \right),
\end{align}
where $P_1^{\alpha\beta}$ and $P_2^{\alpha\beta}$ are defined in Eq.~\eqref{eq:stress_projector}.

As we are concentrating on the $q \rightarrow 0$ case, we will omit $(\bm{q},i\epsilon_n)$ from the arguments and use a superscript/subscript $L$ to denote that we are working in the long-wavelength limit. Together with the stress-tensor vertex from the Maxwell-Chern-Simons theory, which arises from the heavy Dirac fermion, we can now express the general form of  $\Gamma^{0 0; \alpha \beta}_{(\text{AL}),L} (\bm{p},i\Omega_n; T)$ at finite temperature, derived from the Ward identity, as follows:
\begin{align}
\label{eq:Gamma_00_kubo_def}
    \Gamma^{0 0; \alpha \beta}_{(\text{AL}),L} (\bm{p},i\Omega_n; T) = {} & A_{0;1}^L P_1^{\alpha\beta} (p) + A_{0;2}^L P_2^{\alpha\beta}(p) \nonumber\\
    & + A_{0;3}^L P_3^{\alpha\beta}(p).
\end{align}

We begin by taking the long-wavelength limit $\Gamma^{i 0; \alpha \beta}_{f,(\text{AL})} (\bm{p},i\Omega_n;\bm{q},i\epsilon_n; T)$ to extract the zero-temperature contribution, which is given by:
\begin{align}
\label{eq:Gamma_i0_kubo_def}
    \Gamma^{x 0; \alpha \beta}_{f,(\text{AL})} (\bm{p},i\Omega_n;\bm{q}=0,i\epsilon_n \rightarrow 0; T=0) & = \frac{i}{12\pi\abs{m}} \mathcal{X}^{\alpha\beta}(p), \nonumber\\
    \Gamma^{y 0; \alpha \beta}_{f,(\text{AL})} (\bm{p},i\Omega_n;\bm{q}=0,i\epsilon_n \rightarrow 0; T=0) & = \frac{i}{12\pi\abs{m}} \mathcal{Y}^{\alpha\beta}(p),
\end{align}
where $\mathcal{X}^{\alpha\beta}$ and $\mathcal{Y}^{\alpha\beta}$ are defined in Eq.~\eqref{eq:stress_projector}. Unfortunately, the above tensor structure can not be decomposed into other independent tensors. However, we obtain the general form of $\Gamma^{i 0; \alpha \beta}_{(\text{AL}), L} (\bm{p},i\Omega_n; T)$ at finite temperature, which can be derived from the Ward identity and is given by:
\begin{align}
    \Gamma^{x 0; \alpha \beta}_{(\text{AL}),L} (\bm{p},i\Omega_n; T) = & A_{x;1}^L P_1^{\alpha\beta}(p) + A_{x;2}^L P_2^{\alpha\beta}(p) \nonumber\\
    & + A_{x;3}^L P_3^{\alpha\beta}(p) + B_x^L \mathcal{X}^{\alpha\beta}(p), \nonumber\\
    \Gamma^{y 0; \alpha \beta}_{(\text{AL}),L} (\bm{p},i\Omega_n; T) = & A_{y;1}^L P_1^{\alpha\beta}(p) + A_{y;2}^L P_2^{\alpha\beta}(p) \nonumber\\
    & + A_{y;3}^L P_3^{\alpha\beta}(p) + B_y^L \mathcal{Y}^{\alpha\beta}(p).
\end{align}
Since we have determined the general structure of the dashed triangle diagram at finite-$T$, the calculation of the coefficients $A_{\mu;1}^L, A_{\mu;2}^L, A_{\mu;3}^L$, and $B_i^L$ at finite-$T$ will closely follow the procedure for the fermion polarization tensor, as detailed in Appendix \ref{app:effective_action}. In this case, too, the behavior depends on the ratio $b=\Omega_n/\abs{\bm{p}}$.

We are now ready to compute $\Pi^{x 0; y 0}_{(\text{AL})}(\bm{q},i \epsilon_n)$ in Eq.~\eqref{eq:Pi_AL_def}, utilizing the dashed triangle diagram at finite-$T$ from Eqs.~\eqref{eq:Gamma_00_kubo_def} and \eqref{eq:Gamma_i0_kubo_def}, along with the gauge propagator at finite-$T$. To obtain the thermal Hall conductivity, we need to extract the component that is antisymmetric in $xy$. Additionally, to derive the contribution from the Kubo formula, we must first take the limit $\bm{q} \rightarrow 0$. By doing so, we obtain
\begin{align}
    \Pi^{x 0; y 0}_{\text{AS},(\text{AL})}(\bm{q} \rightarrow 0,i \epsilon_n) = 0,
\end{align}
and therefore, $\kappa^{\text{Kubo}}_{xy} = 0$.

\subsection{Transport contribution from the energy magnetization at finite-$T$}

The computation of energy magnetization is carried out using a procedure similar to that of Kubo conductivity, with the added consideration of the static limit. To evaluate $ \Gamma^{\mu 0; \alpha \beta}_{(\text{AL})}$, we first perform the Matsubara sum over internal fermionic frequency $\omega_n$ at $\epsilon_n = 0$, followed by taking the limit as $\bm{q} \rightarrow 0$. We now express the general form of the dashed triangle diagram using the projectors, as outlined in Subsection~\ref{subsec:Kubo_AL}. We calculate the coefficients $A_{\mu;1}^M, A_{\mu;2}^M, A_{\mu;3}^M$, and $B_i^M$ in Eq.~\eqref{eq:Gamma_mu0_mag_def} at finite temperature proceeds similarly to the method used for calculating the fermion polarization tensor, as shown in Appendix~\ref{app:effective_action}. Here as well, the dependence on the parameter ratio $b=\Omega_n/\abs{\bm{p}}$ plays a key role. Taking into account that the stress-tensor vertices for the heavy Dirac fermion are derived from the effective Maxwell-Chern-Simons theory, we obtain the result shown in Eq.~\eqref{eq:stress_coeff_mag}.

With the finite-$T$ stress-tensor vertices in hand, we now evaluate the right-hand side of the differential equation in Eq.~\eqref{eq:EnergyMag_def_AL}. To evaluate, $\Pi^{0 0; i 0}_{\text{AS},(\text{AL})}$, we sum over $\Omega_{n}$ at $\epsilon_n = 0$. In the $\bm{q} \rightarrow 0$ limit, the full integrand can be expanded in powers of $q_x$ and $q_y$.  We then retain only the terms of  $\mathcal{O}(q_x) \left(\mathcal{O}(q_y)\right)$ in the correlator $\Pi^{0 0; y 0}_{\text{AS},(\text{AL})} \left(\Pi^{0 0; x 0}_{\text{AS},(\text{AL})}\right)$  that are relevant after taking the partial derivative with respect to $q_x(q_y)$. Here, we use the gauge propagator in Eq.~\eqref{eq:D_gauge_T}. By doing so, we obtain
\begin{align}
   \Tilde{M}_Q = &  \frac{1}{2 i} \bigg( \partial_{q_x} \Pi^{0 0; y 0}_{\text{AS},(\text{AL})}  - \partial_{q_y} \Pi^{0 0; x 0}_{\text{AS},(\text{AL})} \bigg) \bigg\vert_{\bm{q} \rightarrow 0} \nonumber\\
   & = T \sum_{i\Omega_n} \int \frac{d^2\bm{p}}{(2\pi)^2} \frac{\Tilde{m}_t}{2 g_a} \left[ \frac{g_b \bm{p}^2 - 2 g_a \Omega_n^2}{\left( \Omega_n^2 + (g_b/g_a) \bm{p}^2 + \Tilde{m}_t^2 \right)^2}  \right]\nonumber\\
   & = \frac{g_a \Tilde{m}_t}{2 g_b} \left[I_1 - 3 I_2 - \Tilde{m}_t^2 I_3  \right],
\end{align}
where
\begin{align}
\label{eq:I_123_ainf}
    I_1 & = \left( \frac{g_b}{g_a}\right)  T \sum_{i\Omega_n} \int \frac{d^2\bm{p}}{(2\pi)^2} \frac{1}{\left( \Omega_n^2 + (g_b/g_a) \bm{p}^2 + \Tilde{m}_t^2  \right)}, \nonumber\\
    I_2 & = \left( \frac{g_b}{g_a}\right)  T \sum_{i\Omega_n} \int \frac{d^2\bm{p}}{(2\pi)^2} \frac{\Omega_n^2}{\left( \Omega_n^2 + (g_b/g_a) \bm{p}^2 + \Tilde{m}_t^2  \right)^2}, \nonumber\\
    I_3 & = \left( \frac{g_b}{g_a}\right)  T \sum_{i\Omega_n} \int \frac{d^2\bm{p}}{(2\pi)^2} \frac{1}{\left( \Omega_n^2 + (g_b/g_a) \bm{p}^2 + \Tilde{m}_t^2  \right)^2}.
\end{align}
In the above expressions, we use the long wavelength limits for $g_{a(b)}$ as explained in Section~\ref{sec:kxy_DCS}.

Let us begin by calculating $I_1$, which can be written as
\begin{align}
\label{eq:I1_ainf}
     I_1  = T \sum_{i\Omega_n} \int \frac{d^2\bm{p}}{(2\pi)^2} \frac{1}{\left( \Omega_n^2 + \bm{p}^2 + \Tilde{m}_t^2  \right)},
\end{align}
where we shift the loop momenta $\bm{p} \rightarrow \sqrt{g_a/g_b}\, \bm{p} $.
Using the following summation identity,
\begin{equation}
    \sum_{n=-\infty}^{\infty} \frac{1}{n^2 + a^2} = \frac{\pi}{a}\, \coth{(\pi a)},
\end{equation}
Eq.~\eqref{eq:I1_ainf} can be written as
\begin{align}
\label{eq:I1_ainf_break}
    I_1  = {} & \frac{T}{(2\pi T)^2} \int \frac{d^2\bm{p}}{(2\pi)^2} \sum_n \frac{1}{n^2 + \left( \sqrt{\bm{p}^2 + \Tilde{m}_t^2} / 2\pi T\right)^2} \nonumber \\
     = {} & \frac{1}{2} \int \frac{d^2\bm{p}}{(2\pi)^2} \frac{1}{\sqrt{\bm{p}^2 + \Tilde{m}_t^2}} \left( \coth{\left(\frac{\sqrt{\bm{p}^2 + \Tilde{m}_t^2}}{2 T}\right)} - 1 \right) \nonumber\\
     & {} + \frac{1}{2} \int \frac{d^2\bm{p}}{(2\pi)^2} \frac{1}{\sqrt{\bm{p}^2 + \Tilde{m}_t^2}}
\end{align}
The first integral is convergent, and the second, which is divergent, will be evaluated using dimensional regularization. The first integral may be evaluated as follows:
\begin{align}
\label{eq:I1_ainf_convergent}
    \frac{1}{2}  \int & \frac{d^2\bm{p}}{(2\pi)^2}  \frac{1}{\sqrt{\bm{p}^2 + \Tilde{m}_t^2}} \left( \coth{\left(\frac{\sqrt{\bm{p}^2 + \Tilde{m}_t^2}}{2 T}\right)} - 1 \right) \nonumber \\
     ={} & \frac{1}{4\pi} \int_{\abs{\Tilde{m}_t}}^\infty dy \left( \coth{\left(\frac{y}{2 T}\right)} - 1 \right) \nonumber\\
     = {} & \frac{1}{4\pi} \left\{ \lim_{\Lambda \to \infty} \Bigl(2 T \ln{\left[\sinh{\left(\frac{\Lambda}{2 T}\right)}\right]} -\Lambda \Bigr) \right. \nonumber \\
  &\left. - 2 T \ln{\left[\sinh{\left(\frac{\abs{\Tilde{m}_t}}{2 T}\right)} + \right]} + \abs{\Tilde{m}_t} \right\} \nonumber \\
    ={} & - \frac{T}{2\pi} \ln{\left[2\sinh{\left(\frac{\abs{\Tilde{m}_t}}{2 T}\right)} \right]} + \frac{\abs{\Tilde{m}_t}}{4\pi},
\end{align}
Using the general result for dimensional regularization,
\begin{align}
    \int \frac{d^d l}{(2\pi)^d} \frac{l^a}{(l^2+\Delta^2)^b} ={}& \frac{1}{(4\pi)^{d/2}} \, \Delta^{(a-2b+d)/2}\, \, \nonumber\\
    & \times \frac{\Gamma{\left(b-\frac{a+d}{2}\right)} \, \Gamma{\left(\frac{a+d}{2}\right)}}{\Gamma{\left(\frac{d}{2}\right)} \, \Gamma{(b)}},
\end{align}
the second divergent integral in Eq.~\eqref{eq:I1_ainf_break} can be expressed as
\begin{align}
\label{eq:I1_ainf_divergent}
   \frac{1}{2} \int \frac{d^2 \bm{p}}{(2\pi)^2}\, \frac{1}{\sqrt{\bm{p}^2+\abs{\Tilde{m}_t}^2}} \rightarrow & \frac{1}{2} \int \frac{d^{2-\epsilon} \bm{p}}{(2\pi)^{2-\epsilon}}\, \frac{1}{\sqrt{\bm{p}^2+\abs{\Tilde{m}_t}^2}} \nonumber\\
   & = - \frac{\abs{\Tilde{m}_t}}{4\pi}.
\end{align}
Thus, combining Eqs.~\eqref{eq:I1_ainf_convergent} and \eqref{eq:I1_ainf_divergent}, we get
\begin{align}
\label{eq:I1_ainf_final}
   I_1 = - \frac{T}{2\pi} \ln{\left[2\sinh{\left(\frac{\abs{\Tilde{m}_t}}{2 T}\right)} \right]}.
\end{align}
The remaining two integrals in Eq.~\eqref{eq:I_123_ainf} can be derived straightforwardly using a similar approach:
\begin{align}
\label{eq:I23_ainf_final}
    I_2 & = -\frac{\abs{\Tilde{m}_t}}{8\pi} \coth{\left(\frac{\abs{\Tilde{m}_t}}{2 T}\right)}, \nonumber\\
    I_3 & = \frac{1}{8\pi \abs{\Tilde{m}_t}} \coth{\left(\frac{\abs{\Tilde{m}_t}}{2 T}\right)}.
\end{align}

\bibliography{bibliography}

\end{document}